\documentclass{article}

\usepackage[preprint, nonatbib]{neurips_2024}

\usepackage[utf8]{inputenc} % allow utf-8 input
\usepackage[T1]{fontenc}    % use 8-bit T1 fonts
\usepackage{hyperref}       % hyperlinks
\usepackage{url}            % simple URL typesetting
\usepackage{booktabs}       % professional-quality tables
\usepackage{amsfonts}       % blackboard math symbols
\usepackage{nicefrac}       % compact symbols for 1/2, etc.
\usepackage{microtype}      % microtypography
\usepackage{xcolor}         % colors
\usepackage{graphicx}
\usepackage{booktabs}
\usepackage{array}
\usepackage{longtable}
\usepackage{tabularx}
\usepackage{multirow}
\usepackage{makecell}
\usepackage{pifont}
\usepackage{fontawesome5}
\usepackage{subfigure}
\usepackage{CJK}
\usepackage{amsmath}
\usepackage[normalem]{ulem}
\useunder{\uline}{\ul}{}
\usepackage{graphicx,wrapfig}
\usepackage{colortbl} 
\usepackage[normalem]{ulem}
\useunder{\uline}{\ul}{}
\newcommand{\yj}[1]{\textcolor[rgb]{1,0,0}{#1}}

\usepackage[preprint]{neurips_2024}
\usepackage[utf8]{inputenc} % allow utf-8 input
\usepackage[T1]{fontenc}    % use 8-bit T1 fonts
\usepackage{hyperref}       % hyperlinks
\usepackage{url}            % simple URL typesetting
\usepackage{booktabs}       % professional-quality tables
\usepackage{amsfonts}       % blackboard math symbols
\usepackage{nicefrac}       % compact symbols for 1/2, etc.
\usepackage{microtype}      % microtypography
\usepackage{xcolor}         % colors
\usepackage{graphicx}
\usepackage{booktabs}
\usepackage{array}
\usepackage{float}
\usepackage{supertabular}
\usepackage{adjustbox}
\usepackage{nicematrix}
\usepackage{xcolor}
\usepackage{hyperref}
\usepackage{tcolorbox}
\usepackage{caption}
\usepackage{colortbl}
\usepackage[normalem]{ulem}

\useunder{\uline}{\ul}{}
\newcolumntype{C}[1]{>{\centering\arraybackslash}p{#1}}
\newcolumntype{L}[1]{>{\raggedright\hangindent=1em\arraybackslash}p{#1}}
\newcommand{\boxedgreen}[1]{\fcolorbox{green}{green}{#1}}
\newcommand{\boxedyellow}[1]{\fcolorbox{yellow}{yellow}{#1}}

\title{GMAI-MMBench: A Comprehensive Multimodal Evaluation Benchmark Towards General Medical AI}

\author{%
    Pengcheng Chen\textsuperscript{1,2}\thanks{These authors contributed equally to this work.}\quad
    Jin Ye\textsuperscript{1,3}$^\ast$\thanks{Corresponding authors: jin.ye@monash.edu, hejunjun@pjlab.org.cn, qiaoyu@pjlab.org.cn}\quad
    Guoan Wang\textsuperscript{1,4}\footnotemark[1]\quad
    Yanjun Li\textsuperscript{1,4}\quad \\
    \textbf{Zhongying Deng}\textsuperscript{\textbf{5}}\quad
    \textbf{Wei Li}\textsuperscript{\textbf{1,6}}\quad
    \textbf{Tianbin Li}\textsuperscript{\textbf{1}}\quad 
    \textbf{Haodong Duan}\textsuperscript{\textbf{1}}\quad \\
    \textbf{Ziyan Huang}\textsuperscript{\textbf{1,6}}\quad 
    \textbf{Yanzhou Su}\textsuperscript{\textbf{1}}\quad
    \textbf{Benyou Wang}\textsuperscript{\textbf{7,8}}\quad
    \textbf{Shaoting Zhang}\textsuperscript{\textbf{1}}\quad \\
    \textbf{Bin Fu}\textsuperscript{\textbf{9}}\quad
    \textbf{Jianfei Cai}\textsuperscript{\textbf{3}}\quad
    \textbf{Bohan Zhuang}\textsuperscript{\textbf{3}}\quad
    \textbf{Eric J Seibel}\textsuperscript{\textbf{2}}\quad
    \textbf{Junjun He}\textsuperscript{\textbf{1}}$^\dagger$\quad
    \textbf{Yu Qiao}\textsuperscript{\textbf{1}}$^\dagger$ \\
    \textsuperscript{1}Shanghai AI Laboratory\quad
    \textsuperscript{2}University of Washington\quad
    \textsuperscript{3}Monash University\\
    \textsuperscript{4}East China Normal University\quad
    \textsuperscript{5}University of Cambridge\quad
    \textsuperscript{6}Shanghai Jiao Tong University\\
    \textsuperscript{7}The Chinese University of Hong Kong, Shenzhen\quad
    \textsuperscript{8}Shenzhen Research Institute of Big Data\\
    \textsuperscript{9}Shenzhen Institute of Advanced Technology (SIAT), Chinese Academy of Sciences\\
}

\hypersetup{
    colorlinks=true,
    linkcolor=blue,
    urlcolor=blue,
}

\begin{document}

%\tableofcontents
\maketitle

% \begin{bibunit}

\begin{figure*}[htbp]
\centering
\includegraphics[width=1.0\textwidth]{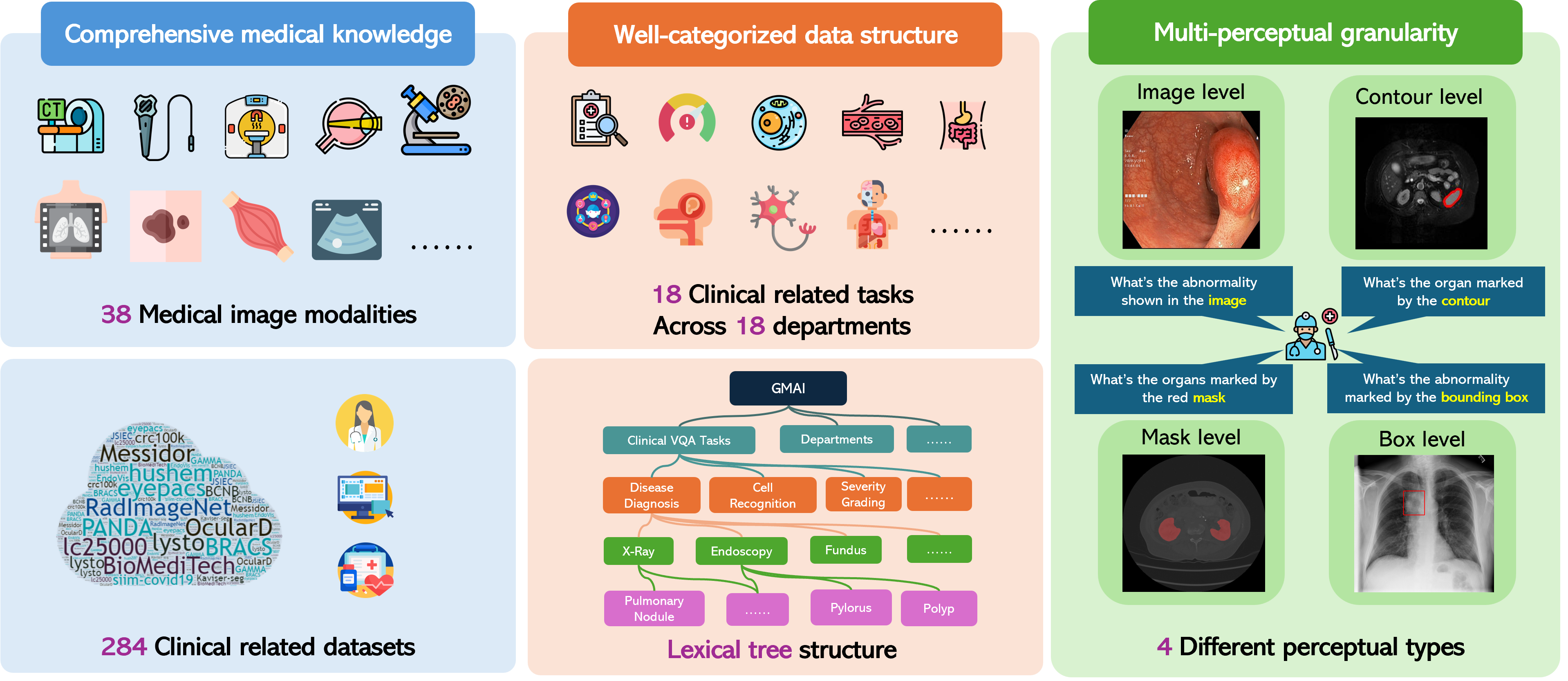}
\caption{Overview of the GMAI-MMBench. The benchmark is meticulously designed for testing LVLMs' abilities in real-world clinical scenarios with three key features: (1) Comprehensive medical knowledge: It consists of 284 diverse clinical-related datasets from worldwide sources, covering 38 modalities. (2) Well-categorized data structure: It features 18 clinical VQA tasks and 18 clinical departments, meticulously organized into a lexical tree. (3) Multi-perceptual granularity: Interactive methods span from image to region level, offering varying degrees of perceptual details.}
\label{fig:cover}
\end{figure*}

% \newpage
% \tableofcontents
% \newpage

\begin{abstract}
Large Vision-Language Models (LVLMs) are capable of handling diverse data types such as imaging, text, and physiological signals, and can be applied in various fields. In the medical field, LVLMs have a high potential to offer substantial assistance for diagnosis and treatment. Before that, it is crucial to develop benchmarks to evaluate LVLMs' effectiveness in various medical applications. Current benchmarks are often built upon specific academic literature, mainly focusing on a single domain, and lacking varying perceptual granularities. Thus, they face specific challenges, including limited clinical relevance, incomplete evaluations, and insufficient guidance for interactive LVLMs. To address these limitations, we developed the GMAI-MMBench, the most comprehensive general medical AI benchmark with well-categorized data structure and multi-perceptual granularity to date. It is constructed from 284 datasets across 38 medical image modalities, 18 clinical-related tasks, 18 departments, and 4 perceptual granularities in a Visual Question Answering (VQA) format. Additionally, we implemented a lexical tree structure that allows users to customize evaluation tasks, accommodating various assessment needs and substantially supporting medical AI research and applications. We evaluated 50 LVLMs, and the results show that even the advanced GPT-4o only achieves an accuracy of 53.96\%, indicating significant room for improvement. 
% Moreover, we identified 5 main insufficiencies to be addressed in the next-generation LVLMs. Addressing them can advance the development of cutting-edge LVLMs for medical applications. 
Moreover, we identified five key insufficiencies in current cutting-edge LVLMs that need to be addressed to advance the development of better medical applications. 
We believe that GMAI-MMBench will stimulate the community to build the next generation of LVLMs toward GMAI.

\begin{flushleft}
    \raisebox{-0.25\height}{\includegraphics[height=12pt]{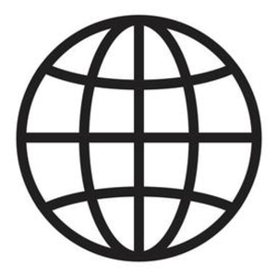}} \href{https://uni-medical.github.io/GMAI-MMBench.github.io/}{Website: https://uni-medical.github.io/GMAI-MMBench.github.io/}\\
    
    \raisebox{-0.25\height}{\includegraphics[height=12pt]{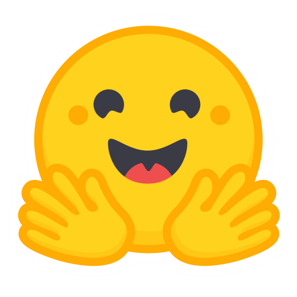}} \href{https://huggingface.co/datasets/OpenGVLab/GMAI-MMBench}{Huggingface: https://huggingface.co/datasets/OpenGVLab/GMAI-MMBench}\\
    
    \raisebox{-0.25\height}{\includegraphics[height=12pt]{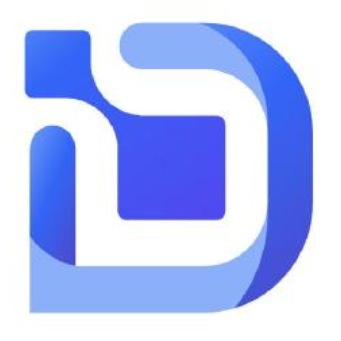}} \href{https://opendatalab.com/GMAI/MMBench}{OpenDataLab: https://opendatalab.com/GMAI/MMBench}\\    
    
    \raisebox{-0.25\height}{\includegraphics[height=12pt]{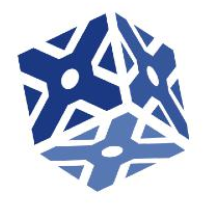}} \href{https://github.com/open-compass/VLMEvalKit}{Evaluation: https://github.com/open-compass/VLMEvalKit~\cite{duan2024vlmevalkit}}\\    
\end{flushleft}

\end{abstract}

\section*{Introduction}
In clinical practice, diverse demands may be proposed by different medical institutions for disease diagnosis and treatment. These demands can be potentially fulfilled by general medical AI which provides general-purpose medical models to tackle a wide range of medical tasks. Such models are typically Large Vision-Language Models (LVLMs) trained on diverse data types, including imaging and clinical texts, to tackle diverse tasks, e.g., disease diagnosis and severity grading. Noticeably, the state-of-the-art LVLMs, including general-purpose ones (e.g., DeepSeek-VL~\cite{lu2024deepseek}, GPT-4V~\cite{achiam2023gpt} and Claude3-Opus~\cite{anthropic2024claude}) and medical purposes (like MedDr~\cite{he2024meddr}, LLaVA-Med~\cite{li2024llava}, and Med-Flamingo~\cite{moor2023med}), have both demonstrated promising performance in some medical visual-textual tasks. However, it remains unclear to what extent these LVLMs can accommodate the diverse demands in real clinical scenarios. To validate their effectiveness and promote their application in clinical practice, it is crucial to establish a comprehensive benchmark to address diverse real-world demands. Therefore, an ideal benchmark should achieve three specific aims:

\textbf{Aim 1. Comprehensive medical knowledge.} Medical knowledge is embedded in medical data, so comprehensive medical knowledge requires diverse medical data of different modalities from various data sources.
In clinical scenarios, various types of imaging modalities, including X-rays, Computed Tomography (CT), Magnetic Resonance Image (MRI), Ultrasound Imaging, Positron Emission Tomography (PET), etc, are employed for diagnostic and therapeutic purposes, reflecting different aspects of medical knowledge~\cite{zhang2023challenges}. 
% These different image modalities have large image-level variations~\cite{zhang2023challenges}. 
% \sout{simulate clinical scenarios}}
Besides, to encompass the diverse medical knowledge from different clinical facilities, the data used in a comprehensive benchmark should cover a range of different clinical institutions and hospitals which are preferably distributed across the world~\cite{qu2024abdomenatlas}. These demands favor benchmarks collected from diverse sources. 
\textbf{Aim 2. Comprehensive evaluation across all clinical aspects.} A comprehensive benchmark should be easily customized to evaluate any specific abilities of LVLMs for each clinical professional. This property is necessary because there are an excessive amount of clinical institutions, departments, and practitioners, each having their own specific demand. 
%The detailed purposes
Their potential demands can be concluded in two sides: 
1) \emph{Evaluation across diverse tasks}. Some clinical practitioners may require MRI data for disease diagnosis while others may need to deal with surgical workflow recognition for computer-assisted or robot-assisted surgery systems. Therefore, a comprehensive benchmark should cover all clinical demands by encompassing a sufficient number of diseases and tasks.
2) \emph{Evaluation for diverse clinical departments}. Some departments may be interested in LVLMs' performance on oncology-related tasks only while others may only focus on urology-related ones. 
As such, a comprehensive benchmark should be easily used for customized evaluation to accommodate the diverse demands of different clinical departments. These demands further require the benchmark to be well-categorized to facilitate ease of use.
\textbf{Aim 3. Interactive ability in multi-perceptual granularity.} Given a specific medical image, doctors need to look through the whole image (image level) for an overview while also requiring comprehensive explanations in a specific position (mask level) or region (box level). This demand requires LVLMs to perceive the granularity range from a specific position to the entire image. Thus, a comprehensive benchmark should also evaluate LVLMs' perceptual granularity. 

As shown in Table~\ref{tab:comparison_exist_benchmark}, there are some medical benchmarks, such as Medical-Diff-VQA~\cite{hu2023expert}, PathVQA~\cite{he2020pathvqa}, Cholec80-VQA~\cite{seenivasan2022surgical}, and Cholec80~\cite{twinanda2016endonet}, dedicated to evaluating specific abilities of LVLMs. These benchmarks effectively assess the performance of LVLMs within a particular modality or task, thereby facilitating the optimization of models for specific applications. Nonetheless, their limited modalities and tasks cannot meet the requirement of modal and task diversity. Other benchmarks including VQA-RAD~\cite{lau2018dataset}, RadBench~\cite{wu2023generalist}, and MMMU (Health \& Medicine)~\cite{yue2023mmmu} address this issue by providing multiple modalities and tasks for evaluation, with data consisting of natural image-text pairs sourced from academic papers, textbooks, and specific databases. Though these benchmarks significantly enhance the breadth and depth of medical assessment, they may not accurately reflect actual clinical requirements, as their sources are distant from clinic practice and prone to data leakage~\cite{chen2024we,fu2024mme}. 
% Other benchmarks like SLAKE~\cite{liu2021slake} and OmniMedVQA~\cite{hu2024omnimedvqa} collect data from clinical 
% Additionally, the number of medical tasks evaluated by these benchmarks is relatively limited (typically fewer than ten). 
% \bohan{the logic here is confusing. You argue that MMMU solves modal and task diversity at the beginning then claim their tasks is limited...}
% As such, their sources and tasks are not sufficiently diverse. 
More importantly, \emph{none of these benchmarks can be customized to evaluate various abilities of LVLMs to accommodate highly diverse clinical demands} because their data are not well categorized. For instance, it is hard to obtain the dimension, modality, and task information of a specific data point in these datasets, which prevents a clinical professional from evaluating LVLMs using the CT (\underline{modality}) of 2D (\underline{dimension}) images for blood vessel recognition (\underline{task}). Due to this, they can hardly be used for customized evaluation. 
%Furthermore, the majority of medical-related benchmarks do not involve annotated data. Typically, these benchmarks only present an image without annotation marking and require the LVLMs to observe and infer conclusions. However, for interactive applications, physicians' input is a fundamental requirement for marking the specific regions for diagnosis or judgment. Consequently, it is imperative to include evaluations that consider various annotation marking conditions. This approach can reflect the model's stability across different annotation marking methods, thereby informing the development of interactive multimodal large models.
In summary, though existing medical multimodal benchmarks provide valuable evaluation frameworks, they present challenges in fully addressing clinical needs. Future developments necessitate more refined and customized benchmarks that are closely aligned with real-world clinical applications. %Such tools are essential for advancing the practical application and development of medical artificial intelligence technologies.

\begin{table*}[t]
\caption{Comparison between GMAI-MMBench and other existing benchmarks in the biomedical field. GMAI-MMBench is sourced from extensive data sources worldwide, offering comprehensive medical knowledge detailed in modalities, clinical tasks, departments, and perceptual granularities. Dept and PG indicate department and perceptual granularity, respectively. In the perceptual granularity types, I, B, M, and C denote image, box, mask, and contour, respectively. $^{*}$ indicates the test set.}
\resizebox{1.0\textwidth}{!}{
\begin{tabular}{l|cccccc}
\hline
\textbf{Benchmark} & \textbf{Modality} & \textbf{Size} & \textbf{Task} & \textbf{Dept} & \textbf{PG} & \textbf{Source} \\ \hline
% General benchmark
% Single domain
Medical-Diff-VQA$^{*}$~\cite{hu2023expert} & 1 &  70K & 7 & \ding{56} & I & MIMIC-CXR~\cite{johnson2019mimic} \\
PathVQA$^{*}$~\cite{he2020pathvqa} & 1 & 6K & 7 & \ding{56} & I & Textbook, PEIR~\cite{peir} \\
Cholec80-VQA$^{*}$~\cite{seenivasan2022surgical} & 1 & 9K & 2 & \ding{56} & I & Cholec80~\cite{twinanda2016endonet} \\
% Limited diversity
VQA-RAD~\cite{lau2018dataset} & 3 & 3K & 11 & \ding{56} & I & Teaching cases from Medpix~\cite{medpix} \\
RadBench~\cite{wu2023generalist} & 6 & 137K & 5 & \ding{56} & I & 13 image-text paired datasets \\
MMMU (H \& M)~\cite{yue2023mmmu} & 6 & 2K & 5 & \ding{56} & I, B & Exam, Quiz, Textbook \\
SLAKE$^{*}$~\cite{liu2021slake} & 3 & 2K & 10 & \ding{56} & I & MSD~\cite{simpson2019large}, Chestx-ray8~\cite{wang2017chestx}, CHAOS~\cite{kavur2021chaos} \\
OmniMedVQA~\cite{hu2024omnimedvqa} & 12 & 128K & 5 & \ding{56} & I &  73 classification datasets \\
% MIMIC-CXR$^{*}$~\cite{johnson2019mimic} & 1 & 5K  & 1 & \ding{56} & I & hospital, \href{https://www.ncbi.nlm.nih.gov/pmc/articles/PMC1718393/}{PACS} \\
% RP3D-DiagDS$^{*}$~\cite{zheng2023large} & 9 & 7K & 1 & \ding{56} & I & \href{https://radiopaedia.org/search?scope=cases}{Radiopedia}) \\
% ROCO$^{*}$~\cite{pelka2018radiology} & 9 & 8K & 2 & \ding{56} & I & \href{https://www.ncbi.nlm.nih.gov/pmc/}{PubMed Central}, \href{https://www.imageclef.org/}{ImageCLEF} \\
% EndoVis-18-VQA$^{*}$~\cite{seenivasan2022surgical} & 1 & 2K & 2 & \ding{56} & I & \href{https://opencas.dkfz.de/endovis/challenges/2018/}{EndoVis-18} \\
% PathMMU$^{*}$~\cite{sun2024pathmmu} & 1 & 9K & 1 & \ding{56} & I & \href{https://medpix.nlm.nih.gov/home}{WebPathology}, Textbook, Twitter, YouTube\\
% MultiMedEval~\cite{royer2024multimedeval} & 11 & & -             &                                   & Repurposed                           & M                & P        \\
% VQA-Med~\cite{abacha2019vqa} & 36 & 15K & - & \ding{56}  & I & Textbook \\
% Medmnist~\cite{yang2023medmnist} & 10                & G-Medicine & 1             & 0/718067                          & Repurposed                                   & L                & P \\
% MMT-Bench (MedU)~\cite{ying2024mmt} & - & 1K & 5 & \ding{56} & I & Datasets \\
% BenchMD$^{*}$~\cite{wantlin2023benchmd} & 7 & G-Medicine & 7             & \textgreater{}1M/\textgreater{}1M & Repurposed                                   & L                & N           \\ 
\hline
GMAI-MMBench & 38 & 26K & 18 & \ding{52} & I, B, M, C & 284 datasets from both public and hospital \\ \hline
\end{tabular}
\label{tab:comparison_exist_benchmark}
}
\vspace{-0.4cm}
\end{table*}

To address these challenges, we introduce the General Medical AI MultiModal Benchmark (GMAI-MMBench), a comprehensive multimodal benchmark that is well-categorized for medical image understanding and reasoning in real-world clinical scenarios. As shown in Figure~\ref{fig:cover}, its comprehensiveness can be concluded in three aspects: 1) \textbf{comprehensive medical knowledge from diverse modalities, tasks, and data sources}, 2) \textbf{well-categorized in lexical tree structures}, and 3) \textbf{multiple perceptual granularity}.

\begin{figure}[htbp]
    \centering
    \includegraphics[width=1\linewidth]{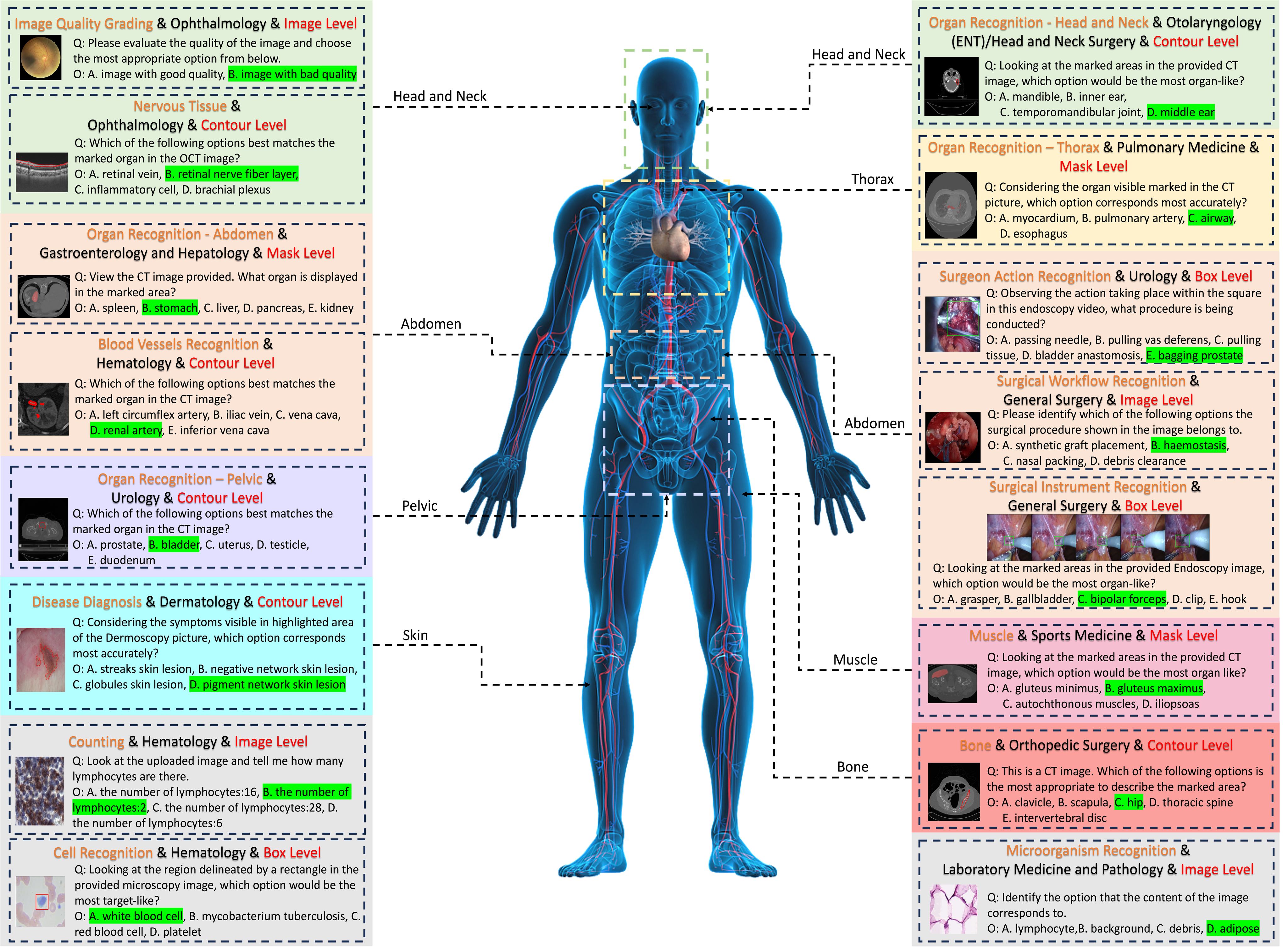}
    % \caption{Overview of GMAI-MMBench: A benchmark for comprehensive medical knowledge derived from diverse modalities, tasks, and data sources, systematically categorized in lexical tree structures, and encompassing multiple perceptual granularities.}
    \caption{Examples of GMAI-MMBench. The benchmark covers a variety of clinical tasks, departments, and perceptual granularities from worldwide data sources.}
    \label{fig:body_med}
\end{figure}

First, GMAI-MMBench has diverse modalities and data sources because it is built upon 284 high-quality datasets collected across the world. These 284 datasets cover various medical image tasks, including 2D detection, 2D classification, and 2D/3D segmentation, to ensure the diversity of tasks. Using these foundational visual-based tasks has two advantages over using off-the-shelf image-text pair data. 1) It minimizes the risk of data leakage since the data in our benchmark are mostly image-label pairs rather than image-text pairs. The image-label pairs are not directly convertible to LVLMs training samples (usually image-text pairs), thus less likely to be used to train LVLMs; 2) It ensures high clinical relevance, as the images are sourced from hospitals and annotated by professional doctors. We then carefully selected approximately 26K cases with 38 different modalities to construct the GMAI-MMBench, thus meeting the modal diversity goal. 

Second, GMAI-MMBench is a well-categorized medical benchmark that can comprehensively evaluate the pros and cons of various aspects of LVLMs, benefiting both model developers and users with specific needs. Specifically, we develop a categorization system, called lexical tree structure, which categorizes all cases into 18 clinical VQA tasks, 18 departments, 38 modalities, etc. The `clinical VQA tasks' / `departments' / `modalities' are the lexicons that can be used to retrieve desired cases for evaluation. For instance, the oncology department can select cases related to oncology to evaluate LVLMs' performance for oncology tasks, thus greatly enhancing flexibility and usability for specific demands. 

Third, GMAI-MMBench can evaluate LVLMs' abilities to perceive different granularity, such as understanding the local image content in a mask or bounding box as well as recognizing the entire image content. This ability is important for detection, segmentation, and classification tasks as these tasks need different perceptual granularity for better performance. Furthermore, the perception of bounding boxes or masks is vital for interactive LVLMs~\cite{kirillov2023segment}, so the perceptual granularity evaluation in our benchmark can possibly be used to improve interactive LVLMs.

We assess 44 publicly available LVLMs (38 general purpose and 6 medical-specific models) as well as advanced proprietary LVLMs such as GPT-4o, GPT-4V, Claude3-Opus, Gemini 1.0, Gemini 1.5, and Qwen-VL-Max on our GMAI-MMBench. We summarize the key findings as follows:

(1) GMAI-MMBench presents significant challenges in clinical practice. Even the best proprietary GPT-4o only achieves an accuracy of 53.96\%, which demonstrates the deficiencies of cutting-edge LVLMs in tackling medical professional issues, thus they can hardly fulfill diverse clinical demands.

(2) Open-source LVLMs, such as MedDr and DeepSeek-VL-7B, achieve approximately 44\% accuracy, making them very competitive compared to proprietary models. For instance, they surpass Claude3-Opus and Qwen-VL-Max and achieve comparable performance to Gemini 1.5 and GPT-4V. However, they still exhibit a clear performance disparity compared to the top-performing GPT-4o.

(3) Most medical-specific models have difficulty reaching a general performance level (approximately 30\% accuracy) achieved by general LVLMs, except MedDr with 43.69\% accuracy.

(4) Most LVLMs exhibit unbalanced performance across different clinical VQA tasks, departments, and perceptual granularity. Notably, in the experiments on different perceptual granularity, box-level annotation consistently results in the worst accuracy, even worse than image-level annotation. 

(5) The major factors leading to performance bottlenecks include perceptual errors (e.g., misrecognition of image content), lack of medical domain knowledge, irrelevant responses, and rejection of answering questions due to safety protocols. 

In summary, our contributions are three-fold. 
(a) We introduce a comprehensive benchmark, GMAI-MMBench, to evaluate existing LVLMs in clinical practice. GMAI-MMBench covers 38 modalities, 18 clinical VQA tasks, 18 departments, and 4 different perceptual granularity from 284 medical-related datasets, 
%thus diverse in modality, task, and data sources.
thereby offering a diverse range of modalities, tasks, and data sources.
(b) GMAI-MMBench organizes each data point in lexical tree structures, with lexicons used to select desired data points to evaluate various aspects of LVLMs’ abilities. Thus, GMAI-MMBench facilitates customized evaluation to meet highly diverse demands in clinical practice. \textbf{See Supplementary C.2}.
(c) We evaluate 44 representative general-purpose LVLMs, including both open-source and proprietary models, as well as 6 medical-specific LVLMs on GMAI-MMBench. The comprehensive evaluation reveals the pros and cons of different LVLMs from diverse perspectives, providing insights to improve these models to accommodate real-world clinical applications.

\begin{figure*}[t]
\centering
\includegraphics[width=1.0\textwidth]{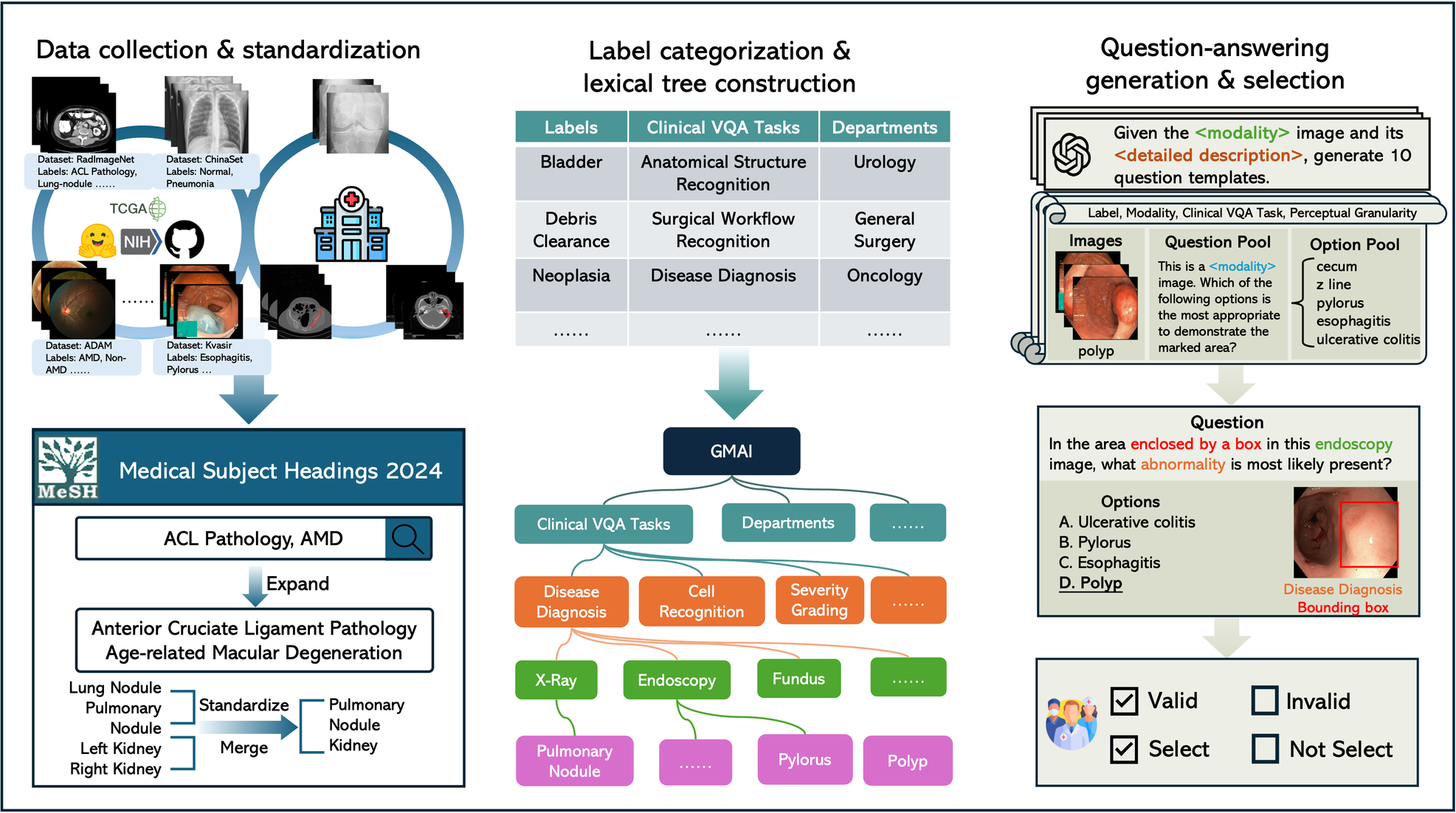}
\caption{Overall illustration of GMAI-MMBench. The data collection can be divided into three main steps: 1) We search hundreds of datasets from both the public and hospitals, then keep 284 datasets with highly qualified labels after dataset filtering, uniforming image format, and standardizing label expression. 2) We categorize all labels into 18 clinical VQA tasks and 18 clinical departments, then export a lexical tree for easily customized evaluation. 3) We generate QA pairs for each label from its corresponding question and option pool. Each question must include information about image modality, task cue, and corresponding annotation granularity. The final benchmark is obtained through additional validation and manual selection.}
\label{fig:workflow}
\end{figure*}

\section*{GMAI-MMBench}

\subsection*{Overview}
We propose GMAI-MMBench, an innovative benchmark meticulously designed for the medical field, capable of providing comprehensive evaluations of LVLMs across various aspects of healthcare. (shown in the Figure~\ref{fig:body_med}) We collect 284 datasets from public sources and hospitals, covering medical imaging tasks of detection, classification, and segmentation, to form the data fuel for establishing such a benchmark. The detailed datasets are listed in the supplementary. Based on the data foundation, we design a reliable pipeline to generate question-answering pairs and organize them from different perspectives with manual validation. Finally, we carefully select approximately 26K questions with varying levels of perceptual granularity from the manually validated cases to construct the final GMAI-MMBench.
% \yj{TODO: GMAI-MMBench features three aspects: xxxx}

% comprehensive and highly clinical relevance, well-categorized, heterogeneous granularitical perception.

% GMAI-MMBench features three aspects: (1) comprehensive and highly clinical relevance, Compared to current benchmarks, 

\subsection*{Benchmark Construction}
The detailed steps of constructing our GMAI-MMBench can be divided into three main steps as shown in Figure~\ref{fig:workflow}.

\textbf{Dataset collection and standardization.} 
As our aim is to build a large-scale benchmark for the comprehensive evaluation of LVLMs, the first and most important step is data collection. In contrast to benchmarks that directly use multimodal paired datasets, we source the datasets in two ways to minimize the data leakage problem and ensure the diversity and clinical property: First, we conduct thorough Internet searches to collect as many 2D/3D medical-related datasets as possible, retaining those that involve classification, detection, and segmentation tasks. Second, we collaborate with several hospitals that have agreed to share their ethically approved data. This process has enabled us to curate 284 datasets with highly qualified labels. Following data collection, we standardize both images and labels. For images, we adhere to the SA-Med2D-20M~\cite{ye2023sa} protocol, transforming all 2D/3D medical images into 2D RGB images for further evaluation. For labels, we refer to the Medical Subject Headings (MeSH)\footnote{\url{https://www.ncbi.nlm.nih.gov/mesh/1000048}} to ensure every label is unique, clear, and free from conflict or ambiguity within each task. Specifically, we focus on three main situations: (1) expanding all abbreviations, such as changing ``AMD'' to ``Age-related macular degeneration''; (2) unifying different expressions for the same target, such as standardizing both ``lung nodule'' and ``pulmonary nodule'' to ``pulmonary nodule''; (3) merging labels with left and right distinctions, such as combining ``left kidney'' and ``right kidney'' into ``kidney'', since our goal is to evaluate the abilities of understanding and reasoning rather than directional judgment. 

\textbf{Label categorization and lexical tree construction.}
We construct a well-categorized lexical tree to ensure GMAI-MMBench can be easily customized to evaluate the specific abilities of LVLMs for each clinical professional. The overview of the tree is shown in Figure~\ref{fig:workflow}, and the complete version is in supplementary. First, we integrate data properties and real applications to propose three subjects tailored for the biomedical fields: clinical VQA tasks, departments, and perceptual granularities. Specialized options are generated for each subject individually: For clinical VQA tasks, we extract keywords according to the original dataset descriptions and then lead to 18 categories. For departments, we refer to the Mayo Clinic\footnote{\url{https://www.mayoclinic.org/departments-centers}} and assign all labels to 18 departments. For perceptual granularity, we construct 4 types based on annotation methods (see the rightmost panel in Figure~\ref{fig:cover}). We then recruit several biomedical engineering university students (including coauthors) to tag labels from the constructed options in these subjects. Specifically, each label is randomly assigned to 3 people, and their tagging results are merged by voting. After label categorization, the lexical tree can be directly exported for customized evaluation. An example of customized evaluation is presented in Supplementary C.2.

\textbf{QA generation and selection.}
% Despite the correlation between an image and its accompanying label is tight, 
Following the label categorization, all labels are assigned to specific modalities, clinical VQA tasks, departments, and perceptual granularities. Based on the well-organized structure, we generate the VQA pairs for every label with three steps. 
First, questions and options generation. For question generation, a question must include three key pieces of information in GMAI-MMBench: modality, clinical task hint, and perceptual granularity information. For each combination of the three elements, we randomly pick 10 labels and generate 10 candidate questions with GPT-4o for each selected label. These questions are then manually reviewed to meet the following criteria: (1) they must include necessary information on modality, clinical task, and perceptual granularity; (2) they do not include any hints that would allow the question to be answered without viewing the image. After manual review, the modality is replaced with a placeholder for standardization. For example, a valid question template for \underline{Disease Diagnosis} in \underline{segmentation task} is: ``\textit{This is a <modality> image. Which of the following options is the most appropriate to demonstrate \underline{symptoms} in the \underline{marked area}?}'' Once the question pool is generated, each category has its question pool based on its tags of modality, clinical VQA task, and perceptual granularity. For options generation, the global view (image level) and local view (mask level, bounding box level, and contour level) of perceptual granularity are handled separately. For the global view, the option pool for each answer is sourced from the remaining categories within the answer's dataset to avoid introducing multiple correct answers. For instance, a fundus image dataset may focus solely on pathological myopia, but the images might also contain other diseases like diabetic retinopathy. Including other categories could render the question invalid. For the local view, we construct a shared option pool for the answers with the combination of modality, clinical VQA task, and perceptual granularity. 
Second, as each answer with corresponding images has its own question and option pool, we generate all QA pairs for all images. For each image, we randomly select a question from its question pool and replace the placeholder with its modality. Along with the correct answer, we randomly select $n$ options (where $n=\mathbf{randint}(\mathbf{max}(1, \textbf{len(option pool)}), \mathbf{min}(4, \textbf{len(option pool)}$) from the corresponding option pool to create the set of options. 
Third, to ensure data quality and balanced distribution, we perform additional manual validation and selection. In the validation stage, we assess the QA pairs based on the following criteria: (1) We drop cases whose questions do not contain the three key components and can be answered without the image. (2) We filter out cases with incorrect answers. (3) We drop cases where images have unclear targets or poor image quality. In the selection stage, we choose 30 cases per answer to ensure balance across all tasks (all cases are included if the number is less than 30). The selection rule is based on the consideration of diversity: Selecting images with large differences in appearance, data source, age, gender, etc. As a result, we finalize 25831 QA pairs for the GMAI-MMBench (4550 in the validation set and 21281 in the test set).  

\section*{Experiments}
\label{headings}
\subsection*{Experiment setup}

In this study, we evaluated various LVLMs, including medical-specific, open-source, and proprietary API general models. We selected versions with approximately 7 billion parameters for testing, and the model weights were sourced from their respective official Hugging Face repositories. Our evaluation was conducted using the VLMEvalKit\footnote{\url{https://github.com/open-compass/VLMEvalKit}} framework and Multi-Modality-Arena\footnote{\url{https://github.com/OpenGVLab/Multi-Modality-Arena/tree/main/MedicalEval/Question-answering\_Score}}.

The assessment was performed in a ``zero-shot'' setting. Specifically, our evaluation prompts did not include any example cues, and the models were required to perform inference on tasks without prior training or examples related to those tasks. This approach better tests the models' generalization capabilities and comprehension, examining their performance when confronted with novel problems. All tests were executed using NVIDIA A100 GPUs with 80GB of memory.

% \vspace{-1em}

\subsection*{Models}

For completeness, we conducted evaluations using several state-of-the-art LVLMs to benchmark their performance on GMAI-MMBench, including both general models that have extended capabilities in the biomedical domain and medical-specific models that are meticulously trained for clinical medicine. By default, we use the latest, largest, and best-performing available checkpoint for each model family to ensure optimal performance. We picked 29 out of 50 models for demonstration in the main text, additional results are provided in the supplementary material.
For medical-specific models, we include 5 latest powerful LVLMs: MedDr~\cite{he2024meddr}, LLaVA-Med~\cite{li2024llava}, Med-Flamingo~\cite{moor2023med}, RadFM~\cite{wu2023generalist}, and Qilin-Med-VL-Chat~\cite{liu2023qilin}. 
For general models, we test 18 representative LVLMs:  TransCore-M~\cite{transcorem}, VisualGLM-6B~\cite{ding2021cogview}, mPLUG-Owl2~\cite{ye2023mplug}, OmniLMM-12B~\cite{yu2024rlaifv}, Mini-Gemini-7B~\cite{li2024mini}, Emu2-Chat~\cite{sun2023generative}, MMAlaya~\cite{datacanvas2024mmalaya}, CogVLM-Chat~\cite{wang2023cogvlm}, InstructBLIP-7B~\cite{dai2024instructblip}, DeepSeek-VL-7B~\cite{lu2024deepseek}, Idefics-9B-Instruct~\cite{laurencon2023obelics}, XComposer2~\cite{internlmxcomposer2}, Yi-VL-6B~\cite{ai2024yi}, InternVL-Chat-V1.5~\cite{chen2024far}, LLAVA-V1.5-7B~\cite{liu2023visual}, LLAVA-InternLM2-7b~\cite{2023xtuner}, MiniCPM-V2~\cite{xu2024llava-uhd}, and Qwen-VL-Chat~\cite{Qwen-VL}. 
In addition, we also evaluate 6 proprietary LVLMs via API: Qwen-VL-Max~\cite{Qwen-VL}, Claude3-Opus~\cite{anthropic2024claude}, GPT-4V~\cite{achiam2023gpt}, GPT-4o~\cite{achiam2023gpt}, Gemini 1.0~\cite{team2023gemini}, and Gemini 1.5~\cite{reid2024gemini}. 
% These models encompass a variety of architectures and training paradigms, providing comprehensive evaluation results for our benchmark.

% \begin{figure*}[htbp]
% \centering
% \includegraphics[width=0.6\textwidth]{imgs/mcls.pdf}
% \caption{The overall accuracy and recall results for multiple-choice questions across different models.}
% \label{fig:mcls_comparison}
% \end{figure*}

% \vspace{-1em}

\subsection*{Metrics}
To evaluate the model's performance, we use macro-averaged accuracy (ACC) as the evaluation metric for single-choice questions. For multiple-choice questions, we first count the number of correct predictions for each case, then calculate accuracy (\(\mathrm{ACC_{mcq}}\)) and recall (\(\mathrm{Recall_{mcq}}\)) based on the proportion of correct matches to the prediction length and the length of the ground-truth options, respectively. More details are shown in supplementary materials. If a model's output does not include clearly followed instructions to select an answer or letter options, we use ChatGPT-3.5-turbo-0613 to extract the answer. If an answer cannot be extracted, it is treated as an error.

\section*{Results}

% \begin{table}[h!]
% \centering
% \begin{tabular}{ll}
% \toprule
% \textbf{Model Series} & \textbf{Models} \\
% \midrule
% GPT Series & - \\
% \midrule
% Claude Series & - \\
% \midrule
% Gemini Series & - \\
% \midrule
% LLaMA Series & llava\_v1.5\_7b \quad llava\_next\_vicuna\_7b \quad llava\_next\_mistral\_7b \\
%  & llava-internlm2-7b \quad llava-internlm-7b \quad llava-v1.5-7b-xtuner \\
%  & llava-v1.5-13b-xtuner \\
% \midrule
% InternVL-Chat Series & InternVL-Chat-V1-1 \quad InternVL-Chat-V1-2 \quad InternVL-Chat-V1-2-Plus \\
%  & InternVL-Chat-V1-5 \\
% \midrule
% XComposer Series & XComposer \quad XComposer2 \quad XComposer2\_4KHD \\
% \midrule
% QWEN Series & qwen\_max \quad qwen\_base \quad qwen\_chat \\
% \midrule
% Monkey Series & monkey \quad monkey-chat \\
% \midrule
% DeepSeek Series & deepseek\_vl\_7b \quad deepseek\_vl\_1.3b \\
% \midrule
% Other Models & MiniCPM-V \quad MiniCPM-V2 \quad flamingov2 \quad VisualGLM\_6b \\
%  & mPLUG-Owl2 \quad cogvlm-grounding-generalist \quad cogvlm-chat \\
%  & MMAlaya \quad sharecaptioner \quad OmniLMM\_12B \quad TransCore\_M \\
%  & Yi\_VL\_6B \\
% \bottomrule
% \end{tabular}
% \caption{General-Purpose Models}
% \label{tab:general_models}
% \end{table}

% \begin{table}[h!]
% \centering
% \begin{tabular}{ll}
% \toprule
% \textbf{Model Series} & \textbf{Models} \\
% \midrule
% Medical Models & TBD \\
% \bottomrule
% \end{tabular}
% \caption{Medical Models}
% \label{tab:medical_models}
% \end{table}

\begin{figure*}[t]
\centering
\includegraphics[width=1.0\textwidth]{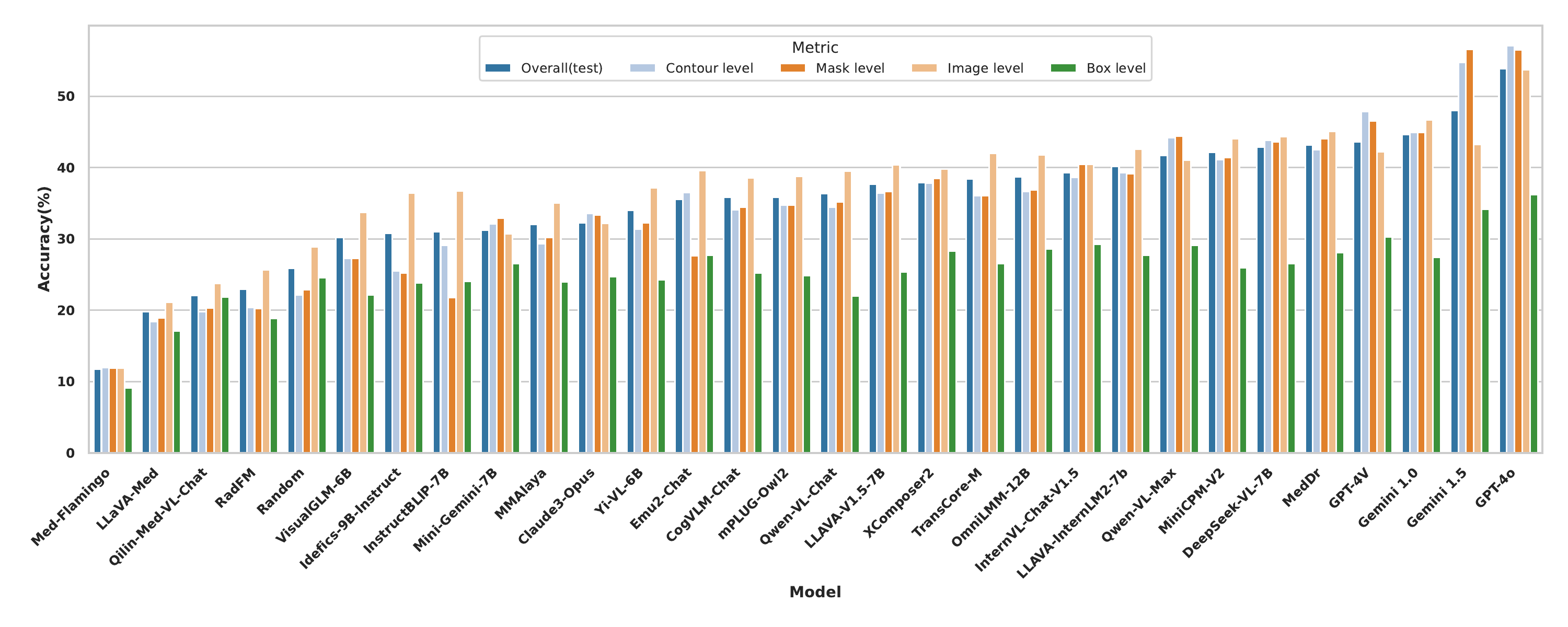}
% \caption{The results of different models for single-choice questions under various annotation methods, including Contour level, Mask level, Image level, and Box level.}
\vspace{-0.7cm}
\caption{Results for single-choice questions of different models on different perceptual granularities, including Contour level, Mask level, Image level, and Box level.}
\label{fig: different perceptual types results}
\end{figure*}

\begin{table}[]
% \caption{The results of different models for single-choice questions across various clinical VQA tasks (All results are expressed as percentages. The best-performing model in each category is \textbf{in-bold}, and the second best is \underline{underlined}). \textbf{Full names of the clinical VQA tasks are demonstrated in supplementary material}}
\caption{Results for single-choice questions of different LVLMs on clinical VQA tasks. The best-performing model in each category is \textbf{in-bold}, and the second best is \underline{underlined}. Abbreviations: the full terms of all clinical VQA tasks are listed in Table~\yj{5} of supplementary material.}
\label{tab:metatasksresults}
\resizebox{1.0\textwidth}{!}{
\begin{tabular}{l|cc|cccccccccccccccccc}
\hline
Model name                  & \begin{tabular}[c]{@{}c@{}}Overall\\ (val)\end{tabular} & \begin{tabular}[c]{@{}c@{}}Overall\\ (test)\end{tabular} & AR               & BVR              & B                & CR               & C                & DD               & IQG              & MR               & M                & NT               & OR-A             & OR-HN            & OR-P             & OR-T             & SG               & SAR              & SIR              & SWR              \\ \hline
Random                      & 25.70                                                 & 25.94                                                  & 38.20          & 22.73          & 22.92          & 22.72          & 24.06          & 26.66          & 27.13          & 27.00          & 20.00          & 24.75          & 21.37          & 22.93          & 22.33          & 21.18          & 32.43          & 24.23          & 21.39          & 23.71          \\
\multicolumn{21}{c}{\cellcolor[HTML]{EFEFEF}Medical Special Model}                                                                                                                                                                                                                                                                                                                                                                                                                                                             \\
% MedVInT ~\cite{zhang2023pmcvqa}                     & 2.29                                                  & 1.96                                                   & 5.75           & 0.00           & 0.00           & 0.00           & 2.56           & 2.11           & 4.05           & 0.00           & 0.00           & 0.00           & 0.11           & 0.00           & 0.00           & 0.12           & 7.36           & 0.00           & 1.88           & 0.00           \\
Med-Flamingo ~\cite{moor2023med}                & 12.74                                                 & 11.64                                                  & 6.67           & 10.14          & 9.23           & 11.27          & 6.62           & 13.43          & 12.15          & 6.38           & 8.00           & 18.18          & 9.26           & 18.27          & 11.00          & 11.53          & 12.16          & 5.19           & 8.47           & 11.43          \\
LLaVA-Med ~\cite{li2024llava}                   & 20.54                                                 & 19.60                                                  & 24.51          & 17.83          & 17.08          & 19.86          & 15.04          & 19.81          & 20.24          & 21.51          & 13.20          & 15.15          & 20.42          & 23.73          & 17.67          & 19.65          & 21.70          & 19.81          & 14.11          & 20.86          \\
Qilin-Med-VL-Chat ~\cite{liu2023qilin}           & 22.34                                                 & 22.06                                                  & 29.57          & 19.41          & 16.46          & 23.79          & 15.79          & 24.19          & 21.86          & 16.62          & 7.20           & 13.64          & 24.00          & 14.67          & 12.67          & 15.53          & 26.13          & 24.42          & 17.37          & 25.71          \\
RadFM ~\cite{wu2023generalist}                      & 22.95                                                 & 22.93                                                  & 27.16          & 20.63          & 13.23          & 19.14          & 20.45          & 24.51          & 23.48          & 22.85          & 15.60          & 16.16          & 14.32          & 24.93          & 17.33          & 21.53          & 29.73          & 17.12          & 19.59          & 31.14          \\
MedDr~\cite{he2024meddr}                       & 41.95                                                 & 43.69                                                  & 41.20          & 50.70          & 37.85          & 29.87          & 28.27          & 52.53          & 36.03          & 31.45          & 29.60          & 47.47          & 33.37          & 51.33          & 32.67          & 44.47          & 35.14          & 25.19          & 25.58          & 32.29          \\
\multicolumn{21}{c}{\cellcolor[HTML]{FFF3E4}Open-Source LVLMs}                                                                                                                                                                                                                                                                                                                                                                                                                                                                  \\
VisualGLM-6B~\cite{ding2021cogview}                & 29.58                                                 & 30.45                                                  & 40.16          & 33.92          & 24.92          & 25.22          & 24.21          & 32.99          & 29.96          & 29.53          & 21.20          & 37.88          & 30.32          & 24.80          & 13.33          & 29.88          & 33.11          & 19.62          & 19.16          & 37.43          \\
Idefics-9B-Instruct~\cite{laurencon2023obelics}          & 29.74                                                 & 31.13                                                  & 40.39          & 30.59          & 26.46          & 33.63          & 22.56          & 34.38          & 25.51          & 26.71          & 21.60          & 27.78          & 27.47          & 32.80          & 24.67          & 23.41          & 32.66          & 23.08          & 21.39          & 30.57          \\
InstructBLIP-7B ~\cite{dai2024instructblip}            & 31.80                                                 & 30.95                                                  & 42.12          & 26.92          & 24.92          & 28.09          & 21.65          & 34.58          & 31.58          & 29.23          & 22.40          & 30.30          & 28.95          & 27.47          & 23.00          & 24.82          & 32.88          & 19.81          & 21.64          & 26.57          \\
Mini-Gemini-7B~\cite{li2024mini}              & 32.17                                                 & 31.09                                                  & 29.69          & 39.16          & 31.85          & 28.26          & 10.38          & 35.58          & 29.96          & 28.78          & 20.80          & 34.34          & 29.58          & 36.53          & 24.00          & 31.76          & 22.45          & 25.96          & 18.56          & 29.43          \\
MMAlaya~\cite{datacanvas2024mmalaya}                     & 32.19                                                 & 32.30                                                  & 41.20          & 35.14          & 32.15          & 34.17          & 27.82          & 35.09          & 28.34          & 30.27          & 18.00          & 46.97          & 20.21          & 31.20          & 16.00          & 34.59          & 32.28          & 23.65          & 22.93          & 30.29          \\
% Qwen-VL ~\cite{bai2023qwenvl}                    & 34.80                                                 & 36.05                                                  & 37.05          & 37.24          & 35.85          & 28.98          & 24.81          & 43.60          & 24.70          & 30.12          & 19.20          & 44.44          & 29.68          & 31.87          & 25.00          & 31.18          & 30.26          & 21.54          & 20.10          & 26.86          \\
Yi-VL-6B ~\cite{ai2024yi}                   & 34.82                                                 & 34.31                                                  & 41.66          & 39.16          & 26.62          & 30.23          & 31.88          & 38.01          & 26.72          & 24.93          & 25.20          & 37.37          & 29.58          & 31.20          & 32.33          & 30.59          & 36.71          & 24.81          & 23.18          & 31.43          \\
% LLaVA-NeXT-vicuna-7B ~\cite{liu2024llavanext}       & 34.86                                                 & 35.42                                                  & 40.62          & 38.64          & 21.08          & 35.42          & 23.91          & 41.22          & 32.39          & 28.04          & 20.53          & 44.95          & 27.92          & 34.98          & 20.22          & 32.82          & 33.63          & 23.08          & 25.06          & 34.86          \\
Qwen-VL-Chat~\cite{Qwen-VL}               & 35.07                                                 & 36.96                                                  & 38.09          & 40.56          & 38.00          & 32.20          & 25.71          & 44.07          & 24.70          & 30.56          & 24.00          & 40.91          & 29.37          & 36.53          & 26.00          & 27.29          & 35.14          & 16.54          & 20.10          & 34.00          \\
CogVLM-Chat ~\cite{wang2023cogvlm}                & 35.23                                                 & 36.08                                                  & 40.97          & 30.77          & 27.69          & 32.74          & 19.40          & 41.10          & 36.84          & 34.72          & 24.00          & 40.91          & 36.74          & 37.33          & 26.00          & 33.65          & 36.56          & 20.19          & 23.95          & 26.57          \\
% Monkey ~\cite{li2024monkey}                      & 35.48                                                 & 36.39                                                  & 38.32          & 35.31          & 35.54          & 34.53          & 23.16          & 43.40          & 31.98          & 30.12          & 19.20          & 33.33          & 30.00          & 32.53          & 25.33          & 31.65          & 34.46          & 20.00          & 20.27          & 30.29          \\
mPLUG-Owl2~\cite{ye2023mplug}                  & 35.62                                                 & 36.21                                                  & 37.51          & 41.08          & 30.92          & 38.10          & 27.82          & 41.59          & 28.34          & 32.79          & 22.40          & 40.91          & 24.74          & 38.27          & 23.33          & 36.59          & 33.48          & 20.58          & 23.01          & 32.86          \\
% ShareCaptioner ~\cite{chen2023sharegpt4v}             & 36.37                                                 & 36.19                                                  & 42.35          & 32.69          & 31.08          & 27.19          & 30.83          & 41.19          & 30.36          & 33.23          & 28.40          & 42.93          & 27.79          & 33.73          & 28.33          & 40.71          & 29.58          & 20.96          & 28.83          & 30.00          \\
Emu2-Chat ~\cite{sun2023generative}                  & 36.50                                                 & 37.59                                                  & 43.27          & 47.73          & 26.31          & 40.07          & 28.12          & 44.00          & 36.44          & 28.49          & 20.40          & 31.82          & 26.74          & 37.60          & 26.67          & 29.76          & 33.63          & 23.27          & 26.43          & 29.43          \\
OmniLMM-12B  ~\cite{yu2024rlaifv}                & 37.89                                                 & 39.30                                                  & 39.82          & 40.56          & 32.62          & 37.57          & 24.81          & 46.68          & 35.63          & 35.01          & 27.60          & 57.58          & 28.42          & 34.00          & 25.00          & 29.18          & 34.46          & 24.42          & 27.54          & 40.29          \\
% InternVL-Chat-V1.1~\cite{chen2023internvl}          & 38.16                                                 & 39.41                                                  & 42.46          & 43.88          & 35.23          & 45.08          & 23.31          & 45.96          & 38.87          & 29.23          & 29.60          & 40.40          & 31.68          & 41.87          & 26.67          & 38.82          & 32.13          & 19.42          & 25.58          & 30.29          \\
LLAVA-V1.5-7B~\cite{liu2023visual}               & 38.23                                                 & 37.96                                                  & 45.45          & 34.27          & 30.92          & 41.32          & 21.65          & 44.68          & 34.01          & 27.74          & 23.60          & 43.43          & 28.00          & 42.13          & 29.00          & 35.06          & 33.41          & 22.12          & 23.61          & 29.14          \\
% Monkey-Chat ~\cite{li2024monkey}                & 38.39                                                 & 39.50                                                  & 40.62          & 41.43          & 37.08          & 35.24          & 23.76          & 47.73          & 29.96          & 32.94          & 26.00          & 37.88          & 34.84          & 32.67          & 24.67          & 33.18          & 34.91          & 21.73          & 22.24          & 34.00          \\
% LLAVA-V1.5-7B-xtuner~\cite{2023xtuner}        & 38.68                                                 & 38.22                                                  & 38.90          & 40.03          & 28.00          & 40.25          & 30.08          & 44.08          & 33.60          & 32.49          & 21.20          & 40.91          & 29.47          & 40.40          & 30.33          & 38.59          & 31.46          & 23.85          & 26.95          & 36.86          \\
XComposer2 ~\cite{internlmxcomposer2}                 & 38.68                                                 & 39.20                                                  & 41.89          & 37.59          & 33.69          & 40.79          & 22.26          & 45.87          & 36.44          & 32.94          & 27.20          & 58.59          & 26.11          & 36.40          & 43.67          & 37.29          & 32.06          & 23.46          & 27.80          & 32.86          \\
% LLAVA-InternLM-7b~\cite{2023xtuner}           & 38.71                                                 & 39.11                                                  & 36.36          & 36.54          & 32.62          & 38.10          & 30.68          & 46.53          & 34.82          & 28.19          & 25.20          & 48.99          & 28.11          & 40.53          & 33.33          & 36.00          & 34.08          & 26.73          & 24.12          & 29.71          \\
TransCore-M ~\cite{transcorem}                & 38.86                                                 & 38.70                                                  & 40.74          & 41.78          & 20.77          & 35.06          & {\ul 34.74}    & 45.69          & 32.39          & 32.94          & 24.40          & 44.95          & 31.05          & 38.93          & 27.00          & 33.76          & 33.86          & 23.46          & 25.49          & 31.14          \\
InternVL-Chat-V1.5~\cite{chen2024far}           & 38.86                                                 & 39.73                                                  & 43.84          & 44.58          & 34.00          & 33.99          & 31.28          & 45.59          & 33.20          & 38.28          & 32.40          & 42.42          & 31.89          & 42.80          & 27.00          & 36.82          & 34.76          & 23.27          & 24.72          & 32.57          \\
% InternVL-Chat-V1.2-Plus~\cite{chen2023internvl}     & 39.41                                                 & 40.79                                                  & 42.58          & 42.31          & 32.46          & 37.03          & 31.43          & 47.49          & 42.51          & 35.01          & 21.20          & 50.51          & 34.95          & 42.93          & 22.67          & 42.47          & 35.74          & 22.31          & 24.98          & 28.29          \\
% InternVL-Chat-V1.2 ~\cite{chen2023internvl}         & 39.52                                                 & 40.01                                                  & 41.66          & 44.06          & 27.38          & 38.46          & 34.29          & 46.99          & 33.60          & 34.42          & 21.20          & 47.98          & 30.63          & 42.80          & 27.67          & 35.88          & 35.59          & {\ul 23.85}    & 24.98          & 28.00          \\
LLAVA-InternLM2-7b~\cite{2023xtuner}          & 40.07                                                 & 40.45                                                  & 39.82          & 37.94          & 30.62          & 35.24          & 29.77          & 48.97          & 34.01          & 25.96          & 20.80          & 53.03          & 30.95          & 42.67          & 32.00          & 39.88          & 32.43          & 21.73          & 24.38          & 38.00          \\
% DeepSeek-VL-1.3B ~\cite{lu2024deepseek}           & 40.25                                                 & 40.77                                                  & 38.55          & 35.14          & 38.92          & 40.07          & 27.97          & 48.12          & 35.63          & 31.75          & 22.80          & 46.97          & 40.74          & 44.93          & 31.00          & 40.47          & 33.33          & 22.31          & 21.39          & 31.71          \\
% MiniCPM-V ~\cite{hu2024large}                  & 40.95                                                 & 41.05                                                  & 39.70          & 46.50          & 36.31          & 39.36          & 22.26          & 48.09          & 34.82          & 35.76          & 24.00          & 45.45          & 34.11          & 44.80          & 23.00          & 44.47          & 36.19          & 21.15          & 23.95          & 35.14          \\
DeepSeek-VL-7B ~\cite{lu2024deepseek}             & 41.73                                                 & 43.43                                                  & 38.43          & 47.03          & 42.31          & 37.03          & 26.47          & 51.11          & 33.20          & 31.16          & 26.00          & 44.95          & 36.00          & 58.13          & 36.33          & 47.29          & 34.91          & 18.08          & 25.49          & {\ul 39.43}    \\
MiniCPM-V2~\cite{xu2024llava-uhd}                  & 41.79                                                 & 42.54                                                  & 40.74          & 43.01          & 36.46          & 37.57          & 27.82          & 51.08          & 28.74          & 29.08          & 26.80          & 47.47          & 37.05          & 46.40          & 25.33          & 46.59          & 35.89          & 22.31          & 23.44          & 31.71          \\
\multicolumn{21}{c}{\cellcolor[HTML]{FFF0F0}Proprietary LVLMs}                                                                                                                                                                                                                                                                                                                                                                                                                                                                  \\
Claude3-Opus~\cite{anthropic2024claude}                & 32.37                                                 & 32.44                                                  & 1.61           & 39.51          & 34.31          & 31.66          & 12.63          & 39.26          & 28.74          & 30.86          & 22.40          & 37.37          & 25.79          & 41.07          & 29.33          & 33.18          & 31.31          & 21.35          & 23.87          & 4.00           \\
Qwen-VL-Max ~\cite{Qwen-VL}                & 41.34                                                 & 42.16                                                  & 32.68          & 44.58          & 31.38          & 40.79          & 10.68          & 50.53          & 32.79          & 44.36          & 29.20          & 51.52          & 41.37          & 58.00          & 30.67          & 41.65          & 26.95          & 25.00          & 24.64          & 39.14          \\
GPT-4V ~\cite{achiam2023gpt}                     & 42.50                                                 & 44.08                                                  & 29.92          & 48.95          & 44.00          & 37.39          & 12.93          & 52.88          & 32.79          & 44.21          & {\ul 32.80}    & 63.64          & 39.89          & 54.13          & 37.00          & 50.59          & 27.55          & 23.08          & 25.75          & 37.43          \\
Gemini 1.0 ~\cite{team2023gemini}                 & 44.38                                                 & 44.93                                                  & {\ul 42.12}    & 45.10          & 46.46          & 37.57          & 20.45          & 53.29          & 35.22          & 36.94          & 25.20          & 51.01          & 34.74          & 59.60          & 34.00          & 50.00          & \textbf{36.64} & 23.65          & 23.87          & 35.43          \\
Gemini 1.5 ~\cite{reid2024gemini}                 & {\ul 47.42}                                           & {\ul 48.36}                                            & \textbf{43.50} & {\ul 56.12}    & {\ul 51.23}    & {\ul 47.58}    & 2.26           & {\ul 55.33}    & {\ul 38.87}    & {\ul 48.07}    & 30.00          & \textbf{76.26} & {\ul 51.05}    & \textbf{75.87} & {\ul 46.33}    & {\ul 62.24}    & 20.57          & \textbf{27.69} & \textbf{30.54} & \textbf{40.57} \\
GPT-4o ~\cite{achiam2023gpt}                     & \textbf{53.53}                                        & \textbf{53.96}                                         & 38.32          & \textbf{61.01} & \textbf{57.08} & \textbf{49.02} & \textbf{46.62} & \textbf{61.45} & \textbf{46.56} & \textbf{56.38} & \textbf{34.00} & {\ul 75.25}    & \textbf{53.79} & {\ul 69.47}    & \textbf{48.67} & \textbf{65.88} & {\ul 33.93}    & 22.88          & {\ul 29.51}    & {\ul 39.43}    \\ \hline
\end{tabular}
}
\vspace{-0.2cm}
\end{table}

\begin{table}[]
 \caption{Results for single-choice questions of different LVLMs on departments. The best-performing model in each category is \textbf{in-bold}, and the second best is \underline{underlined}. Abbreviations: the full terms of all departments are listed in Table \yj{6} of supplementary material}
\resizebox{1.0\textwidth}{!}{
\begin{tabular}{l|cc|cccccccccccccccccc}
\hline
\multicolumn{1}{c|}{Model name}                  & \begin{tabular}[c]{@{}c@{}}Overall\\ (val)\end{tabular} & \multicolumn{1}{c|}{\begin{tabular}[c]{@{}c@{}}Overall\\ (test)\end{tabular}} & CS               & D                & E                & GH               & GS               & H                & ID               & LMP              & NH               & N                & OG               & OM               & O                & OS               & ENT/HNS          & PM               & SM               & U                \\ \hline
\multicolumn{1}{l|}{Random}                      & 25.70                                                 & \multicolumn{1}{c|}{25.94}                                                  & 22.82          & 25.19          & 21.00          & 25.97          & 22.24          & 24.45          & 31.13          & 28.99          & 22.86          & 24.00          & 29.15          & 27.77          & 30.36          & 25.92          & 22.53          & 24.74          & 22.87          & 29.19          \\
\multicolumn{21}{c}{\cellcolor[HTML]{EFEFEF}Medical Special Model}                                                                                                                                                                                                                                                                                                                                                                                                                                                                                                       \\
% \multicolumn{1}{l|}{MedVInT~\cite{zhang2023pmcvqa} }                     & 2.29                                                  & \multicolumn{1}{c|}{1.96}                                                   & 0.24           & 2.50           & 1.00           & 1.94           & 1.09           & 0.88           & 3.31           & 5.23           & 1.14           & 0.73           & 0.00           & 1.40           & 4.44           & 0.56           & 0.00           & 2.24           & 0.64           & 0.86           \\
\multicolumn{1}{l|}{Med-Flamingo~\cite{moor2023med}}                & 12.74                                                 & \multicolumn{1}{c|}{11.64}                                                  & 11.76          & 12.49          & 10.00          & 10.88          & 9.33           & 5.42           & 7.28           & 10.05          & 12.00          & 10.91          & 12.88          & 14.89          & 15.37          & 12.40          & 13.43          & 12.89          & 14.92          & 10.47          \\
\multicolumn{1}{l|}{LLaVA-Med~\cite{li2024llava}}                   & 20.54                                                 & \multicolumn{1}{c|}{19.60}                                                  & 26.12          & 20.20          & 29.00          & 20.31          & 16.30          & 18.46          & 15.23          & 21.84          & 20.86          & 16.73          & 21.69          & 19.23          & 20.18          & 18.38          & 20.99          & 16.87          & 20.49          & 21.55          \\
\multicolumn{1}{l|}{Qilin-Med-VL-Chat~\cite{liu2023qilin}}           & 22.34                                                 & \multicolumn{1}{c|}{22.06}                                                  & 12.94          & 21.06          & 15.50          & 22.09          & 18.98          & 17.33          & 17.88          & 22.92          & 31.14          & 29.82          & 20.00          & 21.83          & 25.55          & 19.07          & 14.81          & 29.42          & 22.17          & 22.29          \\
\multicolumn{1}{l|}{RadFM~\cite{wu2023generalist}}                       & 22.95                                                 & \multicolumn{1}{c|}{22.93}                                                  & 24.24          & 23.02          & 20.00          & 20.59          & 20.83          & 19.49          & 28.48          & 24.42          & 18.00          & 32.00          & 16.95          & 26.90          & 26.25          & 18.26          & 26.54          & 25.19          & 23.74          & 20.20          \\
\multicolumn{1}{l|}{MedDr~\cite{he2024meddr}}                       & 41.95                                                 & \multicolumn{1}{c|}{43.69}                                                  & 53.18          & 45.28          & 33.00          & 44.78          & 28.03          & 29.91          & 47.68          & 35.22          & 38.29          & 78.55          & 25.08          & 49.53          & 45.31          & 52.09          & 48.61          & 52.36          & 54.21          & 39.90          \\
\multicolumn{21}{c}{\cellcolor[HTML]{FFF3E4}Open-Source LVLMs}                                                                                                                                                                                                                                                                                                                                                                                                                                                                                                            \\
\multicolumn{1}{l|}{VisualGLM-6B~\cite{ding2021cogview}}                & 29.58                                                 & \multicolumn{1}{c|}{30.45}                                                  & 52.71          & 25.95          & 14.00          & 31.69          & 22.06          & 25.17          & 30.46          & 25.50          & 30.29          & 59.27          & 15.93          & 29.97          & 37.79          & 30.09          & 23.61          & 32.85          & 38.19          & 23.03          \\
\multicolumn{1}{l|}{Idefics-9B-Instruct~\cite{laurencon2023obelics}}         & 29.74                                                 & \multicolumn{1}{c|}{31.13}                                                  & 19.76          & 33.98          & 21.00          & 30.08          & 24.46          & 26.66          & 50.33          & 28.74          & 36.00          & 58.55          & 36.27          & 29.64          & 36.76          & 36.07          & 24.38          & 31.36          & 32.04          & 29.19          \\
\multicolumn{1}{l|}{InstructBLIP-7B~\cite{dai2024instructblip}}             & 31.80                                                 & \multicolumn{1}{c|}{30.95}                                                  & 27.06          & 28.99          & 17.50          & 34.24          & 21.78          & 25.84          & 43.05          & 29.15          & 19.14          & 53.09          & 27.46          & 28.64          & 31.99          & 34.58          & 30.25          & 30.76          & 41.09          & 31.28          \\
\multicolumn{1}{l|}{Mini-Gemini-7B~\cite{li2024mini}}              & 32.17                                                 & \multicolumn{1}{c|}{31.09}                                                  & 34.59          & 39.63          & 23.50          & 35.74          & 23.46          & 19.80          & 41.06          & 25.91          & 40.86          & 56.00          & 19.32          & 21.63          & 35.73          & 35.83          & 33.95          & 40.57          & 29.14          & 29.56          \\
\multicolumn{1}{l|}{MMAlaya~\cite{datacanvas2024mmalaya}}                     & 32.19                                                 & \multicolumn{1}{c|}{32.30}                                                  & 71.06          & 37.68          & 38.00          & 28.30          & 27.40          & 27.64          & 51.66          & 32.39          & 28.86          & 83.64          & 29.49          & 27.37          & 35.92          & 36.70          & 20.99          & 27.53          & 29.43          & 28.08          \\
% \multicolumn{1}{l|}{Qwen-VL~\cite{bai2023qwenvl}}                     & 34.80                                                 & \multicolumn{1}{c|}{36.05}                                                  & 39.53          & 41.59          & 40.50          & 28.69          & 20.74          & 26.77          & 45.03          & 28.82          & 56.57          & 73.09          & 39.32          & 41.39          & 39.23          & 43.36          & 33.64          & 35.74          & 45.15          & 42.73          \\
\multicolumn{1}{l|}{Yi-VL-6B~\cite{ai2024yi} }                    & 34.82                                                 & \multicolumn{1}{c|}{34.31}                                                  & 39.76          & 43.76          & 56.00          & 27.30          & 25.91          & 27.23          & 45.70          & 32.56          & 44.29          & 65.45          & 47.46          & 36.38          & 39.00          & 35.39          & 25.46          & 29.77          & 39.06          & 35.22          \\
% \multicolumn{1}{l|}{LLaVA-NeXT-vicuna-7B~\cite{liu2024llavanext}}        & 34.86                                                 & \multicolumn{1}{c|}{35.42}                                                  & 40.00          & 37.13          & 51.60          & 31.82          & 29.15          & 26.18          & 49.01          & 31.06          & 32.94          & 65.33          & 28.44          & 35.98          & 43.21          & 38.71          & 26.87          & 40.02          & 36.47          & 32.36          \\
\multicolumn{1}{l|}{Qwen-VL-Chat~\cite{Qwen-VL}}                & 35.07                                                 & \multicolumn{1}{c|}{36.96}                                                  & 36.47          & 39.63          & 36.50          & 27.08          & 20.79          & 27.64          & {\ul 60.93}    & 30.23          & 52.57          & 70.55          & 37.29          & 47.13          & 39.37          & 46.67          & 34.57          & 37.63          & 47.88          & 39.90          \\
\multicolumn{1}{l|}{CogVLM-Chat~\cite{wang2023cogvlm}}                 & 35.23                                                 & \multicolumn{1}{c|}{36.08}                                                  & 30.59          & 38.98          & 42.50          & 31.41          & 26.22          & 23.62          & 47.02          & 34.22          & 51.43          & 56.00          & 32.54          & 44.13          & 38.67          & 37.94          & 30.86          & 41.11          & 45.91          & 29.19          \\
% \multicolumn{1}{l|}{Monkey~\cite{li2024monkey}}                      & 35.48                                                 & \multicolumn{1}{c|}{36.39}                                                  & 38.59          & 39.52          & 35.00          & 29.74          & 20.97          & 25.73          & 52.98          & 28.90          & 48.29          & 68.00          & 34.24          & 41.46          & 40.78          & 45.23          & 31.79          & 39.27          & 45.91          & 42.49          \\
\multicolumn{1}{l|}{mPLUG-Owl2~\cite{ye2023mplug}}                  & 35.62                                                 & \multicolumn{1}{c|}{36.21}                                                  & 47.76          & 40.50          & 41.00          & 33.46          & 27.22          & 28.16          & 51.66          & 33.14          & 38.86          & 68.73          & 16.27          & 38.58          & 43.34          & 35.70          & 27.78          & 41.61          & 39.76          & 30.91          \\
% \multicolumn{1}{l|}{ShareCaptioner~\cite{chen2023sharegpt4v}}              & 36.37                                                 & \multicolumn{1}{c|}{36.19}                                                  & 37.88          & 35.50          & 45.50          & 35.63          & 25.54          & 28.16          & 56.29          & 31.15          & 27.14          & 64.00          & 35.59          & 38.52          & 39.65          & 38.57          & 30.56          & 44.05          & 36.68          & 40.15          \\
\multicolumn{1}{l|}{Emu2-Chat~\cite{sun2023generative}}                   & 36.50                                                 & \multicolumn{1}{c|}{37.59}                                                  & 27.53          & 35.83          & 27.50          & 34.41          & 28.49          & 29.35          & 60.26          & 36.63          & 34.00          & 64.73          & 28.81          & 44.79          & 43.20          & 37.69          & 37.50          & 41.86          & 43.18          & 35.34          \\
\multicolumn{1}{l|}{OmniLMM-12B~\cite{yu2024rlaifv}}                 & 37.89                                                 & \multicolumn{1}{c|}{39.30}                                                  & 39.53          & 37.46          & 41.50          & 36.18          & 27.36          & 28.00          & {\ul 60.93}    & 37.46          & 55.43          & 80.00          & 31.19          & 35.71          & 44.89          & 42.49          & 28.24          & 43.80          & 51.19          & 42.86          \\
% \multicolumn{1}{l|}{InternVL-Chat-V1.1~\cite{chen2023internvl}}          & 38.16                                                 & \multicolumn{1}{c|}{39.41}                                                  & 45.88          & 40.07          & 56.00          & 34.30          & 26.68          & 26.20          & 52.32          & 37.79          & 45.14          & 64.00          & 35.93          & 52.74          & 44.14          & 40.56          & 39.51          & 41.16          & 45.56          & 35.84          \\
\multicolumn{1}{l|}{LLAVA-V1.5-7B~\cite{liu2023visual}}               & 38.23                                                 & \multicolumn{1}{c|}{37.96}                                                  & 42.35          & 37.57          & 44.50          & 36.13          & 27.99          & 24.91          & 49.01          & 31.31          & 34.00          & 68.36          & 27.12          & 45.39          & 42.46          & 42.80          & 33.80          & 44.20          & 41.21          & 38.92          \\
% \multicolumn{1}{l|}{Monkey-Chat~\cite{li2024monkey}}                 & 38.39                                                 & \multicolumn{1}{c|}{39.50}                                                  & 43.53          & 40.28          & 40.00          & 33.30          & 23.28          & 29.09          & 54.97          & 29.73          & 55.71          & 72.36          & 35.25          & 50.53          & 42.41          & 45.98          & 33.49          & 42.66          & 50.15          & 44.83          \\
% \multicolumn{1}{l|}{LLAVA-V1.5-7B-xtuner~\cite{2023xtuner}}        & 38.68                                                 & \multicolumn{1}{c|}{38.22}                                                  & 51.53          & 35.07          & 31.00          & 38.07          & 31.52          & 29.04          & 58.94          & 36.79          & 28.29          & 69.09          & 29.15          & 50.80          & 39.89          & 40.12          & 27.78          & 40.82          & 39.12          & 36.08          \\
\multicolumn{1}{l|}{XComposer2~\cite{internlmxcomposer2}}                  & 38.68                                                 & \multicolumn{1}{c|}{39.20}                                                  & 32.71          & 42.13          & 70.50          & 33.13          & 29.62          & 27.02          & 54.30          & 34.05          & 23.14          & 83.64          & 39.66          & 46.53          & 44.23          & 45.73          & 28.86          & 45.55          & 41.32          & 41.87          \\
% \multicolumn{1}{l|}{LLAVA-InternLM-7b~\cite{2023xtuner}}           & 38.71                                                 & \multicolumn{1}{c|}{39.11}                                                  & 44.94          & 39.85          & 33.50          & 43.06          & 27.54          & 27.08          & 52.98          & 34.22          & 31.14          & 79.64          & 37.97          & 50.67          & 42.41          & 39.69          & 36.73          & 37.63          & 46.72          & 39.78          \\
\multicolumn{1}{l|}{TransCore-M~\cite{transcorem}}                 & 38.86                                                 & \multicolumn{1}{c|}{38.70}                                                  & 39.06          & 43.87          & 24.50          & 40.18          & 29.08          & 30.79          & 52.98          & 32.48          & 38.86          & 66.91          & 42.37          & 42.79          & 44.75          & 40.44          & 36.73          & 34.00          & 47.19          & 35.71          \\
\multicolumn{1}{l|}{InternVL-Chat-V1.5~\cite{chen2024far}}          & 38.86                                                 & \multicolumn{1}{c|}{39.73}                                                  & 36.47          & 44.84          & 53.50          & 37.07          & 26.63          & 31.61          & 60.26          & 34.14          & 36.29          & 67.27          & 37.63          & 55.21          & 47.13          & 38.69          & 41.98          & 39.17          & 37.55          & 41.26          \\
% \multicolumn{1}{l|}{InternVL-Chat-V1.2-Plus~\cite{chen2023internvl}}     & 39.41                                                 & \multicolumn{1}{c|}{40.79}                                                  & 51.06          & 43.54          & 60.00          & 39.07          & 29.39          & {\ul 31.82}    & 50.99          & 37.54          & 54.00          & 79.64          & 30.17          & 50.87          & 43.72          & 37.88          & 36.88          & 42.61          & 43.53          & 38.55          \\
% \multicolumn{1}{l|}{InternVL-Chat-V1.2~\cite{chen2023internvl}}          & 39.52                                                 & \multicolumn{1}{c|}{40.01}                                                  & 40.71          & 46.25          & 77.50          & 31.52          & 26.36          & 31.10          & 50.33          & 36.96          & 52.00          & 80.00          & 31.19          & 45.46          & 43.20          & 40.06          & 34.10          & 44.40          & 46.66          & {\ul 42.36}    \\
\multicolumn{1}{l|}{LLAVA-InternLM2-7b~\cite{2023xtuner}}          & 40.07                                                 & \multicolumn{1}{c|}{40.45}                                                  & 43.53          & 40.72          & 60.50          & 34.74          & 30.12          & 27.44          & 51.66          & 33.39          & 50.86          & 74.55          & 26.44          & 49.13          & 42.74          & 43.12          & 31.94          & 50.87          & 47.01          & 39.04          \\
% \multicolumn{1}{l|}{DeepSeek-VL-1.3B~\cite{lu2024deepseek}}            & 40.25                                                 & \multicolumn{1}{c|}{40.77}                                                  & 56.71          & 37.13          & 27.00          & 45.73          & 28.40          & 27.85          & 52.32          & 35.96          & 45.43          & 71.64          & 45.42          & 50.20          & 41.66          & 47.48          & 37.81          & 43.90          & 45.50          & 33.50          \\
% \multicolumn{1}{l|}{MiniCPM-V~\cite{hu2024large}}                   & 40.95                                                 & \multicolumn{1}{c|}{41.05}                                                  & 28.47          & 42.02          & 40.00          & 42.79          & 28.80          & 28.62          & 46.36          & 36.30          & 40.00          & 67.27          & 31.53          & 42.46          & 44.04          & 50.28          & 37.50          & 51.92          & 52.29          & 27.22          \\
\multicolumn{1}{l|}{DeepSeek-VL-7B~\cite{lu2024deepseek}}              & 41.73                                                 & \multicolumn{1}{c|}{43.43}                                                  & 60.00          & 43.97          & 47.50          & 45.12          & 28.22          & 31.20          & 46.36          & 32.97          & 52.29          & 67.64          & \textbf{61.36} & 49.27          & 44.23          & 49.97          & 52.78          & 45.00          & 53.63          & 38.79          \\
\multicolumn{1}{l|}{MiniCPM-V2~\cite{xu2024llava-uhd}}                  & 41.79                                                 & \multicolumn{1}{c|}{42.54}                                                  & 37.88          & 43.65          & 35.50          & 42.67          & 26.49          & 29.24          & 37.75          & 33.31          & {\ul 59.71}    & 67.27          & 38.64          & 50.87          & 42.64          & 50.59          & 40.90          & 51.07          & 57.81          & 35.10          \\
\multicolumn{21}{c}{\cellcolor[HTML]{FFF0F0}Proprietary LVLMs}                                                                                                                                                                                                                                                                                                                                                                                                                                                                                                             \\
\multicolumn{1}{l|}{Claude3-Opus~\cite{anthropic2024claude}}                & 32.37                                                 & \multicolumn{1}{c|}{32.44}                                                  & 38.59          & 34.42          & 43.50          & 27.97          & 22.96          & 23.62          & 52.32          & 25.42          & 25.14          & 66.91          & 15.93          & 35.25          & 41.06          & 36.07          & 37.50          & 40.67          & 35.40          & 34.24          \\
\multicolumn{1}{l|}{Qwen-VL-Max~\cite{Qwen-VL}}                 & 41.34                                                 & \multicolumn{1}{c|}{42.16}                                                  & 50.59          & 47.23          & \textbf{74.00} & 40.68          & 29.03          & 26.71          & 58.94          & 34.05          & 62.29          & 85.45          & 27.80          & 44.39          & 43.90          & 42.99          & 48.61          & 49.38          & 51.13          & 40.52          \\
\multicolumn{1}{l|}{GPT-4V~\cite{achiam2023gpt}}                      & 42.50                                                 & \multicolumn{1}{c|}{44.08}                                                  & {\ul 64.00}    & 44.95          & 58.50          & 42.45          & 30.03          & 29.40          & 58.28          & 32.31          & 54.57          & 83.27          & 37.63          & 48.26          & 49.04          & 48.41          & 44.60          & 51.87          & 53.98          & 40.89          \\
\multicolumn{1}{l|}{Gemini 1.0~\cite{team2023gemini}}                  & 44.38                                                 & \multicolumn{1}{c|}{44.93}                                                  & 57.41          & 46.25          & 57.50          & 36.40          & 28.67          & 27.80          & 45.03          & {\ul 38.21}    & 58.57          & 86.55          & 40.68          & {\ul 51.74}    & 47.45          & 55.64          & 50.46          & 47.83          & {\ul 61.58}    & 41.87          \\
\multicolumn{1}{l|}{Gemini 1.5~\cite{reid2024gemini}}                  & {\ul 47.42}                                           & \multicolumn{1}{c|}{{\ul 48.36}}                                            & 55.29          & \textbf{50.81} & 54.00          & {\ul 51.05}    & \textbf{36.59} & 29.86          & 56.95          & 36.88          & 58.00          & {\ul 88.00}    & {\ul 47.46}    & 48.13          & {\ul 51.19}    & {\ul 56.88}    & {\ul 64.51}    & {\ul 56.50}    & 59.78          & 31.65          \\
\multicolumn{1}{l|}{GPT-4o~\cite{achiam2023gpt} }                      & \textbf{53.53}                                        & \multicolumn{1}{c|}{\textbf{53.96}}                                         & \textbf{66.82} & {\ul 48.53}    & {\ul 64.50}    & \textbf{55.94} & {\ul 35.10}    & \textbf{48.53} & \textbf{74.17} & \textbf{43.52} & \textbf{64.57} & \textbf{91.64} & 37.63          & \textbf{57.88} & \textbf{55.21} & \textbf{62.80} & \textbf{66.98} & \textbf{58.39} & \textbf{64.60} & \textbf{46.18} \\ \hline
\end{tabular}
\label{tab:departmentsresults}
}
\vspace{-0.2cm}
\end{table}

\subsection*{Analysis}
After reviewing the evaluation results, we have drawn \textbf{2 conclusions} and identified \textbf{5 insufficiencies} that require further improvement in future LVLMs in the medical domain:

\textbf{Conclusion 1. Medical tasks are still challenging for all LVLMs:} Our GMAI-MMBench provides a comprehensive multitask challenge, revealing that even the most advanced model, GPT-4o, is limited to an accuracy of around 54\% (see Table~\ref{tab:metatasksresults} and Table~\ref{tab:departmentsresults}). This does not meet the clinical requirement and indicates that all current LVLMs in the medical domain still require significant improvement. 

\textbf{Conclusion 2. Open-source models are catching up to the commercialized models:} In the comparison between open-source and commercialized models, most open-source models lag behind their commercialized counterparts. Leading open-source models such as MedDr and DeepSeek-VL-7B, although not as accurate as GPT-4o, have surpassed Claude3 Opus and Qwen-VL-Max, approaching the performance of GPT-4V. This suggests that open-source models in the medical field are gradually catching up to the top-performing commercialized models.

\textbf{Insufficiency 1. Performance on different clinical VQA tasks needs improvement:} Table~\ref{tab:metatasksresults} shows that the best-performing clinical VQA tasks are Disease Diagnosis (DD) and Nervous Tissue (NT), with models exceeding the random baseline by an average of over 10\%. However, in clinical VQA tasks such as Severity Grading (SG) and Attribute Recognition (AR), most LVLMs face challenges, and most of them perform worse than the random baseline. Overall, despite the advanced models like GPT-4o and Gemini 1.5 significantly outperforming the random baseline, there remains a substantial gap between their performance and the requirements of real-world applications, indicating that all the models still need more specialized medical knowledge for training.

\textbf{Insufficiency 2. The performance across different departments needs further balancing:} In examining performance across different medical departments, as shown in Table~\ref{tab:departmentsresults}, we found that the Infectious Diseases (ID) and Neurosurgery (N) departments performed the best. In contrast, departments such as General Surgery (GS) and Obstetrics and Gynecology (OG) showed a need for improvement, as the performance of all models in these areas did not significantly exceed the random baseline compared to other departments. This indicates that current large models exhibit specialization biases, suggesting that future development of LVLMs aiming to achieve general medical AI should focus on balancing capabilities across all departments.

\begin{wrapfigure}{r}{0.6\textwidth}
    \centering
    \vspace{-1em}
    \includegraphics[width=0.98\linewidth]{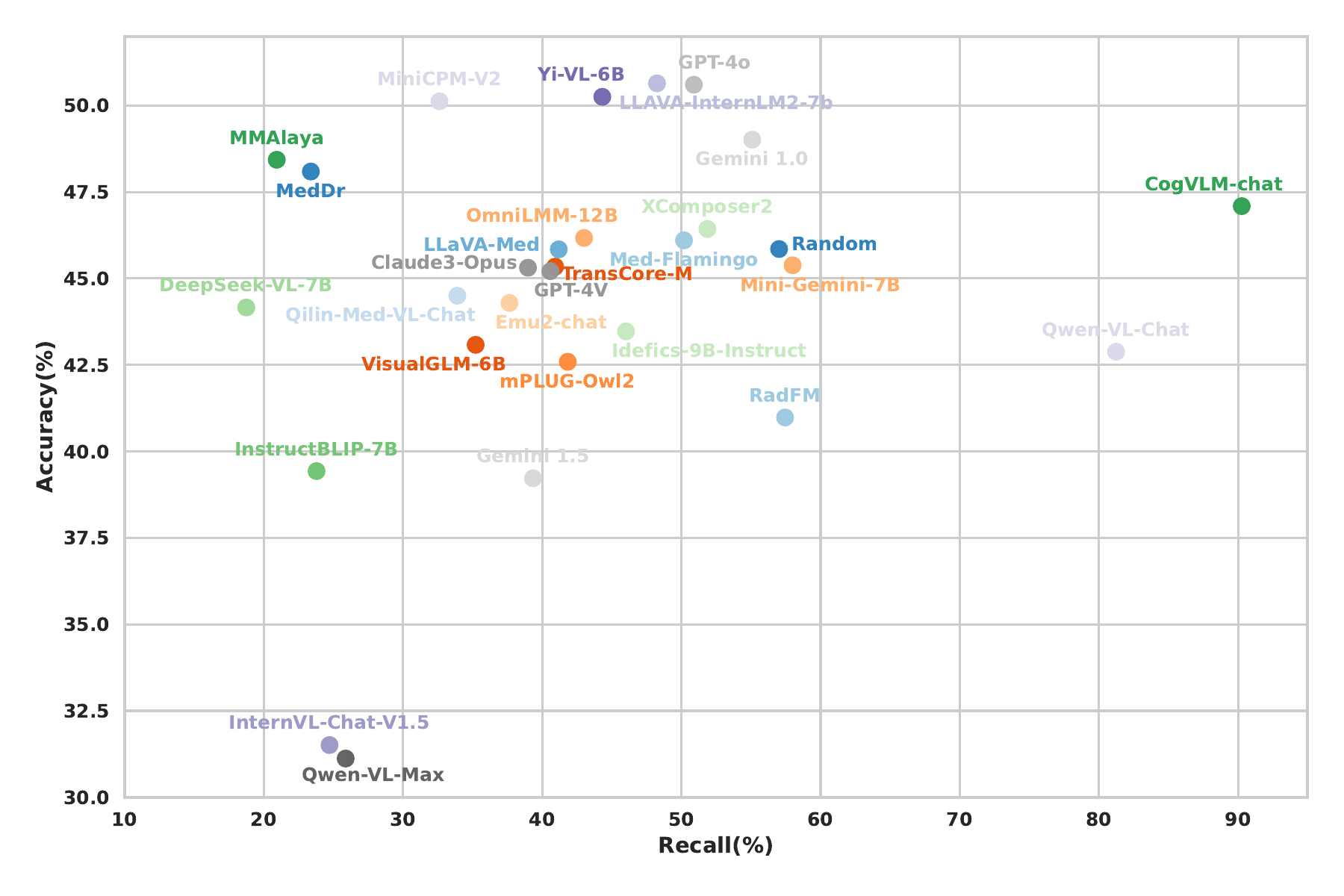}
    \vspace{-1.5em}
    \caption{Overall results for multiple-choice questions of different models.}\label{fig:mcls_comparison}
    % \vspace{-1em}
\end{wrapfigure}

\textbf{Insufficiency 3. The LVLMs are not robust among different perceptual types:} As shown in Figure~\ref{fig: different perceptual types results}, models perform slightly better with contour-level perception compared to mask-level perception, and both outperform image-level perception (without annotation) significantly. However, bounding box-level perception shows the worst performance among all perceptual types, indicating that models are sensitive to this perceptual type. This evaluation underscores the need for LVLMs to address robustness issues across different perceptual types, which is crucial for their effectiveness in interactive applications.

\textbf{Insufficiency 4. Medical-specific models need to enhance their instruction tuning:} Interestingly, medical-specific models significantly underperform compared to general models, despite being trained and fine-tuned directly on relevant medical data. Specifically, LLaVA-Med is fine-tuned from the LLaVA model series in the medical field, but its performance is even worse than LLAVA-V1.5-7B. The primary reason for the poor performance of these medical-specific models is their inability to follow instructions correctly and their failure to understand or answer medical-related questions accurately. Detailed analysis can be found in the case study and supplementary materials sections on medical model analysis. Among these, the best-performing medical-specific model is MedDr, which is fine-tuned from the InternVL series and successfully surpasses the InternVL-Chat-V1.5. Unlike other medical-specific models that derive instruction-tuning data from papers, online sources, and books, MedDr builds its dataset based on high-quality medical image classification datasets. This result suggests that the quality of currently available medical instruction tuning datasets on the internet needs improvement and highlights the effectiveness of MedDr's dataset construction strategy, serving as a valuable reference for future medical-specific models.

\textbf{Insufficiency 5. The performance of most LVLMs on multiple-choice questions needs improvement:} 
Based on our tests, none of the models can totally match the correct answers (they always miss or over-select), so we adopt a relatively loose evaluation method for multiple-choice questions: using multi-choice hit rate (\(\mathrm{ACC_{mcq}}\)) and recall rate (\(\mathrm{Recall_{mcq}}\)). The experimental results are shown in Figure~\ref{fig:mcls_comparison}. Using this method, we found that most models have an accuracy rate of around 40\%-50\% and a recall rate of around 40\%-60\%. Surprisingly, InternVL-Chat-V1.5 and Qwen-VL-Max performed well in single-choice questions but showed very poor recall and accuracy rates in multiple-choice questions. In contrast, Qwen-VL-Chat and CogVLM-Chat, which performed relatively poorly in single-choice questions, achieved very high recall rates and moderate accuracy rates in multiple-choice questions, especially CogVLM-Chat with over 90\% recall rate. Nonetheless, even with this less strict evaluation method, all models had accuracy rates below 55\%, indicating that there is still significant room for improvement in answering multiple-choice questions.

% When evaluating the performances on multiple-choice questions, we applied \(\mathrm{ACC_{mcls}}\) and \(\mathrm{Recall_{mcls}}\) to evaluate the models. This metric can better differentiate model performance when compared with macro-averaged accuracy in single-choice questions because absolute ACC will lead the accuracy below 5\% for all models. %we did not use a simple ratio of correctly answered questions to total questions as with single-choice questions \eqref{eq1}. Instead, we assigned partial points for each selected option and calculated recall accuracy through \ref{eq2}, this adjustment was necessary because applying \ref{eq1} would result in accuracies below 5\% for all models. while using \ref{eq2}, we can better differentiate model performance, 
% Unable to evaluate by absolute ACC highlights that current LVLMs are more robust for "single in, single out" tasks but struggle with "single in, multiple out" tasks. Besides, as shown in Figure~\ref{fig:mcls_comparison}, it is notable that CogVLM-Chat maintains a relatively high \(\mathrm{ACC_{mcq}}\) and nearly 90\% \(\mathrm{Recall_{mcq}}\), even superior to GPT-4o and Gemini 1.5, which demonstrated that open-source models are on par with, or even surpass commercial models in certain areas.

\begin{figure*}[t]
\centering
\includegraphics[width=1.0\textwidth]{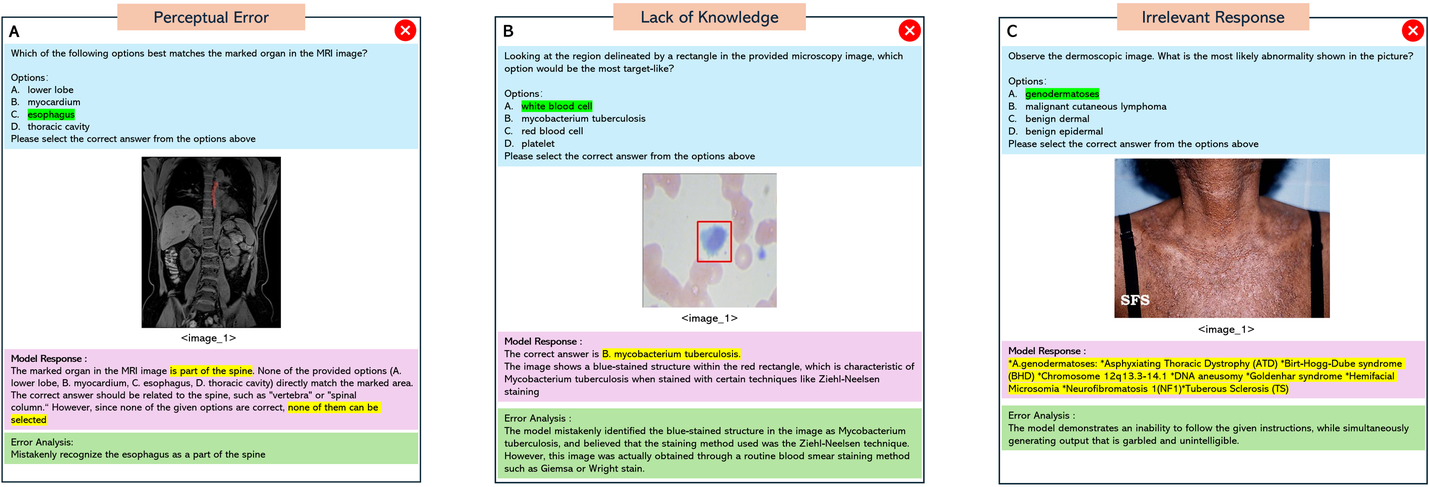}
\caption{Three examples of error cases. \textbf{A:} Question misunderstanding. \textbf{B:} Perceptual Errors. \textbf{C:} Lack of Knowledge. More studies can be found in the appendix.}
\label{fig:casestudy}
\vspace{-1.5em}
\end{figure*}

\subsection*{Case Study}
We further analyze the results by requiring the models to output content beyond the provided options and explain their reasoning process. This approach helps us better understand the causes of errors. Through detailed testing and analysis, we identify 5 typical errors present in the LVLMs:

\textbf{Question misunderstanding:}
This occurs when the model incorrectly understands the purpose of the question, leading to an inability to provide a correct response. As shown in Figure~\ref{fig:casestudy}A, the model is asked to answer a multiple-choice question, but it describes the problem or repeats the options rather than choosing an option.

\textbf{Perceptual Error:} These errors occur when there is a mislocation or misrecognition of image content. This means that the model's understanding or interpretation of the visual content is incorrect, leading to an inaccurate response. As shown in Figure~\ref{fig:casestudy}B, the model mistakenly identifies the esophagus as the spine, suggesting that while the model can locate the target on the image (The annotated esophagus is very close to the spine), it makes an error in perceiving the masked content. 

% In general, there are two types of perceptual errors. One type is the ability to correctly identify the general information of an image, such as identifying the location of the disease or the organs indicated in the image, etc., but overlooking the details present in the image; this is termed \textbf{PE - detail missing}. The other type is the inability to recognize the image's general information, which we call \textbf{PE - misinterpretation}. %The model can interpret textual and image input correctly but fails in accurately perceiving images. 
% %These errors occur when there is a mislocation or misrecognition of image content. This means that the model's understanding or interpretation of the visual content is incorrect, leading to an inaccurate response.
% Here we demonstrate a typical PE - misinterpretation in Figure~\ref{fig:casestudy}A, the annotated content is the esophagus, whereas the position is very close to the spine, consequently, the model mistakenly identifies the esophagus as the spine. Based on the model's output, it is evident that the model recognizes the image but makes an error in perceiving the annotated content, leading to incorrect identification of the structure.

%\begin{figure*}[htbp]
%\centering
%\includegraphics[width=0.6\textwidth]{imgs/perceptionerror.png}
%\caption{Two instances of perceptual errors.}
%\label{fig:perceptionerror}
%\end{figure*}

\textbf{Lack of knowledge:} While the model can recognize text and images, it makes errors in specific areas that require specific knowledge, indicating a deficiency in relevant training or fine-tuning in those areas. For example, in Figure~\ref{fig:casestudy}C, the model incorrectly identifies the staining method as Ziehl-Neelsen and misrecognizes the blue-stained structure as Mycobacterium tuberculosis, where it is actually a white blood cell stained with Giemsa or Wright stain. This error indicates the model's lack of knowledge in experimental medicine.

%\begin{figure*}[htbp]
%\centering
%\includegraphics[width=0.6\textwidth]%{imgs/lackofknowledge.png}
%\caption{Two instances of lack of knowledge errors.}
%\label{fig:lackofknowldege}
%\end{figure*}

\textbf{Irrelevant Responses:} This error indicates the model fails to generate a readable answer, which is easily found in medical-specific models like RadFM. Examples are listed in the appendix.
% Figure~\ref{fig:casestudy}C depicts a garbled, unintelligible output. This example indicates a low performance on the language part, which might be due to inadequate training or fine-tuning on the language processing. The frequent occurrence of such issues highlights deficiencies of medical-specific models in natural language processing, underscoring the need for improving medical-specific models' ability in natural language processing to ensure their effectiveness and reliability in real-world applications.

\textbf{Reject to Answer:} Some models, especially proprietary LVLMs like GPT-4V, GPT-4o, Gemini 1.0, and Gemini 1.5, commonly refuse to provide an answer due to policy reasons, because safety is crucial according to the commercial rules and regulations. Many potentially risky responses are declined to ensure compliance with guidelines. Those models' strict adherence to safety protocols and ethical standards limits response capabilities in certain domains. 

% In addition to the above, other types of error, such as question misunderstanding, and image misinterpretation, can be found in the supplementary material, which includes more comprehensive case analyses. By identifying these typical errors, we can better understand the limitations of current LVLMs and focus on the weaknesses to further improve their performance.

\vspace{-1.2em}

\section*{Conclusion}

\vspace{-1em}

The development of GMAI-MMBench as a benchmark for evaluating LVLMs' capabilities represents a significant advancement in the pursuit of general medical AI. GMAI-MMBench epitomizes the expertise of skilled medical professionals, serving as a pivotal guide for advancing large models toward GMAI by testing the limits of current LVLMs. Owing to the extensive and diverse source of GMAI-MMBench, which comprises medical datasets annotated by professional healthcare providers worldwide, this benchmark can comprehensively evaluate the model's capability across various specific aspects. In this way, GMAI-MMBench can guide the model development at a more fine-grained level, accelerating the development of robust and reliable GMAI systems. Moreover, this benchmark supports the advancement of interactive multimodal medical models by providing more perceptual modes and annotations that are commonly used by physicians in clinical practice, thereby creating a framework for their evaluation and improvement.

However, GMAI-MMBench, like all benchmarks, has its limitations. The manual curation process, despite being thorough, might introduce biases, and focusing solely on medical subjects may not fully meet the criteria for general medical AI as defined. Nevertheless, we assert that high performance on GMAI-MMBench is essential for demonstrating the extensive subject knowledge and expert-level reasoning skills required for general medical AI. Looking ahead, we intend to integrate human evaluations into GMAI-MMBench. This addition will offer a more grounded comparison between model capabilities and expert performance, providing insights into how close current AI systems are achieving general medical AI in the medical field.

% \bibliographystyle{plain}
% \bibliography{citation}

% \putbib
% \end{bibunit}

\bibliographystyle{plain}
\bibliography{citation}

\newpage

% \begin{bibunit}

\begin{center}
  \Large\bfseries GMAI-MMBench: A Comprehensive Multimodal
Evaluation Benchmark Towards General Medical AI
  \\[1em]
  \ Supplementary Materials
\end{center}
\addcontentsline{toc}{section}{Supplementary Materials}

%\section*{Supplementary Materials}

\setcounter{figure}{5}
\setcounter{table}{3}

\maketitle
\appendix

% \newpage
\tableofcontents
% \newpage

%\renewcommand*\contentsname{Contents in Appendix}
%\tableofcontents

\section{Related work}
\subsection{Large Vision-Language Model(LVLMs)}
In contrast to traditional deep learning models, Large Vision-Language Models (LVLMs) offer a broader spectrum of possibilities for AI-assisted healthcare. Their user-friendly and intuitive interaction mechanisms make them one of the most promising paradigms for future AI applications. Among the multitude of LVLMs, prominent proprietary models such as GPT-4o~\cite{achiam2023gpt}, Claude3-opus~\cite{anthropic2024claude}, and Qwen-max~\cite{Qwen-VL} exemplify the pinnacle of contemporary general-purpose large models. Additionally, numerous open-source general-purpose models have emerged, including the InternVL series~\cite{chen2023internvl, chen2024far}, LLAVA series~\cite{liu2024llavanext, liu2023visual, chen2023sharegpt4v}, DeepSeek series~\cite{lu2024deepseek}, CogVLM series~\cite{wang2023cogvlm}, InstructBLIP series~\cite{dai2024instructblip}, Idefics series~\cite{laurencon2023obelics}, XComposer series~\cite{chen2023sharegpt4v, internlmxcomposer, internlmxcomposer2, internlmxcomposer2_4khd}, Yi-VL series~\cite{ai2024yi}, Xtuner series~\cite{2023xtuner}, and MiniCPM series~\cite{hu2024large, xu2024llava-uhd}. These open-source models are rapidly evolving due to their accessibility and collaborative development.

To address specialized medical tasks, researchers have trained and fine-tuned these large models using domain-specific medical data, resulting in specialized large models. Noteworthy examples include LLaVA-Med~\cite{li2024llava} derived from the LLAVA series, and MedDr~\cite{he2024meddr} based on the InternLM framework. The advent of these specialized medical models has laid a solid foundation for the application of LVLMs in the healthcare sector, highlighting their transformative potential and accelerating their development within the medical domain.

\subsection{Benchmarks}
In the swiftly emerging and burgeoning domain of LVLMs, the significance of rigorous evaluation cannot be overstated. Benchmarking serves as a crucial metric for guiding model enhancement, identifying deficiencies, and steering the trajectory of model development. Within the medical domain, benchmarks are typically categorized into specialized and general-purpose benchmarks.

Specialized benchmarks are often concentrated on a particular modality or medical discipline. For instance, VQA-RAD~\cite{lau2018dataset}, SLAKE~\cite{liu2021slake}, and RadBench~\cite{wright2016radbench} focus on radiology, while PathVQA~\cite{he2020pathvqa} and PathMMU~\cite{sun2024pathmmu} are dedicated to pathology. These benchmarks provide a wealth of evaluation data for specific modalities or disciplines, enabling comprehensive assessment of capabilities within targeted fields. However, their limited generalizability constrains their broader applicability.

In addition to these specialized benchmarks, there exist general-purpose medical benchmarks that span multiple medical domains. Prominent examples include MMMU~\cite{yue2024mmmu}, OminimedVQA~\cite{hu2024omnimedvqa}, and MMT-Bench~\cite{ying2024mmt}. These comprehensive benchmarks facilitate a more holistic evaluation of a model's overall competence in the medical field. Nonetheless, these general-purpose benchmarks often exhibit shortcomings in various aspects such as the volume of tasks, number of modalities, data distribution, and granularity of data. Addressing these limitations presents a significant challenge that necessitates prompt resolution.

The development and refinement of benchmarks are indispensable for the progress of LVLMs in healthcare. By elucidating the capabilities and limitations of specialized and general-purpose benchmarks, it becomes evident that while specialized benchmarks excel in evaluating domain-specific performance, their lack of generalizability is a notable drawback. Conversely, general-purpose benchmarks offer a broader assessment across multiple medical fields but often fall short in task diversity, modality coverage, and data granularity. Therefore, there is an urgent need for more comprehensive and robust benchmarks to bridge these gaps and better support the advancement of LVLMs in healthcare.

\section{Dataset Details}
In this section, we provide the detailed datasets used in GMAI-MMBench, including the name of the dataset or challenge, the number of sub-datasets in it, the modality, the dimension of data, the task type, and the number of cases. As shown in Table~\ref{tab:dataset_statistics}, GMAI-MMBench is constructed from 284 datasets across 38 medical image modalities. These datasets are derived from the public (268) and several hospitals (16) that have agreed to share their ethically approved data.

\small
\begin{longtable}{L{0.34\textwidth}|C{0.01\textwidth}|C{0.3\textwidth}|C{0.04\textwidth}|C{0.06\textwidth}|C{0.05\textwidth}}
% \centering
% \captionsetup{
%         width=0.95\textwidth,
%         font=small,
%         labelformat=simple,
%         labelsep=colon,
%         justification=centering
%     }
\caption{Detailed datasets information in GMAI-MMBench. As one challenge/dataset may contain several sub-tasks or sub-challenges in the medical area, we count them in the ``N'' (second column). In the dimension (Dim) column, 2D and 3D denote the dimensions of the original data, respectively. In the task type (Task) column, Cls, MCls, Seg, and Det indicate classification, multi-label classification, segmentation, and detection, respectively. The count represents the number of cases used in GMAI-MMBench.}
\label{tab:dataset_statistics} \\
    \hline
    Challenge / Dataset                                           & N & Modality                                                            & Dim & Task        & Count \\
    \hline
    \endfirsthead
    % \caption[]{(Continued)} \\
    % \hline
    % Challenge / Dataset                                           & N & Modality                                                            & Dim & Task        & Count \\
    % \hline
    \endhead
    \hline
    \endfoot
    \hline
    \endlastfoot

% \begin{tabular}{llllll}

 \href{https://github.com/kc-santosh/medical-imaging-datasets}{5K+ CT Images on Fractured Limbs}~\cite{ruikar20215k+}                              & 1                 & CT                                                                  & 2D  & Cls   & 60         \\
 \href{https://www.aapm.org/GrandChallenge/RT-MAC/}{AAPM RT-MAC 2019}~\cite{cardenas2019rtmac}                                              & 1                 & T2 weighted MRI                                                     & 3D  & Seg     & 68         \\
\href{https://github.com/MrGiovanni/AbdomenAtlas}{Abdomenatlas 1.0}~\cite{qu2024abdomenatlas}                                              & 1                 & CT                                                                  & 3D  & Seg     & 52         \\
\href{https://github.com/JunMa11/AbdomenCT-1K}{AbdomenCT-1K}~\cite{ma2021abdomenct}                                                  & 1                 & CT                                                                  & 3D  & Seg     & 28         \\
\href{https://paperswithcode.com/dataset/acdc}{ACDC 2017}~\cite{bernard2018deep}                                                     & 1                 & MRI                                                                 & 3D  & Seg     & 10         \\
\href{https://www.kaggle.com/datasets/toaharahmanratul/acrima-dataset}{ACRIMA}~\cite{diaz2019cnns}                                                        & 1                 & Fundus Photography                                                  & 2D  & Cls   & 1          \\
\href{https://ieee-dataport.org/documents/adam-automatic-detection-challenge-age-related-macular-degeneration}{ADAM 2020}~\cite{fang2022adam}                                                     & 1                 & Fundus Photography                                                  & 2D  & Cls   & 1          \\
\hline
\end{longtable}
% \end{tabular}%
% }
% \end{table}
\normalsize

\clearpage

\section{Details of Well-categorized Data Structure}
\label{headings}
% \begin{table*}[ht]
% \caption{Statistics of clinical VQA tasks and its sub-task abbreviations mentioned in the paper with their corresponding full names.}
% \begin{center}
% %\resizebox{1.0\textwidth}{!}{
% \begin{tabular}{c|c|ccc}
% \hline
% Abbreviation & Full name & \#Modalities & \#Labels & \#Cases \\ 
% \hline
% ASR & Anatomical Structure Recognition  & 10 & 20 & 1000\\
% \hline 
% \end{tabular}
% \label{tab:stat_clinical_vqa_task}
% %}
% \end{center}
% % \vspace{-2em}
% \end{table*}

\subsection{Data Statistics}
\begin{figure}[H]
    \centering
    \includegraphics[width=1\linewidth]{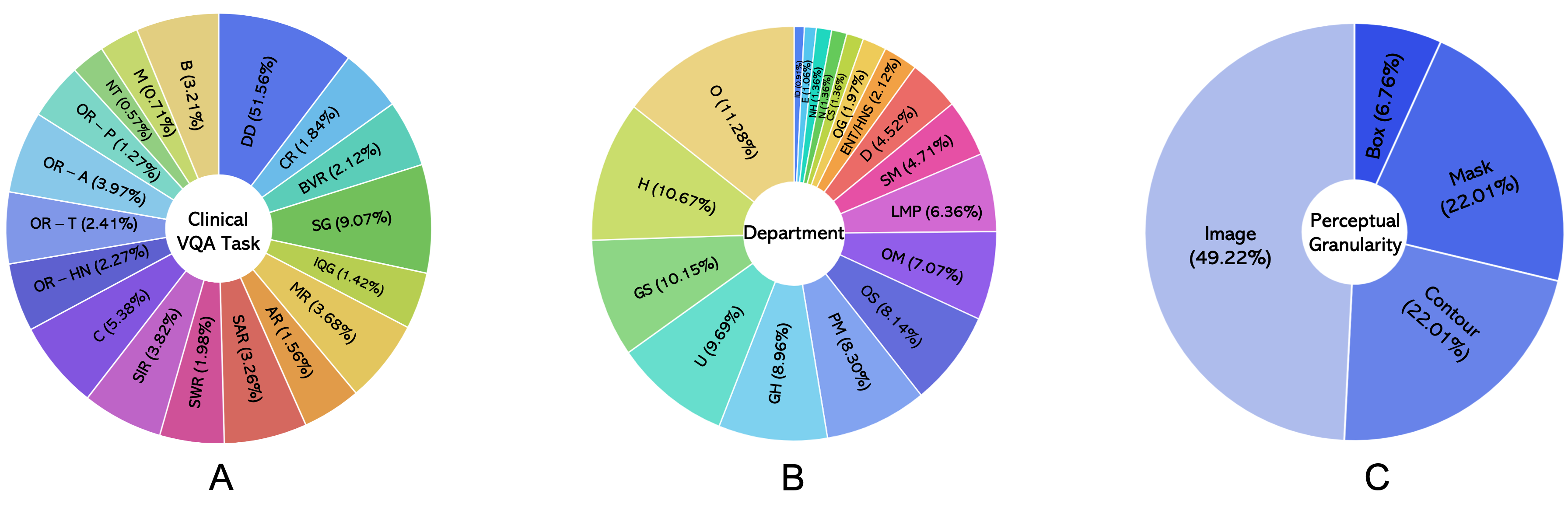}
    \caption{Label distribution for clinical VQA tasks, departments, and perceptual granularities.}
    \label{fig:stat_different_tasks}
\end{figure}

% TODO: lexical Tree（画HTML系统）；3个饼状图；（Departments、Clinical VQA Tasks、Granularity）的类别数量和case数量
In this section, we present the comprehensive statistical information of GMAI-MMBench. Figure~\ref{fig:stat_different_tasks} offers a global view of the label distribution proportions for different clinical VQA tasks, departments, and perceptual granularities. The left pie chart (A) shows the distribution of clinical VQA tasks, with Disease Diagnosis (DD) being the most prevalent at 51.6\%, followed by  Severity Grading (SG) at 9.1\%, Counting (C) at 5.4\%, and Organ Recognition – Abdomen (OR-A) at 4.0\%. The middle pie chart (B) depicts the distribution of cases across various departments, where Ophthalmolog (O) has the highest proportion at 11.3\%, followed by Hematology (H) at 10.7\%, General Surgery (GS) at 10.2\%, and Urolog (U) at 9.7\%. The right pie chart (C) represents the distribution of perceptual granularities, with Image Level accounting for the largest share at 49.2\%, followed by Mask Level at 22.0\%, and Contour Level at 22.0\%.
Specifically, Table~\ref{tab:stat_clinical_vqa_task} provides the statistical details for different clinical VQA tasks, including their full terms, abbreviations, and the number of questions associated with each task. Table~\ref{tab:stat_department} presents the statistical information for different departments, including each department's full term, abbreviation, and the number of questions contained within each department. Table~\ref{tab:granularity} shows the statistical information for different granularity. In the detailed tables, the statistical information for multiple-choice questions is also included, \textbf{specially, for multiple-choice questions, we count the frequency of choice appearances rather than the actual number of cases.}

\clearpage

\begin{table*}[ht]
\caption{Statistics of the clinical VQA tasks and their sub-task abbreviations mentioned in the paper with their corresponding full terms. }
\centering
\resizebox{\columnwidth}{!}{%
\begin{tabular}{l|c|ccc|ccc}
\hline
\multirow{2}{*}{Full Name}        & \multirow{2}{*}{Abbreviation} & \multicolumn{3}{c|}{Single Choice}               & \multicolumn{3}{c}{Multiple Choice} \\
                                  &                               & Modalities & Labels & Cases & Modalities    & Labels    & Cases   \\
\hline
Attribute Recognition             & AR                            & 5          & 26     & 780   & 1             & 4         & 40       \\
Blood Vessels Recognition         & BVR                           & 7          & 15     & 436   & -             & -         & -       \\
Bone                              & B                             & 6          & 22     & 655   & -             & -         & -       \\
Cell Recognition                  & CR                            & 4          & 13     & 383   & 1             & 18         & 7614       \\
Counting                          & C                             & 1          & 38     & 853   & -             & -         & -       \\
Disease Diagnosis                 & DD                            & 29         & 364    & 10167 & 3             & 26         & 8037       \\
Image Quality Grading             & IQG                           & 2          & 10     & 300   & -             & -         & -       \\
Microorganism Recognition         & MR                            & 3          & 26     & 779   & -             & -         & -       \\
Muscle                            & M                             & 1          & 5      & 150   & -             & -         & -       \\
Nervous Tissue                    & NT                            & 2          & 4      & 120   & -             & -         & -       \\
Organ Recognition - Abdomen       & OR-A                          & 7          & 28     & 838   & -             & -         & -       \\
Organ Recognition - Head and Neck & OR-HN                         & 5          & 16     & 480   & -             & -         & -       \\
Organ Recognition - Pelvic        & OR-P                          & 6          & 9      & 270   & -             & -         & -       \\
Organ Recognition - Thorax        & OR-T                          & 9          & 17     & 510   & -             & -         & -       \\
Severity Grading                  & SG                            & 5          & 64     & 1678  & -             & -         & -       \\
Surgeon Action Recognition        & SAR                           & 1          & 23     & 635   & -             & -         & -       \\
Surgical Instrument Recognition   & SIR                           & 1          & 27     & 790   & -             & -         & -       \\
Surgical Workflow Recognition     & SWR                           & 1          & 14     & 420   & -             & -         & -       \\
\hline
\end{tabular}%
}
\label{tab:stat_clinical_vqa_task}
\end{table*}

\begin{table*}[ht]
\caption{Statistics of the departments and their sub-task abbreviations mentioned in the paper with their corresponding full terms. }
\centering
\resizebox{\columnwidth}{!}{%
\begin{tabular}{l|c|ccc|ccc}
\hline
\multirow{2}{*}{Full Name}                 & \multirow{2}{*}{Abbreviation} & \multicolumn{3}{c|}{Single Choice} & \multicolumn{3}{c}{Multiple Choice} \\
                                           &                               & Modalities   & Labels   & Cases   & Modalities    & Labels    & Cases   \\
\hline
Cardiovascular Surgery                     & CS                            & 9            & 9        & 270     & 1             & 1         & 424       \\
Dermatology                                & D                             & 1            & 30       & 894     & -             & -         & -       \\
Endocrinology                              & E                             & 3            & 7        & 210     & -             & -         & -       \\
Gastroenterology and Hepatology            & GH                            & 7            & 60       & 1774    & -             & -         & -       \\
General Surgery                            & GS                            & 6            & 68       & 2009    & -             & -         & -       \\
Hematology                                 & H                             & 6            & 80       & 2112    & -             & -         & -       \\
Infectious Diseases                        & ID                            & 2            & 7        & 180     & -             & -         & -       \\
Laboratory Medicine and Pathology          & LMP                           & 2            & 45       & 1259    & 1             & 18         & 7614       \\
Nephrology and Hypertension                & NH                            & 4            & 9        & 270     & -             & -         & -       \\
Neurosurgery                               & N                             & 8            & 9        & 270     & -             & -         & -       \\
None (Attributes that do not belong to any department)                                       & N/A                          & 2            & 15       & 450     & -             & -         & -       \\
Obstetrics and Gynecology                  & OG                            & 5            & 14       & 389     & -             & -         & -       \\
Oncology (Medical)                         & OM                            & 20           & 51       & 1399    & -             & -         & -       \\
Ophthalmology                              & O                             & 6            & 97       & 2232    & 2             & 11         & 218       \\
Orthopedic Surgery                         & OS                            & 8            & 54       & 1611    & -             & -         & -       \\
Otolaryngology (ENT)/Head and Neck Surgery & ENT/HNS                       & 5            & 14       & 420     & 1             & 6         & 1015       \\
Pulmonary Medicine                         & PM                            & 2            & 55       & 1643    & 1             & 12         & 6420       \\
Sports Medicine                            & SM                            & 3            & 64       & 1919    & -             & -         & -       \\
Urology                                    & U                             & 8            & 33       & 933     & -             & -         & -      \\
\hline
\end{tabular}%
}
\label{tab:stat_department}
\end{table*}

\begin{table*}[ht]
\caption{Statistics of the perceptual granularities. $^*$ and $^\#$ denote the case for single choice and multiple choice, respectively.}
\centering
\resizebox{0.5\textwidth}{!}{%
\begin{tabular}{l|ccc}
\hline
Full Name     & Modalities & Labels & Cases \\
\hline
Mask Level    & 36         & 188    & 5587 \\
Contour Level & 36         & 188    & 5587 \\
Box Level     & 3          & 59     & 1715  \\
Image Level$^*$  & 13         & 474    & 12942 \\
Image Level$^\#$ & 5          & 48     & 15691 \\
\hline
\end{tabular}%
}
\label{tab:granularity}
\end{table*}

\subsection{Lexical Tree}
\begin{figure}[H]
    \centering
    \includegraphics[width=1.0\linewidth]{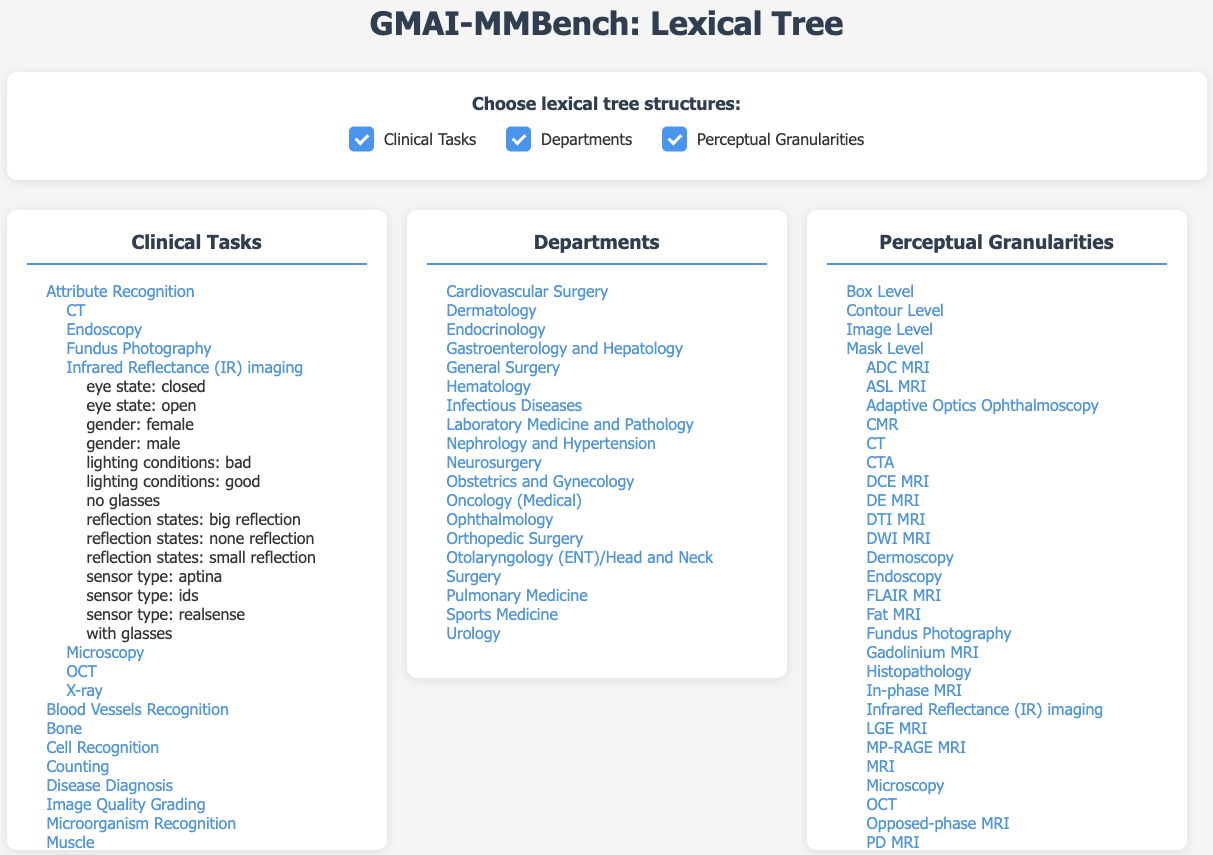}
    \caption{Overview of the lexical tree. The whole tree is provided in the attached HTML file named ``Lexical tree.html''. }
    \label{fig:lexical_tree}
\end{figure}

To make the GMAI-MMBench more intuitive and user-friendly, we have systematized our labels and structured the entire dataset into a lexical tree, which is presented in HTML format as shown in Figure~\ref{fig:lexical_tree}. Users can freely select the test contents based on this lexical tree. We believe that this customizable benchmark will effectively guide the improvement of models in specific areas. For instance, as mentioned in the main text, most models perform poorly at the bounding box level perception. Users can then update their models and test the accuracy at the bounding box level using this lexical tree, thereby achieving targeted improvements in model performance.

Here, we specifically demonstrate how to customize the use of the lexical tree. First, select the data we need to test based on the users' requirements. In this example, we will focus on \textbf{ophthalmology} department and only \textbf{fundus photography} modality.

\begin{figure}[H]
    \centering
    \includegraphics[width=1.0\linewidth]{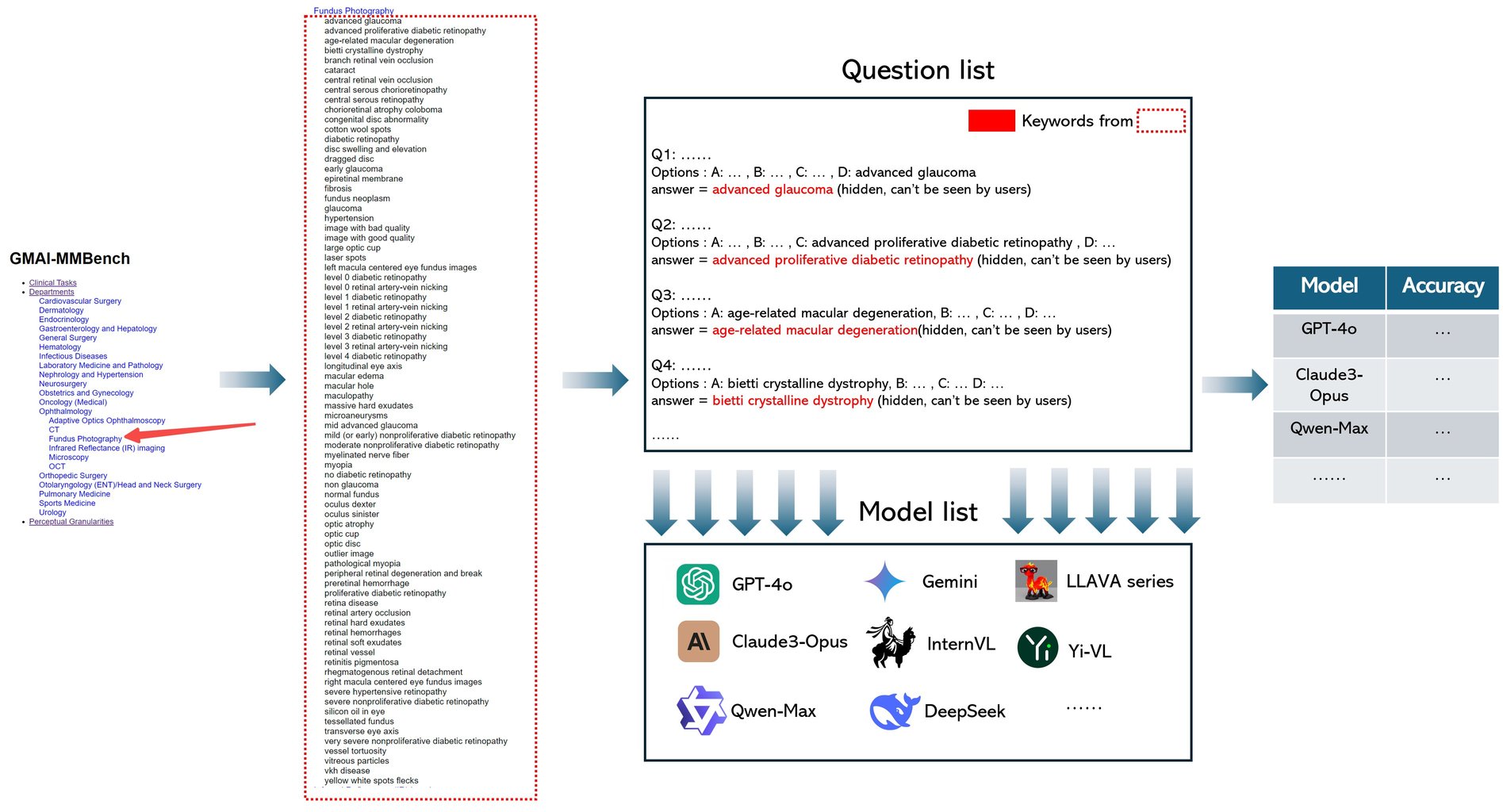}
    \caption{Example of how to use the Lexical Tree for customizing evaluations for the \textbf{ophthalmology} department and \textbf{fundus photography} modality. The process involves selecting the department (ophthalmology), choosing the modality (fundus photography), filtering questions using relevant keywords, and evaluating different models based on their accuracy in answering the filtered questions.}
    \label{fig:lexical_tree_use}
\end{figure}

\textbf{Step-by-Step Process:}
\begin{enumerate}
    \item \textbf{Select the Department:}
    First, navigate to the Lexical Tree interface and select the department relevant to our testing. In our case, we choose the ``Ophthalmology'' department from the available clinical tasks, as shown in Figure~\ref{fig:lexical_tree_use}.
    \item \textbf{Choose the Modality:}
    Within the ophthalmology department, several modalities related to eye conditions are listed. We specifically select the ``Fundus Photography'' modality. This selection allows us to access all the keywords associated with fundus images, which are crucial for the next step.
    \item \textbf{Keyword Filtering:}
    After selecting the fundus photography modality, a comprehensive list of keywords appears. These keywords are critical as they will be used to filter the relevant questions for the evaluation. Examples of keywords include ``advanced glaucoma'', ``age-related macular degeneration'', and ``diabetic retinopathy'' among others.
    \item \textbf{Retrieve Question List:}
    The system filters and retrieves questions from the pre-prepared question list using the selected keywords. Each question includes multiple options, and the correct answer corresponds to the keyword used for filtering. However, the correct answers are hidden from the users during the evaluation process. For instance, a question may ask about identifying a condition shown in an image, with options like ``A. advanced glaucoma'', ``B. early glaucoma'', ``C. non glaucoma'', etc. The correct answer, such as ``advanced glaucoma'' is derived from the keyword used for filtering.
    \item \textbf{Model Evaluation:}
    The filtered question list is then used to evaluate various models. In this example, models such as GPT-4, Claude3-Opus, Qwen-Max, and others are assessed for their accuracy in answering the questions. The results are compiled and displayed in a tabular format, showcasing each model's performance.   
\end{enumerate}

In addition to the provided example, this method allows for the independent testing of \textbf{any other departments, modalities, clinical tasks, and their combinations.} For instance, if the objective is to evaluate only ophthalmology, fundus photographs, and disease diagnosis tasks, further refinement of the keywords can be achieved following the initial selection. By accessing the disease diagnosis task and selecting the fundus photography modality, we can intersect the keywords from the department-fundus photography section with those from the clinical tasks-disease diagnosis section. The resulting keywords will represent those relevant exclusively to disease diagnosis tasks within the context of fundus photographs in ophthalmology. 

%This approach is illustrated in Figure~\ref{fig}.

In summary, the lexical tree provides a versatile framework for customizing evaluation processes across various medical domains, ensuring a comprehensive and focused assessment of model performance.

\clearpage
\section{Evaluation}
\label{Metrics}
% If a model's output does not include clearly followed instructions to select an answer or letter options, we use ChatGPT-3.5-turbo-0613 to extract the answer. If an answer cannot be extracted, it is treated as an error.

%If the outputs of a model do not include clearly followed instructions to select an answer and the output contents do not contain letter options, we will use ChatGPT-3.5-turbo-0613 to extract the answer. The final extracted answer will be considered as the model's predicted answer for that case. For all model prediction results where an answer cannot be extracted, we treat them as errors. 
In this section, we will describe the evaluation process in detail. We evaluated various LVLMs, including medical-specific models, open-source general models, and closed-source API general models. We selected versions with approximately 7 billion parameters for testing, and the model weights were sourced from their respective official Hugging Face repositories. Our evaluation was conducted using the VLMEvalKit\footnote{\url{https://github.com/open-compass/VLMEvalKit}} framework. For medical-specific models, we utilized the Multi-Modality-Arena\footnote{\url{https://github.com/OpenGVLab/Multi-Modality-Arena/tree/main/MedicalEval/Question-answering\_Score}} repository for testing.
Specifically, we input the prompt shown in Table ~\ref{prompt_example} into the tested model to for evaluation, the option-only answers are expected. However, it's hard for some models to follow the instructions, if a model neither outputs a clear answer tagged by the letter options nor provides instructions to select an answer, we use ChatGPT-3.5-turbo-0613 to extract the answer from the model's outputs. If the answer cannot be extracted, we treat the outputs as errors. Otherwise, the extracted answers will be considered as the model's predicted answer for that question.

\begin{table}[H]
%\label{prompts}
\caption{Examples of single-choice and multiple-choice question prompts.}
\resizebox{1.0\textwidth}{!}{
\begin{tabular}{l}
\hline  
\multicolumn{1}{c}{\textbf{Prompt example for single-choice questions}} \\ \hline
\emph{Question}: Observe the image. What is the most likely abnormality shown in the picture?\\ 
\emph{Options}: \\ 
A.osteoporotic bone\\ 
B.healthy bone\\ 
Please select the correct answer from the options above.\\ \textless{}image\textgreater{} \\ 
\hline
\multicolumn{1}{c}{\textbf{Prompt example for multiple-choice questions}}  \\ \hline
\emph{Question}: Determine which part(s) is illustrated in the image.\\ \emph{Options}:\\ 
A. cytosol\\ 
B. actin filaments\\ 
C. vesicles and punctate cytosolic patterns\\ 
D. microtubules \\ 
E. plasma membrane \\ 
F. endoplasmic reticulum \\  
Please select all correct answers from the options above. Note that there is more than one correct answer. \\ 
Please output the answer options directly, separated by commas. For example: A,B\\ 
\textless{}image\textgreater{} \\ \hline
\end{tabular}
}
\label{prompt_example}
\end{table}

\subsection{Evaluation Metric for Single-choice Questions}

For all single-choice questions, we denote $n_{\text{correct}}$ as the number of questions for which the model offered the correct answer, and $n_{\text{questions}}$ as the total number of questions. The ACC can be calculated as follows:

\begin{equation}
\textrm{ACC} = \frac{n_{\text{correct}}}{n_{\text{questions}}} .
\label{eq1}
\end{equation}

\subsection{Evaluation Metric for Multiple-choice Questions}

For all multiple-choice questions, we first count the number of correct predictions by the model within the groundtruth for each case, denoted as $n_{\text{match}}$. The length of the prediction is denoted as $l_{\text{prediction}}$, and the length of the groundtruth options are denoted as $l_{\text{truth}}$. The evaluation metrics for multiple-choice questions is calculated as follows:
\begin{gather}
\textrm{ACC}_{mcls} = \frac{n_{\text{match}}}{l_{\text{prediction}}}, \label{eq2} \\
\textrm{Recall}_{mcls} = \frac{n_{\text{match}}}{l_{\text{truth}}} .
\end{gather}

\subsection{Evaluated Models}
In this paper, we evaluate $50$ models on our GMAI-MMBench, and we list them in Table~\ref{Evaluated_Models}.  

% \newpage
% \usepackage{multirow}
\begin{table}[]
\caption{The model architecture of 50 LVLMs evaluated on GMAIMMBench.}
\resizebox{1.0\textwidth}{!}{
\begin{tabular}{c|cccc}
\hline
Series                             & Models                      & \#Params & Vision Encoder     & LLM                     \\ \hline
\multirow{6}{*}{Med model series}  & MedVInT~\cite{zhang2023pmcvqa}                     & -          & -                  & -                       \\
                                   & Med-Flamingo~\cite{moor2023med}                & 8.3B       & CLIP ViT/L-14      & LLaMA-7B                \\
                                   & LLaVA-Med~\cite{li2024llava}                   &-            &CLIP ViT/L-14                    & Mistral-7B                        \\
                                   & RadFM~\cite{wu2023generalist}                       & 14B        & 3D ViT             & MedLLaMA-13B            \\
                                   & Qilin-Med-VL-Chat~\cite{liu2023qilin}           & -          & Clip ViT/L-14      & Chinese-LLaMA2-Chat-13B \\
                                   & MedDr~\cite{he2024meddr}                       & 40B        & InternViT-6B       & Nous-Hermes-2-Yi-34B    \\ \hline
\multirow{9}{*}{Ungroupped series} & TransCore-M~\cite{transcorem}                 & 13.4B      & CLIP ViT/L-14      & PCITransGPT-13B         \\
                                   & VisualGLM-6B~\cite{ding2021cogview}                & 7.8B       & EVA-CLIP           & ChatGLM-6B              \\
                                   & mPLUG-Owl2~\cite{ye2023mplug}                  & 8.2B       & CLIP ViT-L/14      & LLaMA2-7B               \\
                                   & OmniLMM-12B~\cite{yu2024rlaifv}                 & 12B        & EVA02-5B           & Zephyr-7B-$\beta$             \\
                                   & PandaGPT 13B~\cite{su2023pandagpt}                & 13B        & ImageBind ViT-H/14 & Vicuna-v0-13B           \\
                                   & Mini-Gemini-7B~\cite{li2024mini}              & 7B         & CLIP-L             & Vicuna-v1.5-7B          \\
                                   & Emu2-Chat~\cite{sun2023generative}                   & 37B        & EVA-02-CLIP-E-plus & LLaMA-33B               \\
                                   & Flamingo v2~\cite{awadalla2023openflamingo}                 & 9B         & CLIP ViT-L/14      & MPT-7B                  \\
                                   & MMAlaya~\cite{datacanvas2024mmalaya}                     & 7.8B       & EVA-G              & Alaya-7B-Chat           \\ \hline
\multirow{2}{*}{CogVLM series}     & CogVLM-Chat~\cite{wang2023cogvlm}                 & 17B        & EVA-CLIP-E         & Vicuna-v1.5-7B          \\
                                   & CogVLM-grounding-generalist~\cite{wang2023cogvlm} & 17B        & EVA-CLIP-E         & Vicuna-v1.5-7B          \\ \hline
InstructBLIP series                & InstructBLIP-7B~\cite{dai2024instructblip}             & 8B         & EVA-G              & Vicuna-7B               \\ \hline
\multirow{2}{*}{DeepSeek series}   & DeepSeek-VL-1.3B~\cite{lu2024deepseek}            & 1.3B       & SAM-B \& SigLIP-L  & DeekSeek-1B             \\
                                   & DeepSeek-VL-7B~\cite{lu2024deepseek}              & 7.3B       & SAM-B \& SigLIP-L  & DeekSeek-7B             \\ \hline
Idefics series                     & Idefics-9B-Instruct~\cite{laurencon2023obelics}         & 9B         & CLIP ViT-H/14      & LLaMA 7B                \\ \hline
\multirow{4}{*}{XComposer series}  & ShareCaptioner~\cite{chen2023sharegpt4v}              & 8B         & EVA-G              & InternLM-7B             \\
                                   & XComposer~\cite{internlmxcomposer}                   & 8B         & EVA-CLIP-G         & InternLM-7B             \\
                                   & XComposer2~\cite{internlmxcomposer2}                  & 7B         & CLIP ViT-L/14      & InternLM2-7B            \\
                                   & XComposer2-4KHD~\cite{internlmxcomposer2_4khd}             & 7B         & CLIP ViT-L/14      & InernLM2-7B             \\ \hline
Yi-VL series                       & Yi-VL-6B~\cite{ai2024yi}                    & 6.6B       & CLIP ViT-H/14      & Yi-6B                   \\ \hline
\multirow{4}{*}{InternVL series}   & InternVL-Chat-V1.1~\cite{chen2023internvl}          & 19B        & InternViT-6B       & LLaMA2-13B              \\
                                   & InternVL-Chat-V1.2~\cite{chen2023internvl}          & 40B        & InternViT-6B       & Nous-Hermes-2-Yi-34B    \\
                                   & InternVL-Chat-V1.2-Plus~\cite{chen2023internvl}     & 40B        & InternViT-6B       & Nous-Hermes-2-Yi-34B    \\
                                   & InternVL-Chat-V1.5~\cite{chen2024far}          & 25.5B      & InternViT-6B       & InternLM2-Chat-20B      \\ \hline
\multirow{4}{*}{LLaVA series}      & LLaVA-NeXT-mistral-7B~\cite{liu2024llavanext}       & 7.6B       & CLIP ViT-L/14      & Mistral-7B              \\
                                   & LLaVA-NeXT-vicuna-7B~\cite{liu2024llavanext}        & 7.1B       & CLIP ViT-L/14      & Vicuna-v1.5-7B          \\
                                   & LLAVA-V1.5-7B~\cite{liu2023visual}               & 7.2B       & CLIP ViT-L/14      & Vicuna-v1.5-7B          \\
                                   & ShareGPT4V-7B~\cite{chen2023sharegpt4v}               & 7.2B       & CLIP ViT-L/14      & Vicuna-v1.5-7B          \\ \hline
\multirow{4}{*}{Xtuner series}     & LLAVA-InternLM-7b~\cite{2023xtuner}           & 7.6B       & CLIP ViT-L/14      & InternLM-7B             \\
                                   & LLAVA-InternLM2-7b~\cite{2023xtuner}          & 8.1B       & CLIP ViT-L/14      & InternLM2-7B            \\
                                   & LLAVA-V1.5-7B-xtuner~\cite{2023xtuner}        & 7.2B       & CLIP ViT-L/14      & Vicuna-v1.5-7B          \\
                                   & LLAVA-V1.5-13b-xtuner~\cite{2023xtuner}       & 13.4B      & CLIP ViT-L/14      & Vicuna-v1.5-13B         \\ \hline
\multirow{2}{*}{MiniCPM series}    & MiniCPM-V~\cite{hu2024large}                   & 2.8B       & SigLip-400M        & MiniCPM-2.4B            \\
                                   & MiniCPM-V2~\cite{xu2024llava-uhd}                  & 2.8B       & SigLip-400M        & MiniCPM-2.4B            \\ \hline
\multirow{4}{*}{Qwen series}       & Monkey~\cite{li2024monkey}                      & 9.8B       & CLIP-ViT-BigHuge   & Qwen-7B                 \\
                                   & Monkey-Chat~\cite{li2024monkey}                 & 9.8B       & ViT-BigHuge        & Qwen-7B                 \\
                                   & Qwen-VL~\cite{bai2023qwenvl}                     & 9.6B       & CLIP ViT-G/16      & QWen-7B                 \\
                                   & Qwen-VL-Chat~\cite{bai2023qwenvl}                & 9.6B       & CLIP ViT-G/16      & Qwen-7B                 \\ \hline
\multirow{6}{*}{API series}        & Qwen-VL-Max~\cite{Qwen-VL}                 & -          & -                  & QwenLM                  \\
                                   & Claude3-Opus~\cite{anthropic2024claude}                & -          & -                  & -                       \\
                                   & GPT-4o~\cite{achiam2023gpt}                      & -          & -                  & -                       \\
                                   & GPT-4V~\cite{achiam2023gpt}                      & -          & -                  & -                       \\
                                   & Gemini 1.0~\cite{team2023gemini}                  & -          & -                  & -                       \\
                                   & Gemini 1.5~\cite{reid2024gemini}                 & -          & -                  & -                       \\ \hline
\end{tabular}
}
\label{Evaluated_Models}
\end{table}

\section{Results}
In this section, we first provide the complete quantitative results in our experiments, and then perform the case study by analyzing $53$ representative examples of models' outputs.

\subsection{Quantitative Results}

The complete test results are shown in the table below. Table~\ref{clincial_VQA_results} shows the results in different clinical VQA tasks; Table~\ref{departments_results} shows the results across different departments; Table~\ref{perception_results} shows the results in different perceptual granularities.

\clearpage
\begin{table}[]
\caption{Results for single-choice questions of 50 different LVLMs on clinical VQA tasks. The best-performing model in each category is \textbf{in-bold}, and the second best is \underline{underlined}.}
\resizebox{1.0\textwidth}{!}{
\setlength{\tabcolsep}{1mm}{
\begin{tabular}{l|cc|cccccccccccccccccc}
\hline
Model name                  & \begin{tabular}[c]{@{}c@{}}Overall\\ (val)\end{tabular} & \begin{tabular}[c]{@{}c@{}}Overall\\ (test)\end{tabular} & AR               & BVR              & B                & CR               & C                & DD               & IQG              & MR               & M                & NT               & OR-A             & OR-HN            & OR-P             & OR-T             & SG               & SAR              & SIR              & SWR              \\ \hline
Random                      & 25.70                                                 & 25.94                                                  & 38.20          & 22.73          & 22.92          & 22.72          & 24.06          & 26.66          & 27.13          & 27.00          & 20.00          & 24.75          & 21.37          & 22.93          & 22.33          & 21.18          & 32.43          & 24.23          & 21.39          & 23.71          \\
\multicolumn{21}{c}{\cellcolor[HTML]{EFEFEF}Medical Special Model}                                                                                                                                                                                                                                                                                                                                                                                                                                                             \\
MedVInT ~\cite{zhang2023pmcvqa}                     & 2.29                                                  & 1.96                                                   & 5.75           & 0.00           & 0.00           & 0.00           & 2.56           & 2.11           & 4.05           & 0.00           & 0.00           & 0.00           & 0.11           & 0.00           & 0.00           & 0.12           & 7.36           & 0.00           & 1.88           & 0.00           \\
Med-Flamingo ~\cite{moor2023med}                & 12.74                                                 & 11.64                                                  & 6.67           & 10.14          & 9.23           & 11.27          & 6.62           & 13.43          & 12.15          & 6.38           & 8.00           & 18.18          & 9.26           & 18.27          & 11.00          & 11.53          & 12.16          & 5.19           & 8.47           & 11.43          \\
LLaVA-Med ~\cite{li2024llava}                   & 20.54                                                 & 19.60                                                  & 24.51          & 17.83          & 17.08          & 19.86          & 15.04          & 19.81          & 20.24          & 21.51          & 13.20          & 15.15          & 20.42          & 23.73          & 17.67          & 19.65          & 21.70          & 19.81          & 14.11          & 20.86          \\
Qilin-Med-VL-Chat ~\cite{liu2023qilin}           & 22.34                                                 & 22.06                                                  & 29.57          & 19.41          & 16.46          & 23.79          & 15.79          & 24.19          & 21.86          & 16.62          & 7.20           & 13.64          & 24.00          & 14.67          & 12.67          & 15.53          & 26.13          & 24.42          & 17.37          & 25.71          \\
RadFM ~\cite{wu2023generalist}                      & 22.95                                                 & 22.93                                                  & 27.16          & 20.63          & 13.23          & 19.14          & 20.45          & 24.51          & 23.48          & 22.85          & 15.60          & 16.16          & 14.32          & 24.93          & 17.33          & 21.53          & 29.73          & 17.12          & 19.59          & 31.14          \\
MedDr~\cite{he2024meddr}                       & 41.95                                                 & 43.69                                                  & 41.20          & 50.70          & 37.85          & 29.87          & 28.27          & 52.53          & 36.03          & 31.45          & 29.60          & 47.47          & 33.37          & 51.33          & 32.67          & 44.47          & 35.14          & 25.19          & 25.58          & 32.29          \\
\multicolumn{21}{c}{\cellcolor[HTML]{FFF3E4}Open-Source LVLMs}                                                                                                                                                                                                                                                                                                                                                                                                                                                                  \\
CogVLM-grounding-generalist~\cite{wang2023cogvlm} & 5.20                                                  & 5.66                                                   & 3.11           & 4.02           & 2.92           & 3.22           & 10.83          & 7.98           & 9.72           & 0.15           & 0.00           & 11.11          & 8.32           & 1.87           & 1.67           & 2.00           & 1.65           & 0.00           & 4.02           & 0.57           \\
XComposer~\cite{internlmxcomposer}                   & 8.92                                                  & 7.67                                                   & 1.38           & 7.69           & 8.31           & 12.34          & 22.86          & 7.31           & 6.07           & 5.49           & 2.80           & 16.16          & 5.05           & 8.67           & 2.00           & 9.76           & 11.94          & 7.31           & 3.17           & 4.00           \\
PandaGPT 13B~\cite{su2023pandagpt}                & 16.69                                                 & 16.27                                                  & 24.51          & 23.60          & 22.15          & 23.61          & 14.29          & 14.95          & 13.36          & 12.17          & 18.40          & 28.79          & 18.63          & 27.33          & 18.67          & 16.71          & 11.04          & 9.23           & 13.43          & 9.71           \\
Flamingo v2~\cite{awadalla2023openflamingo}                 & 25.58                                                 & 26.34                                                  & 37.74          & 21.50          & 20.62          & 22.00          & 22.41          & 27.29          & 25.91          & 27.45          & 18.00          & 28.79          & 25.16          & 22.13          & 22.00          & 22.00          & 34.61          & 22.88          & 20.44          & 27.43          \\
VisualGLM-6B~\cite{ding2021cogview}                & 29.58                                                 & 30.45                                                  & 40.16          & 33.92          & 24.92          & 25.22          & 24.21          & 32.99          & 29.96          & 29.53          & 21.20          & 37.88          & 30.32          & 24.80          & 13.33          & 29.88          & 33.11          & 19.62          & 19.16          & 37.43          \\
Idefics-9B-Instruct~\cite{laurencon2023obelics}          & 29.74                                                 & 31.13                                                  & 40.39          & 30.59          & 26.46          & 33.63          & 22.56          & 34.38          & 25.51          & 26.71          & 21.60          & 27.78          & 27.47          & 32.80          & 24.67          & 23.41          & 32.66          & 23.08          & 21.39          & 30.57          \\
InstructBLIP-7B ~\cite{dai2024instructblip}            & 31.80                                                 & 30.95                                                  & 42.12          & 26.92          & 24.92          & 28.09          & 21.65          & 34.58          & 31.58          & 29.23          & 22.40          & 30.30          & 28.95          & 27.47          & 23.00          & 24.82          & 32.88          & 19.81          & 21.64          & 26.57          \\
Mini-Gemini-7B~\cite{li2024mini}              & 32.17                                                 & 31.09                                                  & 29.69          & 39.16          & 31.85          & 28.26          & 10.38          & 35.58          & 29.96          & 28.78          & 20.80          & 34.34          & 29.58          & 36.53          & 24.00          & 31.76          & 22.45          & 25.96          & 18.56          & 29.43          \\
MMAlaya~\cite{datacanvas2024mmalaya}                     & 32.19                                                 & 32.30                                                  & 41.20          & 35.14          & 32.15          & 34.17          & 27.82          & 35.09          & 28.34          & 30.27          & 18.00          & 46.97          & 20.21          & 31.20          & 16.00          & 34.59          & 32.28          & 23.65          & 22.93          & 30.29          \\
Qwen-VL ~\cite{bai2023qwenvl}                    & 34.80                                                 & 36.05                                                  & 37.05          & 37.24          & 35.85          & 28.98          & 24.81          & 43.60          & 24.70          & 30.12          & 19.20          & 44.44          & 29.68          & 31.87          & 25.00          & 31.18          & 30.26          & 21.54          & 20.10          & 26.86          \\
Yi-VL-6B ~\cite{ai2024yi}                   & 34.82                                                 & 34.31                                                  & 41.66          & 39.16          & 26.62          & 30.23          & 31.88          & 38.01          & 26.72          & 24.93          & 25.20          & 37.37          & 29.58          & 31.20          & 32.33          & 30.59          & 36.71          & 24.81          & 23.18          & 31.43          \\
LLaVA-NeXT-vicuna-7B ~\cite{liu2024llavanext}       & 34.86                                                 & 35.42                                                  & 40.62          & 38.64          & 21.08          & 35.42          & 23.91          & 41.22          & 32.39          & 28.04          & 20.53          & 44.95          & 27.92          & 34.98          & 20.22          & 32.82          & 33.63          & 23.08          & 25.06          & 34.86          \\
Qwen-VL-Chat~\cite{bai2023qwenvl}                & 35.07                                                 & 36.96                                                  & 38.09          & 40.56          & 38.00          & 32.20          & 25.71          & 44.07          & 24.70          & 30.56          & 24.00          & 40.91          & 29.37          & 36.53          & 26.00          & 27.29          & 35.14          & 16.54          & 20.10          & 34.00          \\
CogVLM-Chat ~\cite{wang2023cogvlm}                & 35.23                                                 & 36.08                                                  & 40.97          & 30.77          & 27.69          & 32.74          & 19.40          & 41.10          & 36.84          & 34.72          & 24.00          & 40.91          & 36.74          & 37.33          & 26.00          & 33.65          & 36.56          & 20.19          & 23.95          & 26.57          \\
Monkey ~\cite{li2024monkey}                      & 35.48                                                 & 36.39                                                  & 38.32          & 35.31          & 35.54          & 34.53          & 23.16          & 43.40          & 31.98          & 30.12          & 19.20          & 33.33          & 30.00          & 32.53          & 25.33          & 31.65          & 34.46          & 20.00          & 20.27          & 30.29          \\
mPLUG-Owl2~\cite{ye2023mplug}                  & 35.62                                                 & 36.21                                                  & 37.51          & 41.08          & 30.92          & 38.10          & 27.82          & 41.59          & 28.34          & 32.79          & 22.40          & 40.91          & 24.74          & 38.27          & 23.33          & 36.59          & 33.48          & 20.58          & 23.01          & 32.86          \\
ShareCaptioner ~\cite{chen2023sharegpt4v}             & 36.37                                                 & 36.19                                                  & 42.35          & 32.69          & 31.08          & 27.19          & 30.83          & 41.19          & 30.36          & 33.23          & 28.40          & 42.93          & 27.79          & 33.73          & 28.33          & 40.71          & 29.58          & 20.96          & 28.83          & 30.00          \\
Emu2-Chat ~\cite{sun2023generative}                  & 36.50                                                 & 37.59                                                  & 43.27          & 47.73          & 26.31          & 40.07          & 28.12          & 44.00          & 36.44          & 28.49          & 20.40          & 31.82          & 26.74          & 37.60          & 26.67          & 29.76          & 33.63          & 23.27          & 26.43          & 29.43          \\
XComposer2-4KHD~\cite{internlmxcomposer2_4khd}             & 36.66                                                 & 38.54                                                  & 41.89          & 39.86          & 28.77          & 40.43          & 20.60          & 44.25          & 35.22          & 33.53          & 22.80          & 42.42          & 34.84          & 29.60          & 44.00          & 39.53          & 35.21          & 21.54          & 27.20          & 38.00          \\
ShareGPT4V-7B ~\cite{chen2023sharegpt4v}              & 36.71                                                 & 36.70                                                  & 43.96          & 37.59          & 21.54          & 37.57          & 18.80          & 43.26          & 32.39          & 27.30          & 22.80          & 43.43          & 29.47          & 37.33          & 22.00          & 31.76          & 34.98          & 24.42          & 25.06          & 30.00          \\
LLaVA-NeXT-mistral-7B ~\cite{liu2024llavanext}      & 37.20                                                 & 37.16                                                  & 38.43          & 27.98          & 20.31          & 29.16          & 20.60          & 47.19          & 30.36          & 32.64          & 22.40          & 55.56          & 32.75          & 25.58          & 17.56          & 34.04          & 28.38          & 23.27          & 24.12          & 37.43          \\
LLAVA-V1.5-13b-xtuner~\cite{2023xtuner}        & 37.82                                                 & 38.74                                                  & 44.65          & 29.02          & 27.08          & 38.28          & 28.87          & 45.32          & 32.79          & 30.12          & 20.40          & 45.96          & 33.47          & 42.53          & 44.33          & 37.53          & 33.48          & 19.62          & 22.58          & 35.43          \\
OmniLMM-12B  ~\cite{yu2024rlaifv}                & 37.89                                                 & 39.30                                                  & 39.82          & 40.56          & 32.62          & 37.57          & 24.81          & 46.68          & 35.63          & 35.01          & 27.60          & 57.58          & 28.42          & 34.00          & 25.00          & 29.18          & 34.46          & 24.42          & 27.54          & 40.29          \\
InternVL-Chat-V1.1~\cite{chen2023internvl}          & 38.16                                                 & 39.41                                                  & 42.46          & 43.88          & 35.23          & 45.08          & 23.31          & 45.96          & 38.87          & 29.23          & 29.60          & 40.40          & 31.68          & 41.87          & 26.67          & 38.82          & 32.13          & 19.42          & 25.58          & 30.29          \\
LLAVA-V1.5-7B~\cite{liu2023visual}               & 38.23                                                 & 37.96                                                  & 45.45          & 34.27          & 30.92          & 41.32          & 21.65          & 44.68          & 34.01          & 27.74          & 23.60          & 43.43          & 28.00          & 42.13          & 29.00          & 35.06          & 33.41          & 22.12          & 23.61          & 29.14          \\
Monkey-Chat ~\cite{li2024monkey}                & 38.39                                                 & 39.50                                                  & 40.62          & 41.43          & 37.08          & 35.24          & 23.76          & 47.73          & 29.96          & 32.94          & 26.00          & 37.88          & 34.84          & 32.67          & 24.67          & 33.18          & 34.91          & 21.73          & 22.24          & 34.00          \\
LLAVA-V1.5-7B-xtuner~\cite{2023xtuner}        & 38.68                                                 & 38.22                                                  & 38.90          & 40.03          & 28.00          & 40.25          & 30.08          & 44.08          & 33.60          & 32.49          & 21.20          & 40.91          & 29.47          & 40.40          & 30.33          & 38.59          & 31.46          & 23.85          & 26.95          & 36.86          \\
XComposer2 ~\cite{internlmxcomposer2}                 & 38.68                                                 & 39.20                                                  & 41.89          & 37.59          & 33.69          & 40.79          & 22.26          & 45.87          & 36.44          & 32.94          & 27.20          & 58.59          & 26.11          & 36.40          & 43.67          & 37.29          & 32.06          & 23.46          & 27.80          & 32.86          \\
LLAVA-InternLM-7b~\cite{2023xtuner}           & 38.71                                                 & 39.11                                                  & 36.36          & 36.54          & 32.62          & 38.10          & 30.68          & 46.53          & 34.82          & 28.19          & 25.20          & 48.99          & 28.11          & 40.53          & 33.33          & 36.00          & 34.08          & 26.73          & 24.12          & 29.71          \\
TransCore-M ~\cite{transcorem}                & 38.86                                                 & 38.70                                                  & 40.74          & 41.78          & 20.77          & 35.06          & {\ul 34.74}    & 45.69          & 32.39          & 32.94          & 24.40          & 44.95          & 31.05          & 38.93          & 27.00          & 33.76          & 33.86          & 23.46          & 25.49          & 31.14          \\
InternVL-Chat-V1.5~\cite{chen2024far}           & 38.86                                                 & 39.73                                                  & 43.84          & 44.58          & 34.00          & 33.99          & 31.28          & 45.59          & 33.20          & 38.28          & 32.40          & 42.42          & 31.89          & 42.80          & 27.00          & 36.82          & 34.76          & 23.27          & 24.72          & 32.57          \\
InternVL-Chat-V1.2-Plus~\cite{chen2023internvl}     & 39.41                                                 & 40.79                                                  & 42.58          & 42.31          & 32.46          & 37.03          & 31.43          & 47.49          & 42.51          & 35.01          & 21.20          & 50.51          & 34.95          & 42.93          & 22.67          & 42.47          & 35.74          & 22.31          & 24.98          & 28.29          \\
InternVL-Chat-V1.2 ~\cite{chen2023internvl}         & 39.52                                                 & 40.01                                                  & 41.66          & 44.06          & 27.38          & 38.46          & 34.29          & 46.99          & 33.60          & 34.42          & 21.20          & 47.98          & 30.63          & 42.80          & 27.67          & 35.88          & 35.59          & {\ul 23.85}    & 24.98          & 28.00          \\
LLAVA-InternLM2-7b~\cite{2023xtuner}          & 40.07                                                 & 40.45                                                  & 39.82          & 37.94          & 30.62          & 35.24          & 29.77          & 48.97          & 34.01          & 25.96          & 20.80          & 53.03          & 30.95          & 42.67          & 32.00          & 39.88          & 32.43          & 21.73          & 24.38          & 38.00          \\
DeepSeek-VL-1.3B ~\cite{lu2024deepseek}           & 40.25                                                 & 40.77                                                  & 38.55          & 35.14          & 38.92          & 40.07          & 27.97          & 48.12          & 35.63          & 31.75          & 22.80          & 46.97          & 40.74          & 44.93          & 31.00          & 40.47          & 33.33          & 22.31          & 21.39          & 31.71          \\
MiniCPM-V ~\cite{hu2024large}                  & 40.95                                                 & 41.05                                                  & 39.70          & 46.50          & 36.31          & 39.36          & 22.26          & 48.09          & 34.82          & 35.76          & 24.00          & 45.45          & 34.11          & 44.80          & 23.00          & 44.47          & 36.19          & 21.15          & 23.95          & 35.14          \\
DeepSeek-VL-7B ~\cite{lu2024deepseek}             & 41.73                                                 & 43.43                                                  & 38.43          & 47.03          & 42.31          & 37.03          & 26.47          & 51.11          & 33.20          & 31.16          & 26.00          & 44.95          & 36.00          & 58.13          & 36.33          & 47.29          & 34.91          & 18.08          & 25.49          & {\ul 39.43}    \\
MiniCPM-V2~\cite{xu2024llava-uhd}                  & 41.79                                                 & 42.54                                                  & 40.74          & 43.01          & 36.46          & 37.57          & 27.82          & 51.08          & 28.74          & 29.08          & 26.80          & 47.47          & 37.05          & 46.40          & 25.33          & 46.59          & 35.89          & 22.31          & 23.44          & 31.71          \\
\multicolumn{21}{c}{\cellcolor[HTML]{FFF0F0}Proprietary LVLMs}                                                                                                                                                                                                                                                                                                                                                                                                                                                                  \\
Claude3-Opus~\cite{anthropic2024claude}                & 32.37                                                 & 32.44                                                  & 1.61           & 39.51          & 34.31          & 31.66          & 12.63          & 39.26          & 28.74          & 30.86          & 22.40          & 37.37          & 25.79          & 41.07          & 29.33          & 33.18          & 31.31          & 21.35          & 23.87          & 4.00           \\
Qwen-VL-Max ~\cite{Qwen-VL}                & 41.34                                                 & 42.16                                                  & 32.68          & 44.58          & 31.38          & 40.79          & 10.68          & 50.53          & 32.79          & 44.36          & 29.20          & 51.52          & 41.37          & 58.00          & 30.67          & 41.65          & 26.95          & 25.00          & 24.64          & 39.14          \\
GPT-4V ~\cite{achiam2023gpt}                     & 42.50                                                 & 44.08                                                  & 29.92          & 48.95          & 44.00          & 37.39          & 12.93          & 52.88          & 32.79          & 44.21          & {\ul 32.80}    & 63.64          & 39.89          & 54.13          & 37.00          & 50.59          & 27.55          & 23.08          & 25.75          & 37.43          \\
Gemini 1.0 ~\cite{team2023gemini}                 & 44.38                                                 & 44.93                                                  & {\ul 42.12}    & 45.10          & 46.46          & 37.57          & 20.45          & 53.29          & 35.22          & 36.94          & 25.20          & 51.01          & 34.74          & 59.60          & 34.00          & 50.00          & \textbf{36.64} & 23.65          & 23.87          & 35.43          \\
Gemini 1.5 ~\cite{reid2024gemini}                 & {\ul 47.42}                                           & {\ul 48.36}                                            & \textbf{43.50} & {\ul 56.12}    & {\ul 51.23}    & {\ul 47.58}    & 2.26           & {\ul 55.33}    & {\ul 38.87}    & {\ul 48.07}    & 30.00          & \textbf{76.26} & {\ul 51.05}    & \textbf{75.87} & {\ul 46.33}    & {\ul 62.24}    & 20.57          & \textbf{27.69} & \textbf{30.54} & \textbf{40.57} \\
GPT-4o ~\cite{achiam2023gpt}                     & \textbf{53.53}                                        & \textbf{53.96}                                         & 38.32          & \textbf{61.01} & \textbf{57.08} & \textbf{49.02} & \textbf{46.62} & \textbf{61.45} & \textbf{46.56} & \textbf{56.38} & \textbf{34.00} & {\ul 75.25}    & \textbf{53.79} & {\ul 69.47}    & \textbf{48.67} & \textbf{65.88} & {\ul 33.93}    & 22.88          & {\ul 29.51}    & {\ul 39.43}    \\ \hline
\end{tabular}
}}
\label{clincial_VQA_results}
\end{table}

\begin{table}[]
% \label{departments_results}
\caption{Results for single-choice questions of 50 LVLMs on different departments. The best-performing model in each category is \textbf{in-bold}, and the second best is \underline{underlined}.}
\resizebox{1.0\textwidth}{!}{
\setlength{\tabcolsep}{1mm}{
\begin{tabular}{l|cc|cccccccccccccccccc}
\hline
\multicolumn{1}{c|}{Model name}                  & \begin{tabular}[c]{@{}c@{}}Overall\\ (val)\end{tabular} & \multicolumn{1}{c|}{\begin{tabular}[c]{@{}c@{}}Overall\\ (test)\end{tabular}} & CS               & D                & E                & GH               & GS               & H                & ID               & LMP              & NH               & N                & OG               & OM               & O                & OS               & ENT/HNS          & PM               & SM               & U                \\ \hline
\multicolumn{1}{l|}{Random}                      & 25.70                                                 & \multicolumn{1}{c|}{25.94}                                                  & 22.82          & 25.19          & 21.00          & 25.97          & 22.24          & 24.45          & 31.13          & 28.99          & 22.86          & 24.00          & 29.15          & 27.77          & 30.36          & 25.92          & 22.53          & 24.74          & 22.87          & 29.19          \\
\multicolumn{21}{c}{\cellcolor[HTML]{EFEFEF}Medical Special Model}                                                                                                                                                                                                                                                                                                                                                                                                                                                                                                       \\
\multicolumn{1}{l|}{MedVInT~\cite{zhang2023pmcvqa} }                     & 2.29                                                  & \multicolumn{1}{c|}{1.96}                                                   & 0.24           & 2.50           & 1.00           & 1.94           & 1.09           & 0.88           & 3.31           & 5.23           & 1.14           & 0.73           & 0.00           & 1.40           & 4.44           & 0.56           & 0.00           & 2.24           & 0.64           & 0.86           \\
\multicolumn{1}{l|}{Med-Flamingo~\cite{moor2023med}}                & 12.74                                                 & \multicolumn{1}{c|}{11.64}                                                  & 11.76          & 12.49          & 10.00          & 10.88          & 9.33           & 5.42           & 7.28           & 10.05          & 12.00          & 10.91          & 12.88          & 14.89          & 15.37          & 12.40          & 13.43          & 12.89          & 14.92          & 10.47          \\
\multicolumn{1}{l|}{LLaVA-Med~\cite{li2024llava}}                   & 20.54                                                 & \multicolumn{1}{c|}{19.60}                                                  & 26.12          & 20.20          & 29.00          & 20.31          & 16.30          & 18.46          & 15.23          & 21.84          & 20.86          & 16.73          & 21.69          & 19.23          & 20.18          & 18.38          & 20.99          & 16.87          & 20.49          & 21.55          \\
\multicolumn{1}{l|}{Qilin-Med-VL-Chat~\cite{liu2023qilin}}           & 22.34                                                 & \multicolumn{1}{c|}{22.06}                                                  & 12.94          & 21.06          & 15.50          & 22.09          & 18.98          & 17.33          & 17.88          & 22.92          & 31.14          & 29.82          & 20.00          & 21.83          & 25.55          & 19.07          & 14.81          & 29.42          & 22.17          & 22.29          \\
\multicolumn{1}{l|}{RadFM~\cite{wu2023generalist}}                       & 22.95                                                 & \multicolumn{1}{c|}{22.93}                                                  & 24.24          & 23.02          & 20.00          & 20.59          & 20.83          & 19.49          & 28.48          & 24.42          & 18.00          & 32.00          & 16.95          & 26.90          & 26.25          & 18.26          & 26.54          & 25.19          & 23.74          & 20.20          \\
\multicolumn{1}{l|}{MedDr~\cite{he2024meddr}}                       & 41.95                                                 & \multicolumn{1}{c|}{43.69}                                                  & 53.18          & 45.28          & 33.00          & 44.78          & 28.03          & 29.91          & 47.68          & 35.22          & 38.29          & 78.55          & 25.08          & 49.53          & 45.31          & 52.09          & 48.61          & 52.36          & 54.21          & 39.90          \\
\multicolumn{21}{c}{\cellcolor[HTML]{FFF3E4}Open-Source LVLMs}                                                                                                                                                                                                                                                                                                                                                                                                                                                                                                            \\
\multicolumn{1}{l|}{CogVLM-grounding-generalist~\cite{wang2023cogvlm}} & 5.20                                                  & \multicolumn{1}{c|}{5.66}                                                   & 6.59           & 7.27           & 4.50           & 4.94           & 3.58           & 4.44           & 5.96           & 2.66           & 19.14          & 17.82          & 7.80           & 7.94           & 5.00           & 5.36           & 5.40           & 7.86           & 4.59           & 2.34           \\
\multicolumn{1}{l|}{XComposer~\cite{internlmxcomposer}}                   & 8.92                                                  & \multicolumn{1}{c|}{7.67}                                                   & 13.18          & 2.71           & 5.00           & 5.33           & 4.35           & 10.88          & 3.31           & 6.40           & 4.00           & 25.09          & 6.44           & 9.15           & 9.95           & 8.91           & 4.01           & 8.11           & 9.87           & 5.54           \\
\multicolumn{1}{l|}{PandaGPT 13B~\cite{su2023pandagpt}}                & 16.69                                                 & \multicolumn{1}{c|}{16.27}                                                  & 17.41          & 12.70          & 17.00          & 17.20          & 12.68          & 15.42          & 23.84          & 14.70          & 14.86          & 10.55          & 8.81           & 14.29          & 24.75          & 16.26          & 17.13          & 18.07          & 12.07          & 13.92          \\
\multicolumn{1}{l|}{Flamingo v2~\cite{awadalla2023openflamingo}}                 & 25.58                                                 & \multicolumn{1}{c|}{26.34}                                                  & 28.47          & 26.06          & 18.50          & 28.58          & 21.11          & 24.24          & 29.14          & 28.07          & 13.43          & 29.45          & 22.37          & 28.17          & 31.85          & 23.12          & 27.78          & 23.54          & 27.57          & 29.19          \\
\multicolumn{1}{l|}{VisualGLM-6B~\cite{ding2021cogview}}                & 29.58                                                 & \multicolumn{1}{c|}{30.45}                                                  & 52.71          & 25.95          & 14.00          & 31.69          & 22.06          & 25.17          & 30.46          & 25.50          & 30.29          & 59.27          & 15.93          & 29.97          & 37.79          & 30.09          & 23.61          & 32.85          & 38.19          & 23.03          \\
\multicolumn{1}{l|}{Idefics-9B-Instruct~\cite{laurencon2023obelics}}         & 29.74                                                 & \multicolumn{1}{c|}{31.13}                                                  & 19.76          & 33.98          & 21.00          & 30.08          & 24.46          & 26.66          & 50.33          & 28.74          & 36.00          & 58.55          & 36.27          & 29.64          & 36.76          & 36.07          & 24.38          & 31.36          & 32.04          & 29.19          \\
\multicolumn{1}{l|}{InstructBLIP-7B~\cite{dai2024instructblip}}             & 31.80                                                 & \multicolumn{1}{c|}{30.95}                                                  & 27.06          & 28.99          & 17.50          & 34.24          & 21.78          & 25.84          & 43.05          & 29.15          & 19.14          & 53.09          & 27.46          & 28.64          & 31.99          & 34.58          & 30.25          & 30.76          & 41.09          & 31.28          \\
\multicolumn{1}{l|}{Mini-Gemini-7B~\cite{li2024mini}}              & 32.17                                                 & \multicolumn{1}{c|}{31.09}                                                  & 34.59          & 39.63          & 23.50          & 35.74          & 23.46          & 19.80          & 41.06          & 25.91          & 40.86          & 56.00          & 19.32          & 21.63          & 35.73          & 35.83          & 33.95          & 40.57          & 29.14          & 29.56          \\
\multicolumn{1}{l|}{MMAlaya~\cite{datacanvas2024mmalaya}}                     & 32.19                                                 & \multicolumn{1}{c|}{32.30}                                                  & 71.06          & 37.68          & 38.00          & 28.30          & 27.40          & 27.64          & 51.66          & 32.39          & 28.86          & 83.64          & 29.49          & 27.37          & 35.92          & 36.70          & 20.99          & 27.53          & 29.43          & 28.08          \\
\multicolumn{1}{l|}{Qwen-VL~\cite{bai2023qwenvl}}                     & 34.80                                                 & \multicolumn{1}{c|}{36.05}                                                  & 39.53          & 41.59          & 40.50          & 28.69          & 20.74          & 26.77          & 45.03          & 28.82          & 56.57          & 73.09          & 39.32          & 41.39          & 39.23          & 43.36          & 33.64          & 35.74          & 45.15          & 42.73          \\
\multicolumn{1}{l|}{Yi-VL-6B~\cite{ai2024yi} }                    & 34.82                                                 & \multicolumn{1}{c|}{34.31}                                                  & 39.76          & 43.76          & 56.00          & 27.30          & 25.91          & 27.23          & 45.70          & 32.56          & 44.29          & 65.45          & 47.46          & 36.38          & 39.00          & 35.39          & 25.46          & 29.77          & 39.06          & 35.22          \\
\multicolumn{1}{l|}{LLaVA-NeXT-vicuna-7B~\cite{liu2024llavanext}}        & 34.86                                                 & \multicolumn{1}{c|}{35.42}                                                  & 40.00          & 37.13          & 51.60          & 31.82          & 29.15          & 26.18          & 49.01          & 31.06          & 32.94          & 65.33          & 28.44          & 35.98          & 43.21          & 38.71          & 26.87          & 40.02          & 36.47          & 32.36          \\
\multicolumn{1}{l|}{Qwen-VL-Chat~\cite{bai2023qwenvl}}                & 35.07                                                 & \multicolumn{1}{c|}{36.96}                                                  & 36.47          & 39.63          & 36.50          & 27.08          & 20.79          & 27.64          & {\ul 60.93}    & 30.23          & 52.57          & 70.55          & 37.29          & 47.13          & 39.37          & 46.67          & 34.57          & 37.63          & 47.88          & 39.90          \\
\multicolumn{1}{l|}{CogVLM-Chat~\cite{wang2023cogvlm}}                 & 35.23                                                 & \multicolumn{1}{c|}{36.08}                                                  & 30.59          & 38.98          & 42.50          & 31.41          & 26.22          & 23.62          & 47.02          & 34.22          & 51.43          & 56.00          & 32.54          & 44.13          & 38.67          & 37.94          & 30.86          & 41.11          & 45.91          & 29.19          \\
\multicolumn{1}{l|}{Monkey~\cite{li2024monkey}}                      & 35.48                                                 & \multicolumn{1}{c|}{36.39}                                                  & 38.59          & 39.52          & 35.00          & 29.74          & 20.97          & 25.73          & 52.98          & 28.90          & 48.29          & 68.00          & 34.24          & 41.46          & 40.78          & 45.23          & 31.79          & 39.27          & 45.91          & 42.49          \\
\multicolumn{1}{l|}{mPLUG-Owl2~\cite{ye2023mplug}}                  & 35.62                                                 & \multicolumn{1}{c|}{36.21}                                                  & 47.76          & 40.50          & 41.00          & 33.46          & 27.22          & 28.16          & 51.66          & 33.14          & 38.86          & 68.73          & 16.27          & 38.58          & 43.34          & 35.70          & 27.78          & 41.61          & 39.76          & 30.91          \\
\multicolumn{1}{l|}{ShareCaptioner~\cite{chen2023sharegpt4v}}              & 36.37                                                 & \multicolumn{1}{c|}{36.19}                                                  & 37.88          & 35.50          & 45.50          & 35.63          & 25.54          & 28.16          & 56.29          & 31.15          & 27.14          & 64.00          & 35.59          & 38.52          & 39.65          & 38.57          & 30.56          & 44.05          & 36.68          & 40.15          \\
\multicolumn{1}{l|}{Emu2-Chat~\cite{sun2023generative}}                   & 36.50                                                 & \multicolumn{1}{c|}{37.59}                                                  & 27.53          & 35.83          & 27.50          & 34.41          & 28.49          & 29.35          & 60.26          & 36.63          & 34.00          & 64.73          & 28.81          & 44.79          & 43.20          & 37.69          & 37.50          & 41.86          & 43.18          & 35.34          \\
\multicolumn{1}{l|}{XComposer2-4KHD~\cite{internlmxcomposer2_4khd}}             & 36.66                                                 & \multicolumn{1}{c|}{38.54}                                                  & 48.00          & 40.17          & 75.50          & 36.46          & 28.80          & 28.11          & 49.67          & 35.96          & 50.29          & 69.45          & 38.64          & 40.45          & 43.86          & 39.63          & 29.94          & 43.26          & 34.13          & 42.86          \\
\multicolumn{1}{l|}{ShareGPT4V-7B~\cite{chen2023sharegpt4v}}               & 36.71                                                 & \multicolumn{1}{c|}{36.70}                                                  & 43.76          & 39.09          & 48.50          & 37.24          & 27.90          & 23.88          & 49.01          & 30.40          & 46.29          & 60.73          & 29.15          & 44.46          & 44.56          & 37.57          & 30.40          & 38.03          & 35.98          & 36.95          \\
\multicolumn{1}{l|}{LLaVA-NeXT-mistral-7B~\cite{liu2024llavanext}}       & 37.20                                                 & \multicolumn{1}{c|}{37.16}                                                  & 42.96          & 40.17          & 46.40          & 37.84          & 28.53          & 23.76          & 52.32          & 31.81          & 46.59          & 73.00          & 21.25          & 47.08          & 42.61          & 33.37          & 22.75          & 46.94          & 37.45          & 33.48          \\
\multicolumn{1}{l|}{LLAVA-V1.5-13b-xtuner~\cite{2023xtuner}}       & 37.82                                                 & \multicolumn{1}{c|}{38.74}                                                  & 43.06          & 39.20          & 43.50          & 42.01          & 26.36          & 26.41          & 48.34          & 35.55          & 38.29          & 70.55          & 38.64          & 51.60          & 42.08          & 34.70          & 34.41          & 43.90          & 39.35          & 41.26          \\
\multicolumn{1}{l|}{OmniLMM-12B~\cite{yu2024rlaifv}}                 & 37.89                                                 & \multicolumn{1}{c|}{39.30}                                                  & 39.53          & 37.46          & 41.50          & 36.18          & 27.36          & 28.00          & {\ul 60.93}    & 37.46          & 55.43          & 80.00          & 31.19          & 35.71          & 44.89          & 42.49          & 28.24          & 43.80          & 51.19          & 42.86          \\
\multicolumn{1}{l|}{InternVL-Chat-V1.1~\cite{chen2023internvl}}          & 38.16                                                 & \multicolumn{1}{c|}{39.41}                                                  & 45.88          & 40.07          & 56.00          & 34.30          & 26.68          & 26.20          & 52.32          & 37.79          & 45.14          & 64.00          & 35.93          & 52.74          & 44.14          & 40.56          & 39.51          & 41.16          & 45.56          & 35.84          \\
\multicolumn{1}{l|}{LLAVA-V1.5-7B~\cite{liu2023visual}}               & 38.23                                                 & \multicolumn{1}{c|}{37.96}                                                  & 42.35          & 37.57          & 44.50          & 36.13          & 27.99          & 24.91          & 49.01          & 31.31          & 34.00          & 68.36          & 27.12          & 45.39          & 42.46          & 42.80          & 33.80          & 44.20          & 41.21          & 38.92          \\
\multicolumn{1}{l|}{Monkey-Chat~\cite{li2024monkey}}                 & 38.39                                                 & \multicolumn{1}{c|}{39.50}                                                  & 43.53          & 40.28          & 40.00          & 33.30          & 23.28          & 29.09          & 54.97          & 29.73          & 55.71          & 72.36          & 35.25          & 50.53          & 42.41          & 45.98          & 33.49          & 42.66          & 50.15          & 44.83          \\
\multicolumn{1}{l|}{LLAVA-V1.5-7B-xtuner~\cite{2023xtuner}}        & 38.68                                                 & \multicolumn{1}{c|}{38.22}                                                  & 51.53          & 35.07          & 31.00          & 38.07          & 31.52          & 29.04          & 58.94          & 36.79          & 28.29          & 69.09          & 29.15          & 50.80          & 39.89          & 40.12          & 27.78          & 40.82          & 39.12          & 36.08          \\
\multicolumn{1}{l|}{XComposer2~\cite{internlmxcomposer2}}                  & 38.68                                                 & \multicolumn{1}{c|}{39.20}                                                  & 32.71          & 42.13          & 70.50          & 33.13          & 29.62          & 27.02          & 54.30          & 34.05          & 23.14          & 83.64          & 39.66          & 46.53          & 44.23          & 45.73          & 28.86          & 45.55          & 41.32          & 41.87          \\
\multicolumn{1}{l|}{LLAVA-InternLM-7b~\cite{2023xtuner}}           & 38.71                                                 & \multicolumn{1}{c|}{39.11}                                                  & 44.94          & 39.85          & 33.50          & 43.06          & 27.54          & 27.08          & 52.98          & 34.22          & 31.14          & 79.64          & 37.97          & 50.67          & 42.41          & 39.69          & 36.73          & 37.63          & 46.72          & 39.78          \\
\multicolumn{1}{l|}{TransCore-M~\cite{transcorem}}                 & 38.86                                                 & \multicolumn{1}{c|}{38.70}                                                  & 39.06          & 43.87          & 24.50          & 40.18          & 29.08          & 30.79          & 52.98          & 32.48          & 38.86          & 66.91          & 42.37          & 42.79          & 44.75          & 40.44          & 36.73          & 34.00          & 47.19          & 35.71          \\
\multicolumn{1}{l|}{InternVL-Chat-V1.5~\cite{chen2024far}}          & 38.86                                                 & \multicolumn{1}{c|}{39.73}                                                  & 36.47          & 44.84          & 53.50          & 37.07          & 26.63          & 31.61          & 60.26          & 34.14          & 36.29          & 67.27          & 37.63          & 55.21          & 47.13          & 38.69          & 41.98          & 39.17          & 37.55          & 41.26          \\
\multicolumn{1}{l|}{InternVL-Chat-V1.2-Plus~\cite{chen2023internvl}}     & 39.41                                                 & \multicolumn{1}{c|}{40.79}                                                  & 51.06          & 43.54          & 60.00          & 39.07          & 29.39          & {\ul 31.82}    & 50.99          & 37.54          & 54.00          & 79.64          & 30.17          & 50.87          & 43.72          & 37.88          & 36.88          & 42.61          & 43.53          & 38.55          \\
\multicolumn{1}{l|}{InternVL-Chat-V1.2~\cite{chen2023internvl}}          & 39.52                                                 & \multicolumn{1}{c|}{40.01}                                                  & 40.71          & 46.25          & 77.50          & 31.52          & 26.36          & 31.10          & 50.33          & 36.96          & 52.00          & 80.00          & 31.19          & 45.46          & 43.20          & 40.06          & 34.10          & 44.40          & 46.66          & {\ul 42.36}    \\
\multicolumn{1}{l|}{LLAVA-InternLM2-7b~\cite{2023xtuner}}          & 40.07                                                 & \multicolumn{1}{c|}{40.45}                                                  & 43.53          & 40.72          & 60.50          & 34.74          & 30.12          & 27.44          & 51.66          & 33.39          & 50.86          & 74.55          & 26.44          & 49.13          & 42.74          & 43.12          & 31.94          & 50.87          & 47.01          & 39.04          \\
\multicolumn{1}{l|}{DeepSeek-VL-1.3B~\cite{lu2024deepseek}}            & 40.25                                                 & \multicolumn{1}{c|}{40.77}                                                  & 56.71          & 37.13          & 27.00          & 45.73          & 28.40          & 27.85          & 52.32          & 35.96          & 45.43          & 71.64          & 45.42          & 50.20          & 41.66          & 47.48          & 37.81          & 43.90          & 45.50          & 33.50          \\
\multicolumn{1}{l|}{MiniCPM-V~\cite{hu2024large}}                   & 40.95                                                 & \multicolumn{1}{c|}{41.05}                                                  & 28.47          & 42.02          & 40.00          & 42.79          & 28.80          & 28.62          & 46.36          & 36.30          & 40.00          & 67.27          & 31.53          & 42.46          & 44.04          & 50.28          & 37.50          & 51.92          & 52.29          & 27.22          \\
\multicolumn{1}{l|}{DeepSeek-VL-7B~\cite{lu2024deepseek}}              & 41.73                                                 & \multicolumn{1}{c|}{43.43}                                                  & 60.00          & 43.97          & 47.50          & 45.12          & 28.22          & 31.20          & 46.36          & 32.97          & 52.29          & 67.64          & \textbf{61.36} & 49.27          & 44.23          & 49.97          & 52.78          & 45.00          & 53.63          & 38.79          \\
\multicolumn{1}{l|}{MiniCPM-V2~\cite{xu2024llava-uhd}}                  & 41.79                                                 & \multicolumn{1}{c|}{42.54}                                                  & 37.88          & 43.65          & 35.50          & 42.67          & 26.49          & 29.24          & 37.75          & 33.31          & {\ul 59.71}    & 67.27          & 38.64          & 50.87          & 42.64          & 50.59          & 40.90          & 51.07          & 57.81          & 35.10          \\
\multicolumn{21}{c}{\cellcolor[HTML]{FFF0F0}Proprietary LVLMs}                                                                                                                                                                                                                                                                                                                                                                                                                                                                                                             \\
\multicolumn{1}{l|}{Claude3-Opus~\cite{anthropic2024claude}}                & 32.37                                                 & \multicolumn{1}{c|}{32.44}                                                  & 38.59          & 34.42          & 43.50          & 27.97          & 22.96          & 23.62          & 52.32          & 25.42          & 25.14          & 66.91          & 15.93          & 35.25          & 41.06          & 36.07          & 37.50          & 40.67          & 35.40          & 34.24          \\
\multicolumn{1}{l|}{Qwen-VL-Max~\cite{Qwen-VL}}                 & 41.34                                                 & \multicolumn{1}{c|}{42.16}                                                  & 50.59          & 47.23          & \textbf{74.00} & 40.68          & 29.03          & 26.71          & 58.94          & 34.05          & 62.29          & 85.45          & 27.80          & 44.39          & 43.90          & 42.99          & 48.61          & 49.38          & 51.13          & 40.52          \\
\multicolumn{1}{l|}{GPT-4V~\cite{achiam2023gpt}}                      & 42.50                                                 & \multicolumn{1}{c|}{44.08}                                                  & {\ul 64.00}    & 44.95          & 58.50          & 42.45          & 30.03          & 29.40          & 58.28          & 32.31          & 54.57          & 83.27          & 37.63          & 48.26          & 49.04          & 48.41          & 44.60          & 51.87          & 53.98          & 40.89          \\
\multicolumn{1}{l|}{Gemini 1.0~\cite{team2023gemini}}                  & 44.38                                                 & \multicolumn{1}{c|}{44.93}                                                  & 57.41          & 46.25          & 57.50          & 36.40          & 28.67          & 27.80          & 45.03          & {\ul 38.21}    & 58.57          & 86.55          & 40.68          & {\ul 51.74}    & 47.45          & 55.64          & 50.46          & 47.83          & {\ul 61.58}    & 41.87          \\
\multicolumn{1}{l|}{Gemini 1.5~\cite{reid2024gemini}}                  & {\ul 47.42}                                           & \multicolumn{1}{c|}{{\ul 48.36}}                                            & 55.29          & \textbf{50.81} & 54.00          & {\ul 51.05}    & \textbf{36.59} & 29.86          & 56.95          & 36.88          & 58.00          & {\ul 88.00}    & {\ul 47.46}    & 48.13          & {\ul 51.19}    & {\ul 56.88}    & {\ul 64.51}    & {\ul 56.50}    & 59.78          & 31.65          \\
\multicolumn{1}{l|}{GPT-4o~\cite{achiam2023gpt} }                      & \textbf{53.53}                                        & \multicolumn{1}{c|}{\textbf{53.96}}                                         & \textbf{66.82} & {\ul 48.53}    & {\ul 64.50}    & \textbf{55.94} & {\ul 35.10}    & \textbf{48.53} & \textbf{74.17} & \textbf{43.52} & \textbf{64.57} & \textbf{91.64} & 37.63          & \textbf{57.88} & \textbf{55.21} & \textbf{62.80} & \textbf{66.98} & \textbf{58.39} & \textbf{64.60} & \textbf{46.18} \\ \hline
\end{tabular}
}}
\label{departments_results}
\end{table}

\clearpage
\begin{table}[]
\caption{Results for single-choice questions of 50 LVLMs on perceptual granularities. The best-performing model in each category is \textbf{in-bold}, and the second best is \underline{underlined}.}
\resizebox{1.0\textwidth}{!}{
\begin{tabular}{cccccccccc}
\hline
Model name                  & Size  & Overall(val)     & Overall(test)    & Seg C            & Seg M            & 2D Cls update    & 2D Det           & 2D Mcls\_acc     & 2D Mcls\_recall  \\ \hline
Random                      & -     & 25.70          & 25.88          & 22.19          & 22.91          & 28.93          & 24.55          & 45.85          & 57.02          \\
\multicolumn{10}{c}{\cellcolor[HTML]{EFEFEF}Medical Special Model}                                                                                                                                                  \\
MedVInT~\cite{zhang2023pmcvqa}                     & -     & 2.29           & 1.98           & 0.82           & 0.25           & 3.48           & 0.12           & 0.05           & 0.02           \\
Med-Flamingo ~\cite{moor2023med}               & 8.3B  & 12.74          & 11.75          & 11.95          & 11.94          & 11.92          & 9.15           & 46.10          & 50.19          \\
LLaVA-Med ~\cite{li2024llava}                  & -     & 20.54          & 19.83          & 18.45          & 18.97          & 21.15          & 17.14          & 45.84          & 41.19          \\
Qilin-Med-VL-Chat~\cite{liu2023qilin}           & -     & 22.34          & 22.06          & 19.84          & 20.30          & 23.80          & 21.87          & 44.50          & 33.90          \\
RadFM ~\cite{wu2023generalist}                      & 14B   & 22.95          & 22.93          & 20.43          & 20.27          & 25.71          & 18.83          & 40.98          & 57.45          \\
MedDr ~\cite{he2024meddr}                      & 40B   & 41.95          & 43.18          & 42.55          & 44.03          & 45.08          & 28.10          & 48.09          & 23.38          \\
\multicolumn{10}{c}{\cellcolor[HTML]{FFF3E4}Open-Source LVLMs}                                                                                                                                                        \\
CogVLM-grounding-generalist~\cite{wang2023cogvlm} & 17B   & 5.20           & 5.39           & 6.80           & 5.51           & 5.11           & 2.57           & 46.24          & 49.82          \\
XComposer~\cite{internlmxcomposer}                   & 8B    & 8.92           & 7.71           & 8.87           & 6.24           & 8.02           & 6.30           & 31.45          & 23.68          \\
PandaGPT 13B ~\cite{su2023pandagpt}               & 13B   & 16.69          & 15.94          & 19.25          & 18.88          & 13.74          & 12.24          & 41.22          & 49.95          \\
Flamingo v2 ~\cite{awadalla2023openflamingo}                & 9B    & 25.58          & 26.23          & 22.52          & 22.48          & 30.12          & 21.17          & 41.80          & 19.17          \\
VisualGLM-6B ~\cite{ding2021cogview}               & 7.8B  & 29.58          & 30.20          & 27.30          & 27.31          & 33.75          & 22.16          & 43.08          & 35.22          \\
Idefics-9B-Instruct~\cite{laurencon2023obelics}         & 9B    & 29.74          & 30.81          & 25.50          & 25.21          & 36.45          & 23.85          & 43.47          & 46.02          \\
InstructBLIP-7B ~\cite{dai2024instructblip}            & 8B    & 31.80          & 31.00          & 29.12          & 21.77          & 36.71          & 24.08          & 39.43          & 23.79          \\
Mini-Gemini-7B ~\cite{li2024mini}             & 7B    & 32.17          & 31.22          & 32.13          & 32.92          & 30.72          & 26.53          & 45.38          & 57.99          \\
MMAlaya~\cite{datacanvas2024mmalaya}                     & 7.8B  & 32.19          & 32.02          & 29.33          & 30.22          & 35.02          & 24.02          & 48.43          & 20.93          \\
Qwen-VL ~\cite{bai2023qwenvl}                    & 9.6B  & 34.80          & 35.55          & 33.20          & 33.43          & 38.95          & 24.49          & 44.95          & 56.97          \\
Yi-VL-6B ~\cite{ai2024yi}                   & 6.6B  & 34.82          & 34.00          & 31.42          & 32.26          & 37.15          & 24.31          & 50.25          & 44.32          \\
LLaVA-NeXT-vicuna-7B ~\cite{liu2024llavanext}       & 7.1B  & 34.86          & 35.59          & 33.06          & 32.95          & 38.96          & 27.06          & 44.75          & 42.45          \\
Qwen-VL-Chat ~\cite{bai2023qwenvl}               & 9.6B  & 35.07          & 36.35          & 34.45          & 35.20          & 39.55          & 22.04          & 42.88          & {\ul 81.23}    \\
CogVLM-Chat ~\cite{wang2023cogvlm}                & 17B   & 35.23          & 35.83          & 34.13          & 34.49          & 38.55          & 25.25          & 47.09          & \textbf{90.26} \\
Monkey ~\cite{li2024monkey}                     & 9.8B  & 35.48          & 35.92          & 33.18          & 34.01          & 39.32          & 25.42          & 44.57          & 42.35          \\
mPLUG-Owl2 ~\cite{ye2023mplug}                 & 8.2B  & 35.62          & 35.89          & 33.68          & 34.74          & 38.80          & 24.90          & 42.59          & 41.84          \\
ShareCaptioner ~\cite{chen2023sharegpt4v}             & 8B    & 36.37          & 36.07          & 34.74          & 35.93          & 38.25          & 24.37          & 40.00          & 16.95          \\
Emu2-Chat ~\cite{sun2023generative}                  & 37B   & 36.50          & 35.54          & 36.54          & 27.62          & 39.57          & 27.76          & 44.29          & 37.65          \\
XComposer2-4KHD~\cite{internlmxcomposer2_4khd}             & 7B    & 36.66          & 37.93          & 36.84          & 38.02          & 39.84          & 26.65          & 48.83          & 44.08          \\
ShareGPT4V-7B ~\cite{chen2023sharegpt4v}              & 7.2B  & 36.71          & 36.52          & 34.74          & 35.15          & 39.24          & 26.18          & 46.11          & 43.52          \\
LLaVA-NeXT-mistral-7B ~\cite{liu2024llavanext}      & 7.6B  & 37.20          & 37.02          & 36.29          & 35.20          & 39.34          & 27.87          & 44.05          & 47.70          \\
LLAVA-V1.5-13b-xtuner~\cite{2023xtuner}       & 13.4B & 37.82          & 38.27          & 38.29          & 36.95          & 40.48          & 25.83          & 47.54          & 33.19          \\
OmniLMM-12B ~\cite{yu2024rlaifv}                & 12B   & 37.89          & 38.74          & 36.70          & 36.86          & 41.77          & 28.57          & 46.17          & 43.01          \\
InternVL-Chat-V1.1 ~\cite{chen2023internvl}         & 19B   & 38.16          & 38.93          & 38.54          & 40.00          & 40.07          & 28.16          & 39.82          & 27.32          \\
LLAVA-V1.5-7B ~\cite{liu2023visual}              & 7.2B  & 38.23          & 37.72          & 36.45          & 36.65          & 40.38          & 25.36          & 14.10          & 57.09          \\
Monkey-Chat~\cite{li2024monkey}                 & 9.8B  & 38.39          & 39.00          & 37.16          & 37.75          & 42.13          & 25.36          & 43.91          & 28.86          \\
LLAVA-V1.5-7B-xtuner~\cite{2023xtuner}        & 7.2B  & 38.68          & 37.96          & 36.75          & 36.34          & 40.55          & 27.52          & 46.78          & 43.06          \\
XComposer2 ~\cite{internlmxcomposer2}                 & 7B    & 38.68          & 38.95          & 37.86          & 38.52          & 41.00          & 28.34          & 46.43          & 51.87          \\
LLAVA-InternLM-7b ~\cite{2023xtuner}          & 7.6B  & 38.71          & 38.84          & 37.57          & 36.65          & 41.84          & 27.46          & 50.02          & 40.21          \\
TransCore-M ~\cite{transcorem}                & 13.4B & 38.86          & 38.43          & 36.09          & 36.06          & 42.04          & 26.53          & 45.34          & 40.93          \\
InternVL-Chat-V1.5 ~\cite{chen2024far}         & 25.5B & 38.86          & 39.32          & 38.61          & 40.48          & 40.45          & 29.27          & 31.51          & 24.72          \\
InternVL-Chat-V1.2-Plus ~\cite{chen2023internvl}    & 40B   & 39.41          & 40.25          & 40.68          & 41.50          & 40.82          & 30.38          & 36.50          & 37.09          \\
InternVL-Chat-V1.2 ~\cite{chen2023internvl}         & 40B   & 39.52          & 39.57          & 39.04          & 39.75          & 41.05          & 29.62          & 41.08          & 46.06          \\
LLAVA-InternLM2-7b~\cite{2023xtuner}          & 8.1B  & 40.07          & 40.15          & 39.30          & 39.14          & 42.60          & 27.76          & \textbf{50.64} & 48.25          \\
DeepSeek-VL-1.3B ~\cite{lu2024deepseek}           & 1.3B  & 40.25          & 40.54          & 40.61          & 40.71          & 42.13          & 27.64          & 48.71          & 21.38          \\
MiniCPM-V ~\cite{hu2024large}                  & 2.8B  & 40.95          & 40.89          & 39.48          & 39.18          & 44.08          & 27.00          & 42.87          & 32.09          \\
DeepSeek-VL-7B ~\cite{lu2024deepseek}             & 7.3B  & 41.73          & 42.90          & 43.87          & 43.60          & 44.32          & 26.59          & 44.16          & 18.74          \\
MiniCPM-V2 ~\cite{xu2024llava-uhd}                 & 2.8B  & 41.79          & 42.13          & 41.11          & 41.41          & 45.03          & 25.95          & 50.12          & 32.62          \\
\multicolumn{10}{c}{\cellcolor[HTML]{FFF0F0}Proprietary LVLMs}                                                                                                                                                         \\
Claude3-Opus ~\cite{anthropic2024claude}               & -     & 32.37          & 32.24          & 33.56          & 33.36          & 32.17          & 24.72          & 45.31          & 38.98          \\
Qwen-VL-Max ~\cite{Qwen-VL}                & -     & 41.34          & 41.70          & 44.23          & 44.42          & 41.09          & 29.10          & 31.12          & 25.88          \\
GPT-4V ~\cite{achiam2023gpt}                     & -     & 42.50          & 43.61          & 47.87          & 46.58          & 42.24          & 30.32          & 45.21          & 40.59          \\
Gemini 1.0 ~\cite{team2023gemini}                 & -     & 44.38          & 44.65          & 44.92          & 44.96          & {\ul 46.67}    & 27.46          & 49.01          & 55.09          \\
Gemini 1.5 ~\cite{reid2024gemini}                 & -     & {\ul 47.42}    & {\ul 48.03}    & {\ul 54.75}    & \textbf{56.59} & 43.25          & {\ul 34.17}    & 39.22          & 39.34          \\
GPT-4o  ~\cite{achiam2023gpt}                    & -     & \textbf{53.53} & \textbf{53.88} & \textbf{57.09} & {\ul 56.49}    & \textbf{53.70} & \textbf{36.21} & {\ul 50.60}    & 50.90          \\ \hline
\end{tabular}
}
\label{perception_results}
\end{table}

\clearpage

\begin{figure}[H]
    \centering
    \includegraphics[width=1\linewidth]{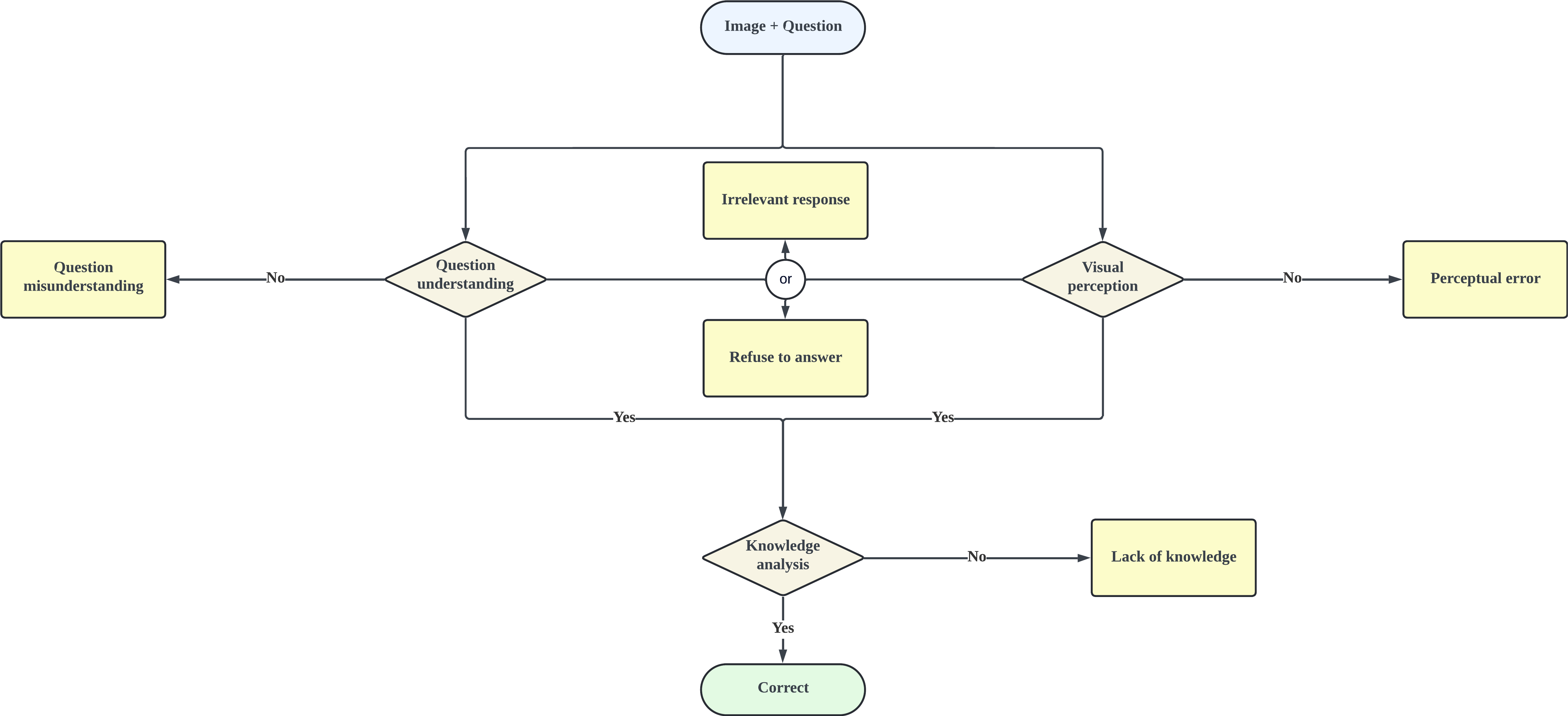}
    \caption{The illustration of the entire logical process from input to output in our case study.}
    \label{case_study_error_list}
\end{figure}
\subsection{Case Study}

%After our evaluation, we find that the model's outputs can be classified into 8 categories, which are correct, perceptual error, lack of knowledge, specifically, question misunderstanding, image misinterpretation, unable to determine, irrelevant responses, and refuse to answer. The definitions of each type are as follows:
In this section, we present a case study analysis of several LVLMs on various cases in GMAI-MMBench. The entire logical process of our study is illustrated in Figure~\ref{case_study_error_list}. Other than \textbf{Correct}, we classify the error types from input to output into five major categories:

\textbf{Correct:} 
LVLMs offer the correct answer. This indicates that the model accurately understands both the image and the question, and provides an appropriate and relevant response. 

\begin{figure}[H]
    \centering
    \includegraphics[width=1\linewidth]{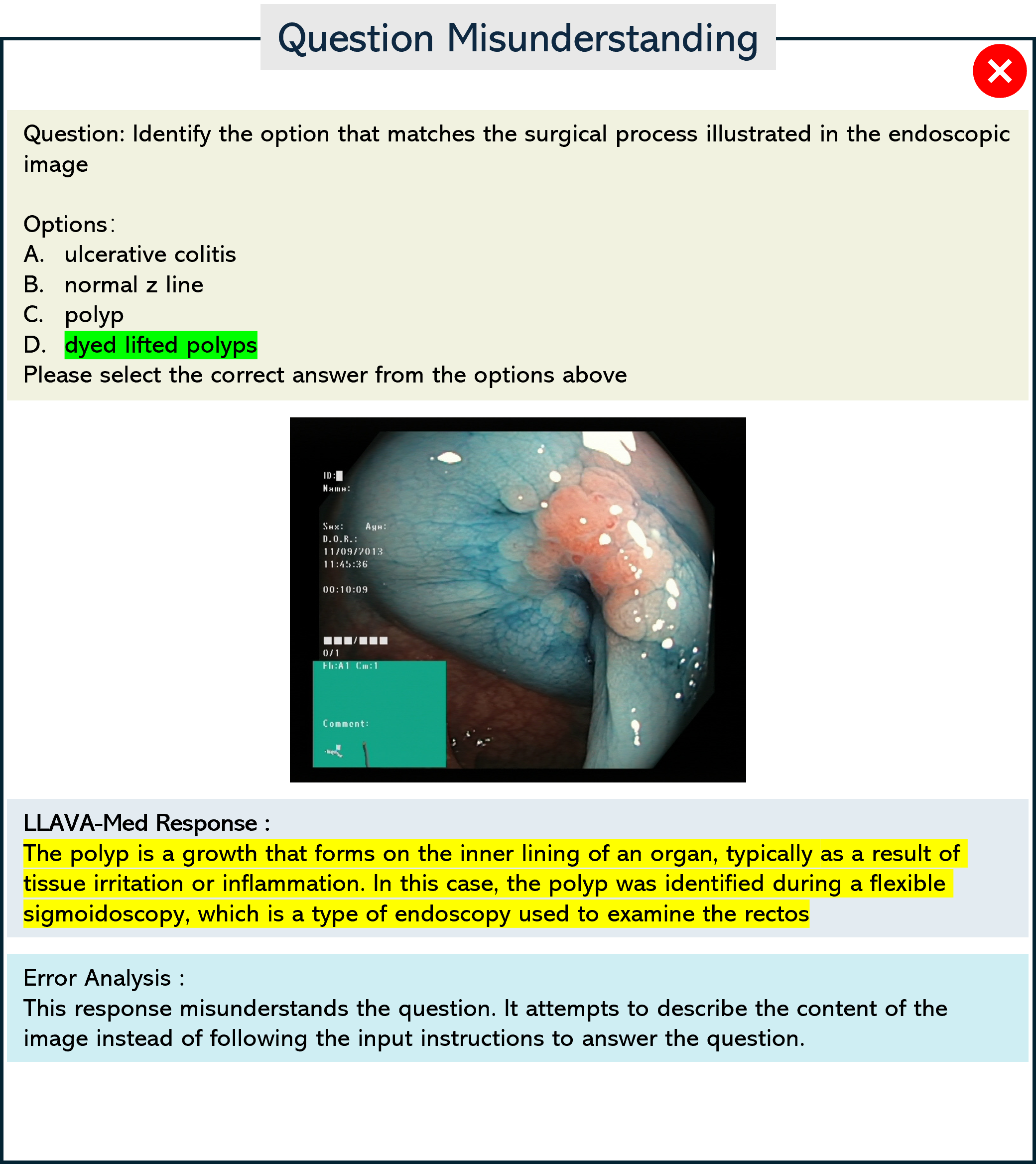}
    \caption{A question misunderstanding example. \boxedgreen{Green} highlights the right answer. \boxedyellow{Yellow} highlights the wrong answer. \hyperref[case_study_table]{\textcolor{red}{Back to Table Index}}.}
    \label{fig:question_misunderstanding_endoscopic_SWR_GS_image_question_misunderstanding}
\end{figure}

\textbf{Question misunderstanding:} 
LVLMs fail to correctly understand the question and generate erroneous answers. For example: LLAVA-Med may not understand the purpose of identifying the surgical process from the question, instead, it describes the image content in detail as shown in Figure~\ref{fig:question_misunderstanding_endoscopic_SWR_GS_image_question_misunderstanding}.

\begin{figure}[H]
    \centering
    \includegraphics[width=1\linewidth]{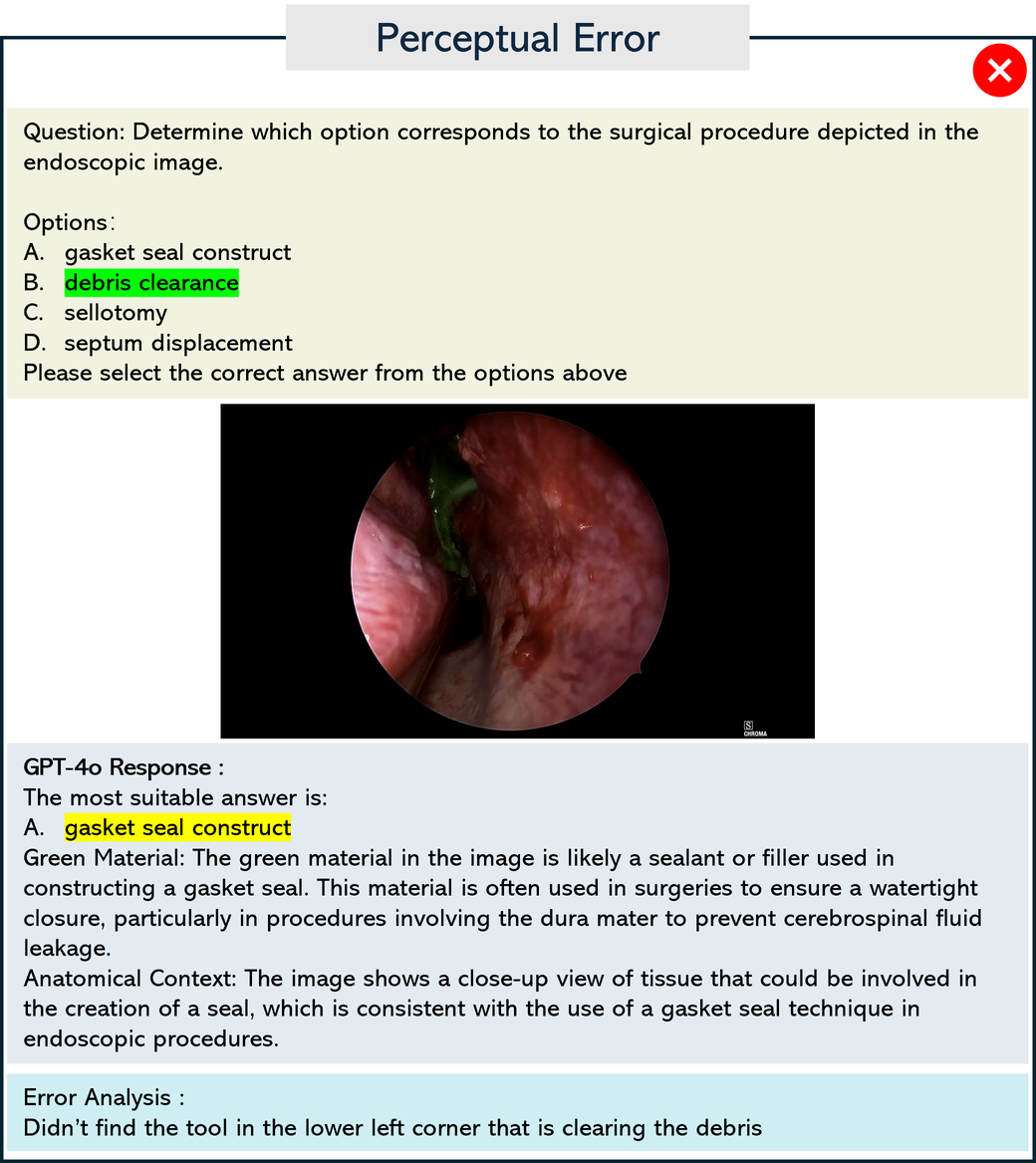}
    \caption{An example of perceptual error - detail missing case. \boxedgreen{Green} highlights the right answer. \boxedyellow{Yellow} highlights the wrong answer. \hyperref[case_study_table]{\textcolor{red}{Back to Table Index}}.}
    \label{fig:PE-D-2_SWR_GS_image}
\end{figure}

\begin{figure}[H]
    \centering
    \includegraphics[width=1\linewidth]{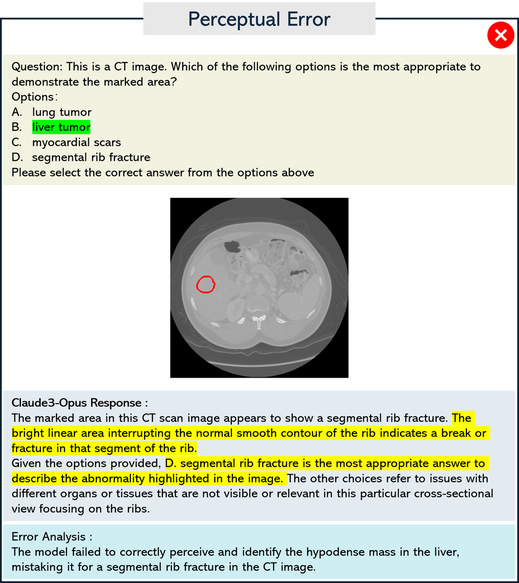}
    \caption{An example of perceptual error - misinterpretation case. \boxedgreen{Green} highlights the right answer. \boxedyellow{Yellow} highlights the wrong answer. \hyperref[case_study_table]{\textcolor{red}{Back to Table Index}}.}
    \label{fig:PE-M-5_DD_GH_contour}
\end{figure}

\textbf{Perceptual error:} 
LVLMs fail to locate, detect, or recognize the content or objects in images, which are necessary for answering the questions. This includes scenarios where the model misses critical details or misinterprets the image's content. For example: GPT-4o may ignore the important tool in the lower left corner that is clearing the debris in Figure~\ref{fig:PE-D-2_SWR_GS_image}. Claude3-Opus chooses the wrong answer as it cannot correctly identify the content in the mask in Figure~\ref{fig:PE-M-5_DD_GH_contour}.

\begin{figure}
    \centering
    \includegraphics[width=1\linewidth]{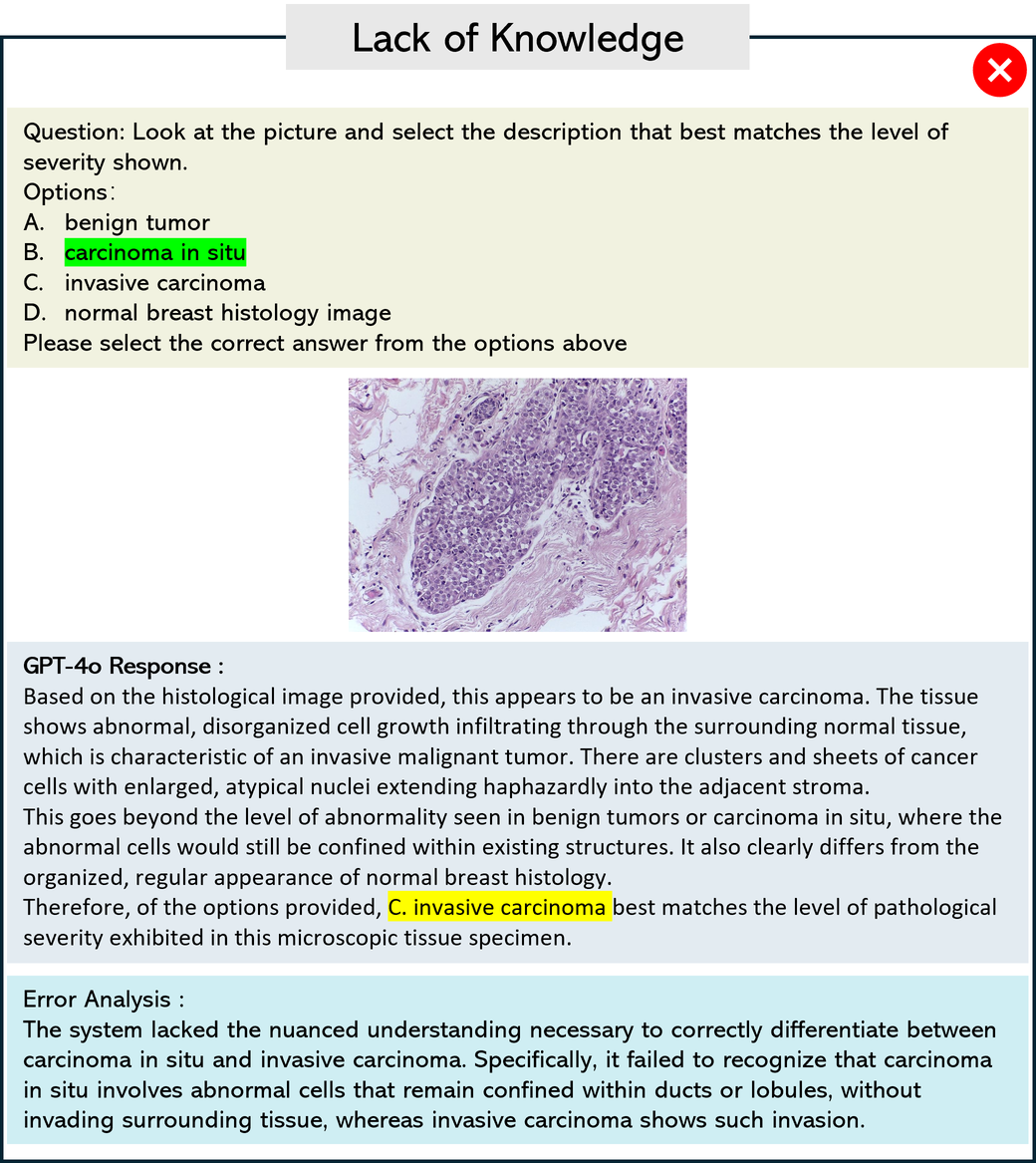}
    \caption{A lack of knowledge example. \boxedgreen{Green} highlights the right answer. \boxedyellow{Yellow} highlights the wrong answer. \hyperref[case_study_table]{\textcolor{red}{Back to Table Index}}.}
    \label{fig:lack_of_knowledge_SG_LMP_image_carcinoma_in_situ}
\end{figure}

\begin{figure}[H]
    \centering
    \includegraphics[width=1\linewidth]{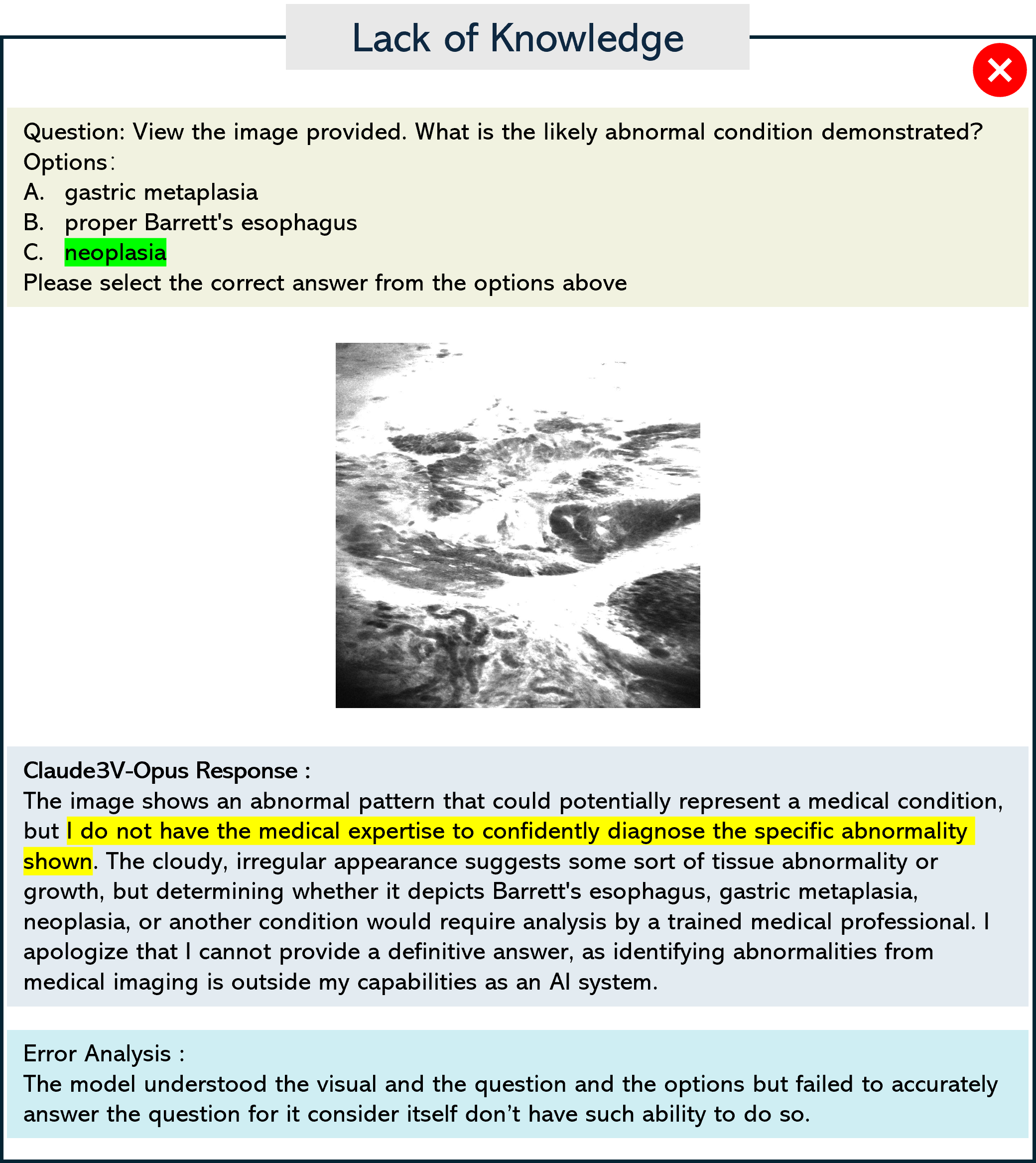}
    \caption{An example of unable to determine. \boxedgreen{Green} highlights the right answer. \boxedyellow{Yellow} highlights the wrong answer. \hyperref[case_study_table]{\textcolor{red}{Back to Table Index}}.}
    \label{fig:Unable1_DD_OM_image}
\end{figure}

\begin{figure}[H]
    \centering
    \includegraphics[width=1\linewidth]{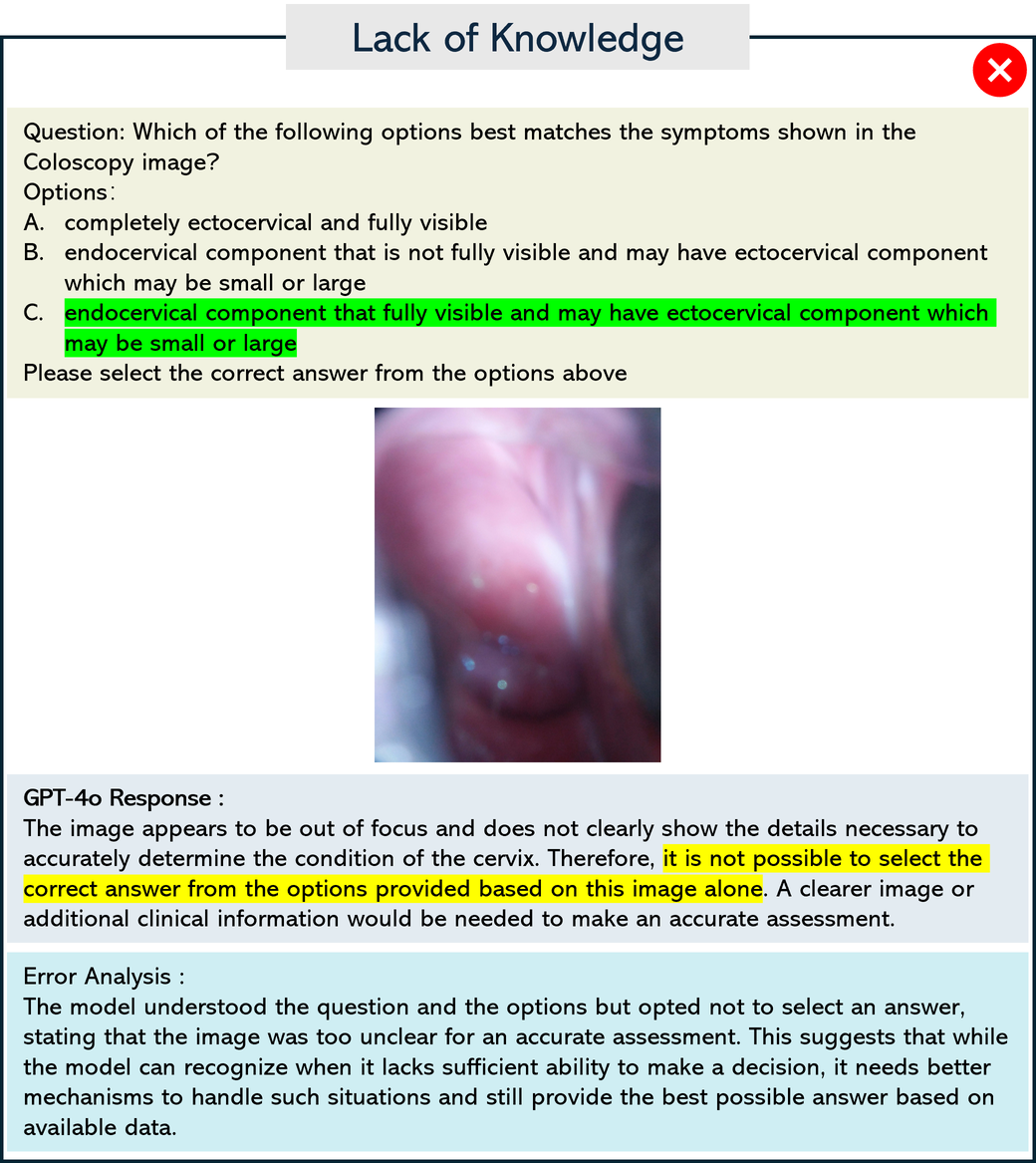}
    \caption{An example of unable to determine. \boxedgreen{Green} highlights the right answer. \boxedyellow{Yellow} highlights the wrong answer. \hyperref[case_study_table]{\textcolor{red}{Back to Table Index}}.}
    \label{fig:unable_to_determine_AR_OG_image_endocervical}
\end{figure}

\textbf{Lack of knowledge:} 
LVLMs can recognize both the image and the question but still make errors in specific cases, suggesting a lack of domain-specific knowledge required to answer specialized questions. For example: Models directly show their insufficient knowledge to answer or fail to respond without additional information as shown in Figure~\ref{fig:Unable1_DD_OM_image}, Figure~\ref{fig:unable_to_determine_AR_OG_image_endocervical}, Figure~\ref{fig:Unable1_DD_OM_image}, etc. Another case in Figure~\ref{fig:lack_of_knowledge_SG_LMP_image_carcinoma_in_situ} shows that GPT-4o correctly describes the image and understands the question but still chooses a wrong answer, suggesting it may lack the ability to distinguish between carcinoma in situ and invasive carcinoma. 

\begin{figure}[H]
    \centering
    \includegraphics[width=1\linewidth]{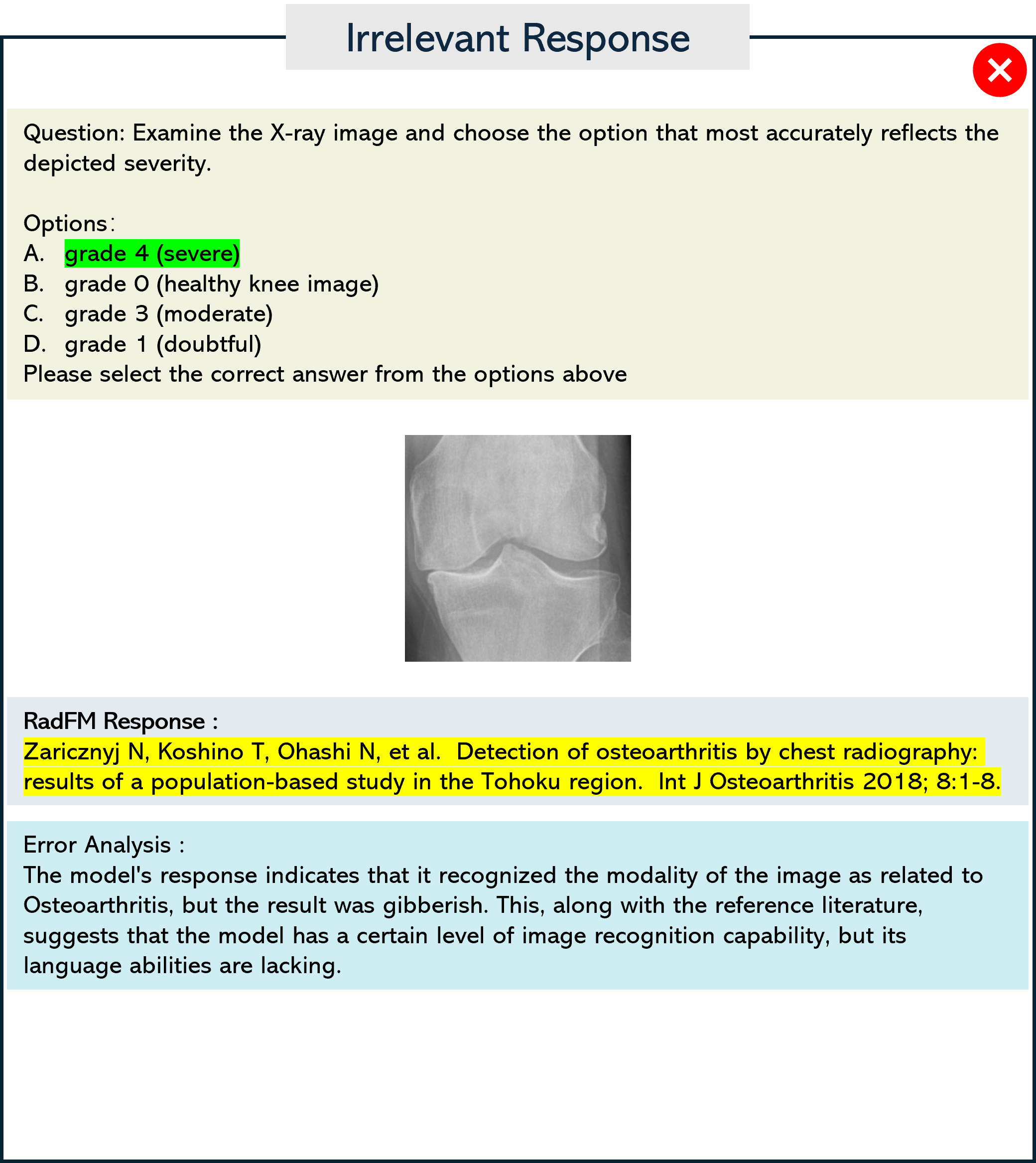}
    \caption{An example of irrelevant response. \boxedgreen{Green} highlights the right answer. \boxedyellow{Yellow} highlights the wrong answer. \hyperref[case_study_table]{\textcolor{red}{Back to Table Index}}.}
    \label{fig:irrelevant_response_knee_osteoarthritis_DD_OS_image_knee}
\end{figure}

\textbf{Irrelevant response:} 
LVLMs do not address the question directly and produce unreadable or unrelated responses. This problem is especially noticeable in open-source models. For example: RadFM only generates a reference paper without any additional outputs in Figure~\ref{fig:irrelevant_response_knee_osteoarthritis_DD_OS_image_knee}.

\textbf{Refuse to answer:} 
LVLMs decline to answer certain questions to keep the system safe for all users, such as those involving sensitive or ethical issues, and refuse to provide medical advice when they determine that human professional assistance is required. This issue only occurs in proprietary models like GPTs and Claudes.

In our test, we randomly select 53 VQA pairs from different clinical VQA tasks, departments, and perceptual granularities. All cases are listed in Table~\ref{case_study_table}. Based on our observations of the evaluation results, we find that proprietary models like GPT-4o and Claude3-Opus rarely encounter difficulties in question understanding. The majority of errors for these models stem from perceptual error and lack of knowledge. In contrast, specialized medical models such as RadFM and LLAVA-Med frequently exhibit language understanding errors, making it difficult to effectively evaluate visual perceptual abilities. As a result, the case study indicates that general models need to enhance their performance on specialized medical images, which may require more medical data for training. Meanwhile, specialized medical models need further training or fine-tuning in language aspects.

\begin{longtable}{ccccc}
    \caption{Table index of our case study figures.}
    \label{case_study_table} \\
    \hline
    Figure & Clinical VQA task & Department & Perceptual granularity & Category \\
    \hline
    \endfirsthead
    \hline
    Figure & Clinical VQA task & Department & Perceptual granularity & Category \\
    \hline
    \endhead
    \hline
    \endfoot
    \hline
    \endlastfoot    
    %\ref{fig:correct_fundus_DD_O_image_correct} & DD & O & Image Level & Correct \\
    \ref{fig:correct_basophil_MR_H_image_correct} & MR & H & Image Level & Correct \\
    \ref{fig:correct_counting_C_H_image_correct} & C & H & Image Level & Correct \\
    \ref{fig::correct_surgicalworkflow_SWR_ENT_image_correct} & SWR & ENT & Image Level & Correct \\
    \ref{fig:ulcerative_colitis_endoscopy_DD_GH_image_correct} & DD & GH & Image Level & Correct \\         
    \ref{fig:NH_kidney_ASR_NH_image_correct} & ASR & NH & Image Level & Correct \\
    \ref{fig:SAR_U_bbox_correct} & SAR & U & Box Level & Correct \\
    \ref{fig:DD_PM_bbox_correct_atelectasis} & DD & PM & Box Level & Correct \\
    \ref{fig:OR-HN_E_mask_correct_thyroid_gland} & OR-NH & E & Mask Level & Correct \\
    \ref{fig:OR-P_U_contour_correct_prostate} & OR-P & U & Contour Level & Correct \\
    \ref{fig:SIR_GS_bbox_correct} & SIR & GS & Box Level & Correct \\ 
    \ref{fig:correct_BVR_H_mask_renal_artery} & BVR & H & Mask Level & Correct \\
    \ref{fig:correct_CR_H_bbox_white_blood_cell} & CR & H & Box Level & Correct \\
    \ref{fig:correct_DD_CS_mask_cardiomegaly} & DD & CS & Mask Level & Correct \\
    \ref{fig:correct_DD_PM_contour_rib_fracture} & DD & OS & Contour Level & Correct \\
    \ref{fig:correct_NT_O_mask_choroidal_layer} & NT & O & Mask Level & Correct \\
    \ref{fig:correct_OR-T_PM_mask_lung} & OR-T & PM & Mask Level & Correct \\
    \ref{fig:correct_SIR_GS_mask_instrument_suction} & SIR & GS & Mask Level & Correct \\
    \ref{fig:question_misunderstanding_endoscopic_SWR_GS_image_question_misunderstanding} & SWR & GS & Image Level & Question misunderstanding \\
    \ref{fig:question_misunderstanding_BVR_O_mask_retinal_vessel} & BVR & O & Mask Level & Question misunderstanding \\
    \ref{fig:question_misunderstanding_ACR_OS_mask_clavicle} & ACR & OS & Mask Level & Question misunderstanding \\  
    \ref{fig:QM_MR_GH_image_normal_colonic_mucosa} & MR & GH & Image Level & Question misunderstanding \\
    \ref{fig:PE-D-1_C_H_image} & C & H & Image Level & Perceptual error \\
    \ref{fig:PE-D-2_SWR_GS_image} & SWR & GS & Image Level & Perceptual error \\
    \ref{fig:PE-D-3_OR-T_PM_mask} & OR-T & PM & Mask Level & Perceptual error \\
    \ref{fig:PE-M-1_AR_LMP_image} & AR & LMP & Image Level & Perceptual error \\
    \ref{fig:PE-M-2_NT_N_mask} & NT & N & Mask Level & Perceptual error \\
    \ref{fig:PE-M-3_DD_CS_bbox} & DD & CS & Box Level & Perceptual error \\
    \ref{fig:PE-M-4_DD_D_mask} & DD & D & Mask Level & Perceptual error \\
    \ref{fig:PE-M-5_DD_GH_contour} & DD & GH & Contour Level & Perceptual error \\
    \ref{fig:PE-M-6_OR-T_PM_mask} & OR-T & PM & Mask Level & Perceptual error \\
    \ref{fig:PE-M-7_NT_N_mask} & NT & N & Mask Level & Perceptual error \\
    \ref{fig:PE-M-8_OR-T_PM_contour} & OR-T & PM & Contour Level & Perceptual error \\
    \ref{fig:lack_of_knowledge_cataract_DD_O_image_lack_of_knowledge} & DD & O & Image Level & Lack of knowledge \\
    \ref{fig:lack_of_knowledge_fundus_bad_quality_IQG_O_image_lack_of_knowledge} & IQG & O & Image Level & Lack of knowledge \\
    \ref{fig:lack_of_knowledge_microorganisms_MR_LMP_image_lack_of_knowledge} & MR & LMP & Image Level & Lack of knowledge \\ 
    \ref{fig:SAR_GS_bbox_lack_of_knowledge} & SAR & GS & Box Level & Lack of knowledge \\
    \ref{fig:SAR_U_bbox_lack_of_knowledge_bladder_anastomosis} & SAR & U & Box Level & Lack of knowledge \\
    \ref{fig:lack_of_knowledge_DD_PM_mask_pneumothorax} & DD & PM & Mask Level & Lack of knowledge \\ 
    \ref{fig:lack_of_knowledge_NT_O_mask_choroidal_layer} & NT & O & Mask Level & Lack of knowledge \\         
    \ref{fig:Lok1_SG_LMP_image} & SG & LMP & Image Level & Lack of knowledge \\
    %\ref{fig:Lok2_CR_LMP_image} & CR & LMP & Image Level & Lack of knowledge \\ 
    \ref{lack_of_knowledge_DD_O_image_diffuse_leakage_diabetic_macular_edema} & DD & O & Image Level & Lack of knowledge \\
    \ref{fig:lack_of_knowledge_SG_LMP_image_carcinoma_in_situ} & SG & LMP & Image Level & Lack of knowledge \\              
    %\ref{fig:image_misinterpretation_OR-T_PM_contour_chest_wall} & OR-T & PM & Contour Level & image misinterpretation \\
    %\ref{fig:image_misinterpretation_OR-T_PM_mask_lung} & OR-T & PM & Mask Level & image misinterpretation \\   
    %\ref{fig:IM_DD_GH_contour_liver_tumor} & DD & GH & Contour Level & image misinterpretation \\  
    %\ref{fig:IM_NT_N_mask_brachial_plexus} & NT & N & Mask Level & image misinterpretation \\
    %\ref{fig:image_misinterpretation_OR-T_OM_mask_breast} & OR-T & OM & Mask Level & image misinterpretation \\ 
    \ref{fig:Unable1_DD_OM_image} & DD & OM & Image Level & Lack of knowledge \\
    \ref{fig:unable_to_determine_male_AR_GS_image_unable_to_determine} & AR & GS & Image Level & Lack of knowledge \\
    \ref{fig:unable_to_determine_AR_OG_image_endocervical} & AR & OG & Image Level & Lack of knowledge \\
    \ref{fig:unable_to_determine_DD_D_image_monkeypox} & DD & D & Image Level & Lack of knowledge \\
    \ref{fig:unable_to_determine_DD_U_image_abnormal_sperm_tail} & DD & U & Image Level & Lack of knowledge \\
    %\ref{fig:UD_DD_OM_image_neoplasia} & DD & OM & Image Level & Lack of knowledge \\    
    \ref{fig:irrelevant_response_knee_osteoarthritis_DD_OS_image_knee} & DD & OS & Image Level & Irrelevant response \\
    \ref{fig:irrelevant_response_AR_ID_image_atypical_appearance_of_COVID-19} & AR & ID & Image Level & Irrelevant response \\
    \ref{fig:irrelevant_response_AR_OS_image_fractures_on_the_left_part_of_lowerlimb.png} & AR & OS & Image Level & Irrelevant response \\
    \ref{fig:irrelevant_response_ASR_OG_image_ovary.png} & ASR & OG & Image Level & Irrelevant response \\         
    \ref{fig:refuse_to_answer_COVIDe-9_negative_DD_PM_image_refuse_to_answer} & DD & PM & Image Level & Refuse to answer \\
    \ref{fig:refuse_to_answer_BVR_O_mask_retinal_vein} & BVR & O & Mask Level & Refuse to answer \\
    \hline
\end{longtable}

\newpage
%%%%%%%%%%%%%%%%%%%%%%%%%%%%%%%%%%%%%%%%%%%%%%%%%%%%%%%%%%%% correct
%\begin{figure}[H]
   %\centering
    %\includegraphics[width=1\linewidth]{Styles//case_study//correct/advanced glaucoma.png}
    %\caption{An example of correct case. \boxedgreen{Green} highlights the right answer.  \hyperref[case_study_table]{\textcolor{red}{Back to Table Index}}. }
    %\label{fig:correct_fundus_DD_O_image_correct}
%\end{figure}

\begin{figure}[H]
    \centering
    \includegraphics[width=1\linewidth]{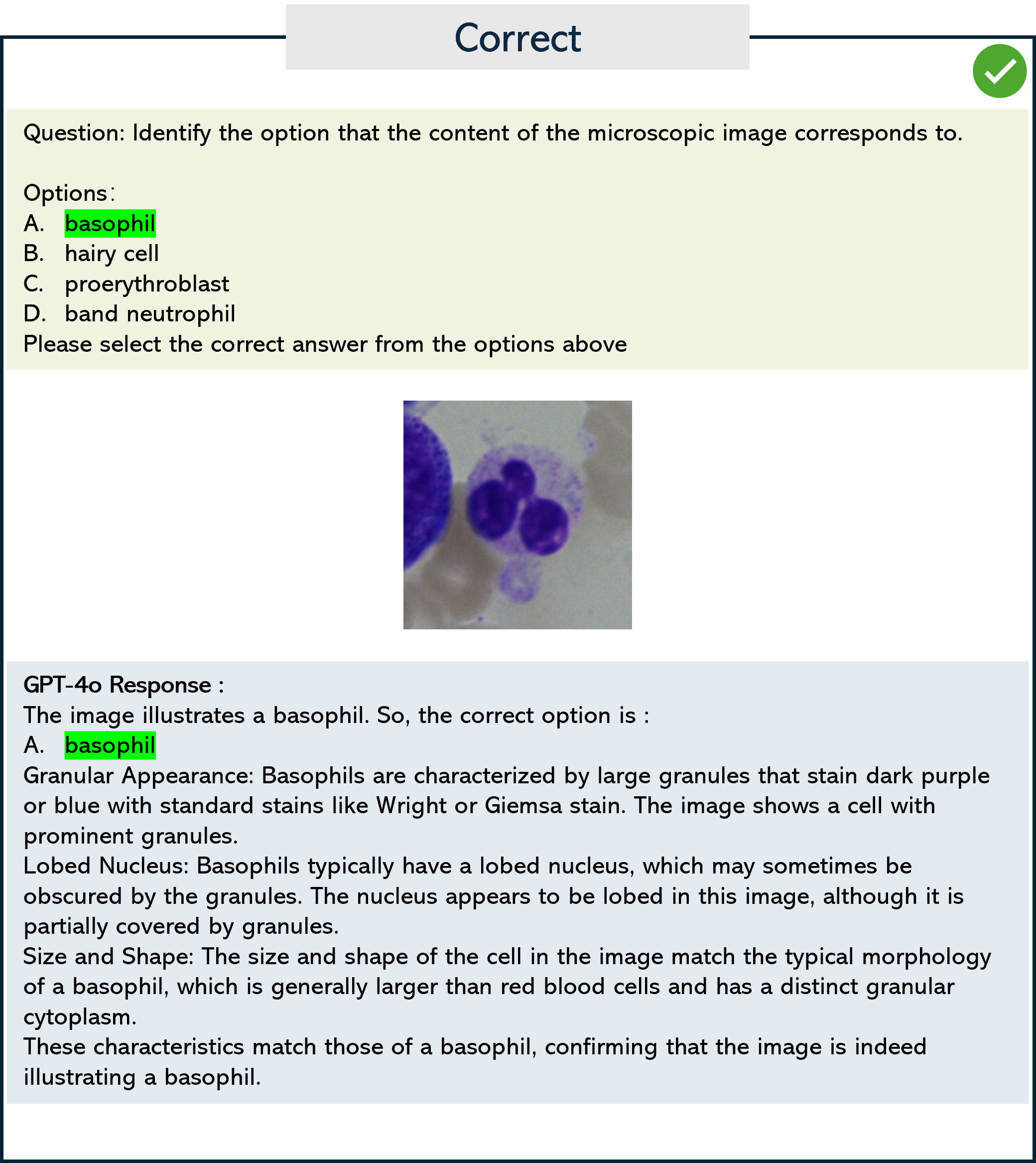}
    \caption{An example of correct case. \boxedgreen{Green} highlights the right answer. \hyperref[case_study_table]{\textcolor{red}{Back to Table Index}}.}
    \label{fig:correct_basophil_MR_H_image_correct}
\end{figure}

\begin{figure}[H]
    \centering
    \includegraphics[width=1\linewidth]{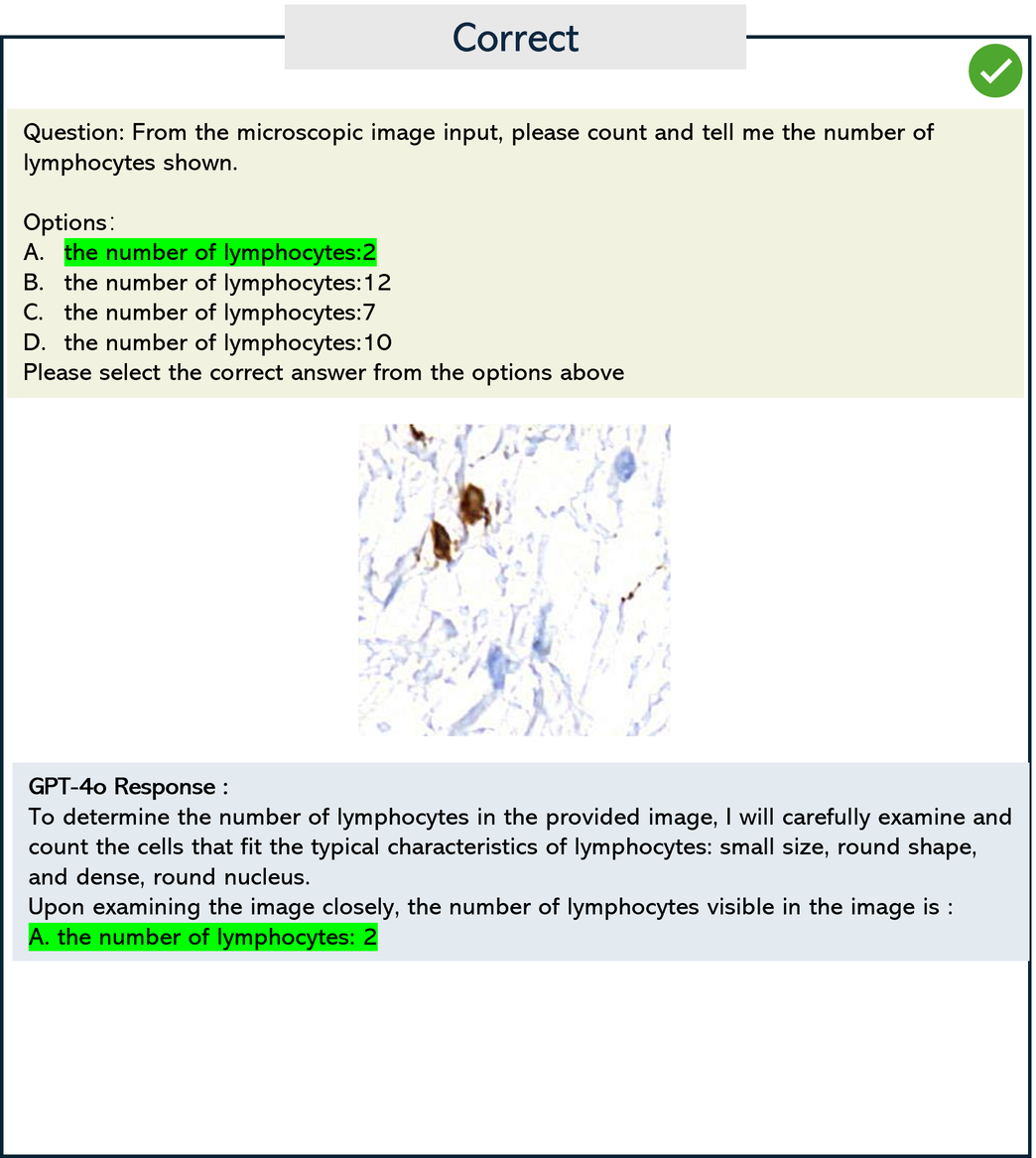}
    \caption{An example of correct case. \boxedgreen{Green} highlights the right answer. \hyperref[case_study_table]{\textcolor{red}{Back to Table Index}}.}
    \label{fig:correct_counting_C_H_image_correct}
\end{figure}

\begin{figure}[H]
    \centering
    \includegraphics[width=1\linewidth]{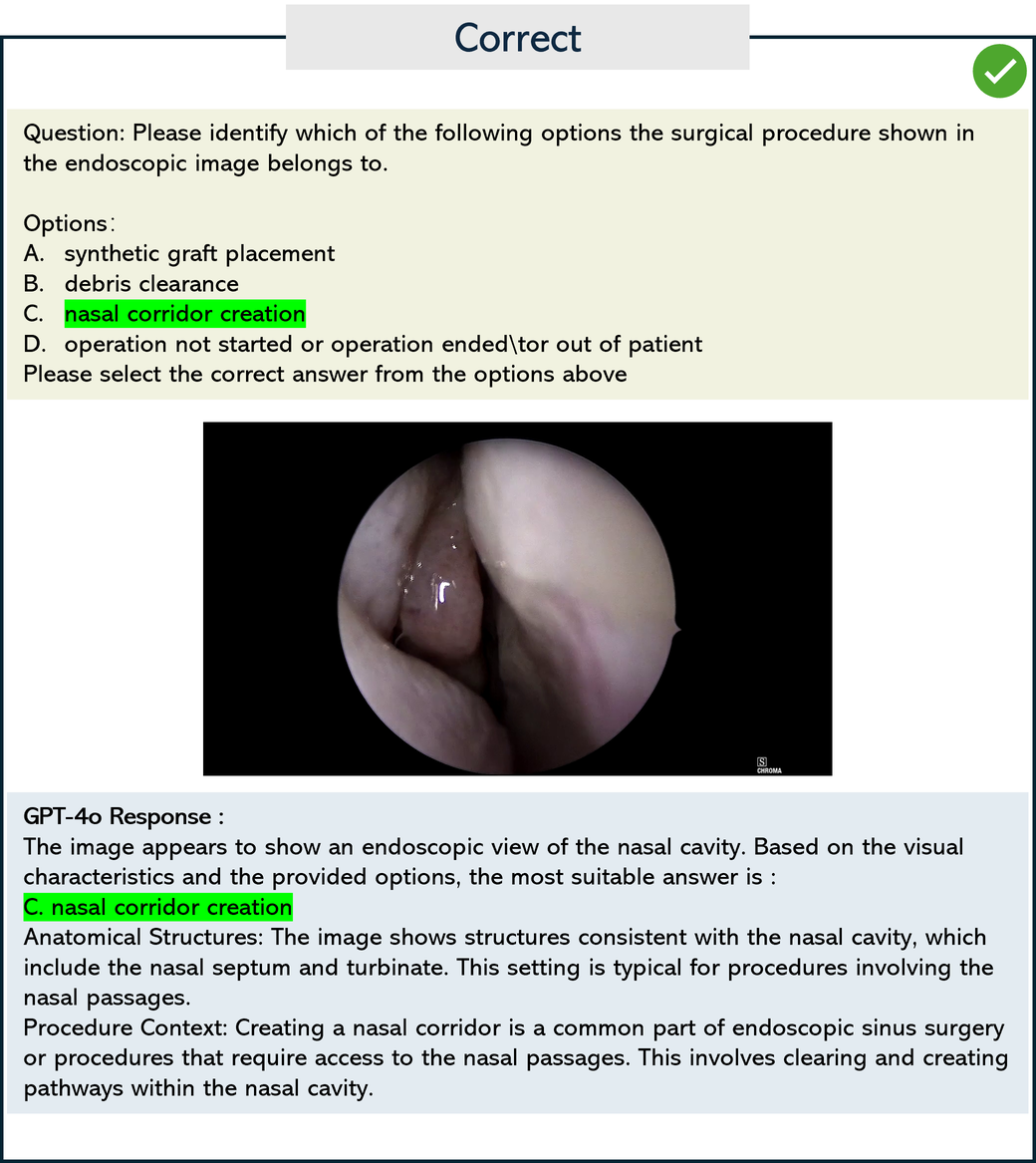}
    \caption{An example of correct case. \boxedgreen{Green} highlights the right answer.  \hyperref[case_study_table]{\textcolor{red}{Back to Table Index}}.}
    \label{fig::correct_surgicalworkflow_SWR_ENT_image_correct}
\end{figure}

\begin{figure}[H]
    \centering
    \includegraphics[width=1\linewidth]{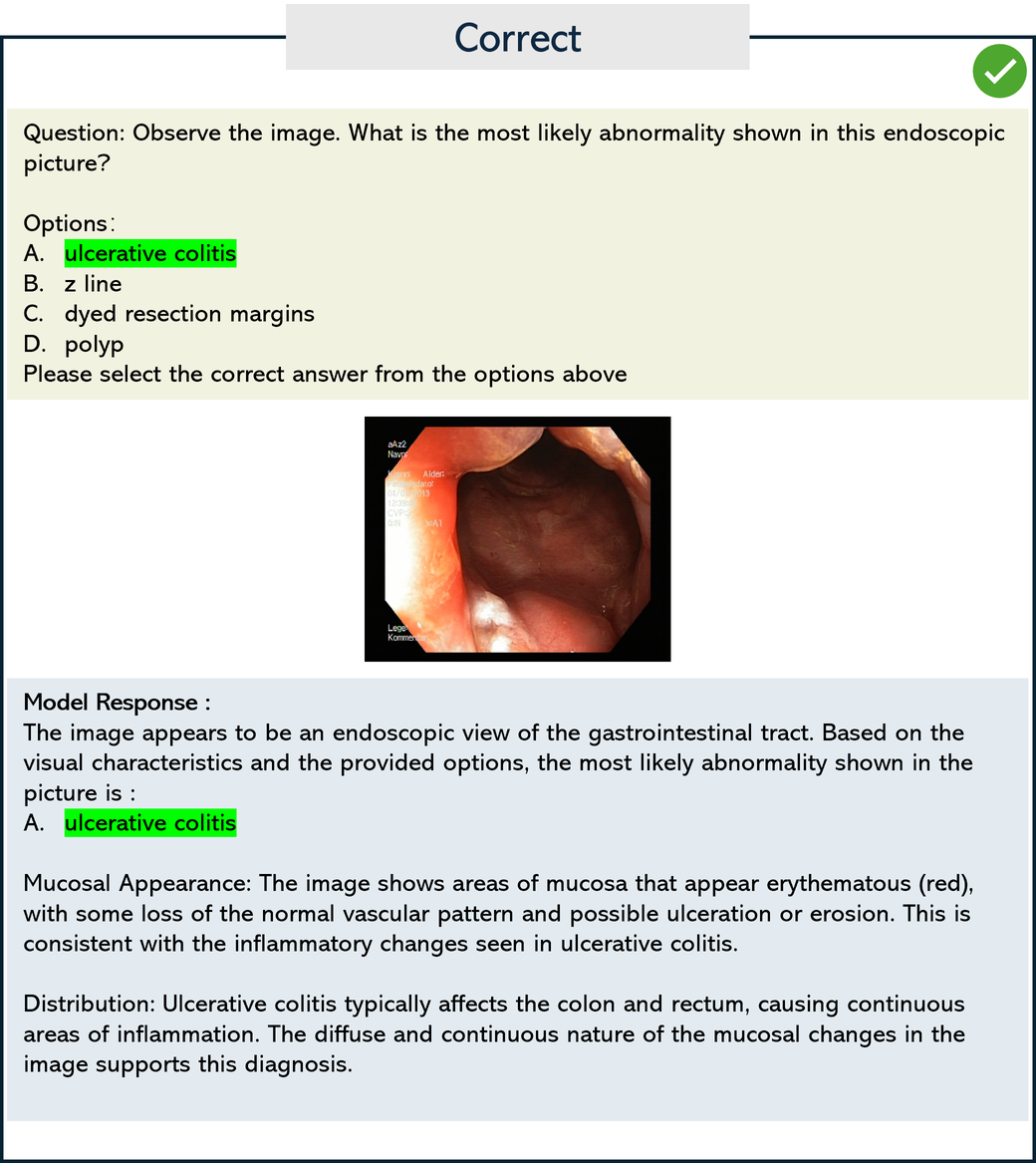}
    \caption{An example of correct case. \boxedgreen{Green} highlights the right answer. \hyperref[case_study_table]{\textcolor{red}{Back to Table Index}}.}
    \label{fig:ulcerative_colitis_endoscopy_DD_GH_image_correct}
\end{figure}

\begin{figure}[H]
    \centering
    \includegraphics[width=1\linewidth]{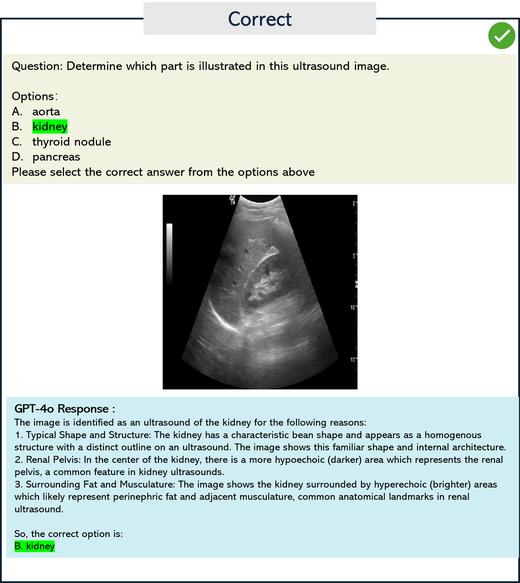}
    \caption{An example of correct case. \boxedgreen{Green} highlights the right answer.  \hyperref[case_study_table]{\textcolor{red}{Back to Table Index}}.}
    \label{fig:NH_kidney_ASR_NH_image_correct}
\end{figure}

\begin{figure}[H]
    \centering
    \includegraphics[width=1\linewidth]{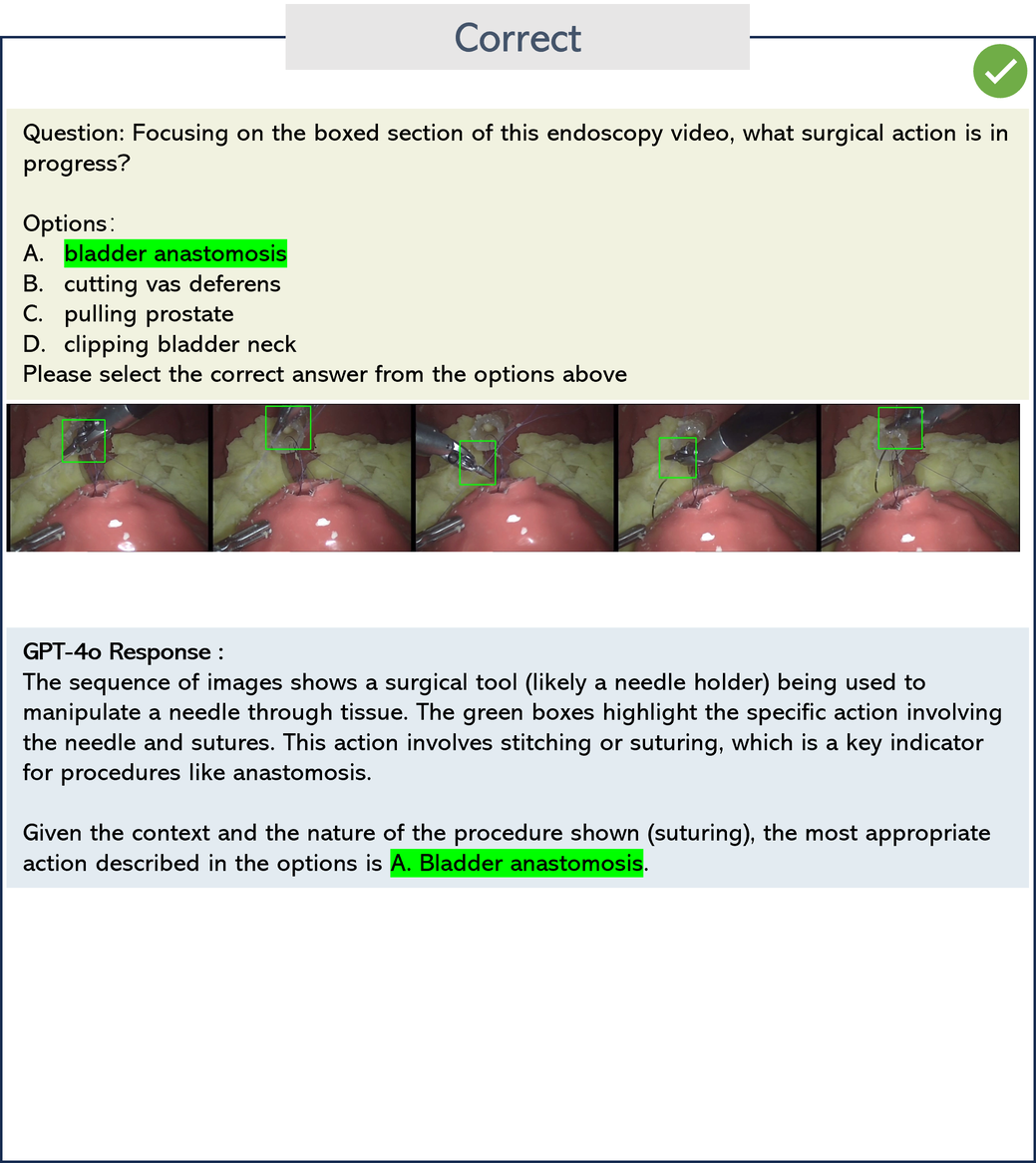}
    \caption{An example of correct case. \boxedgreen{Green} highlights the right answer. \hyperref[case_study_table]{\textcolor{red}{Back to Table Index}}.}
    \label{fig:SAR_U_bbox_correct}
\end{figure}

\begin{figure}[H]
    \centering
    \includegraphics[width=1\linewidth]{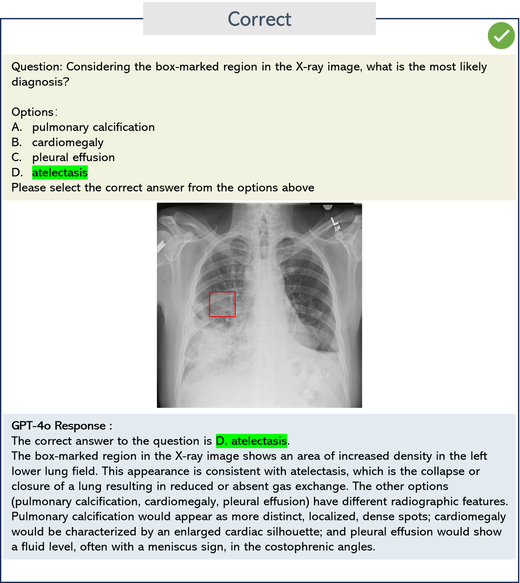}
    \caption{An example of correct case. \boxedgreen{Green} highlights the right answer. \hyperref[case_study_table]{\textcolor{red}{Back to Table Index}}.}
    \label{fig:DD_PM_bbox_correct_atelectasis}
\end{figure}

\begin{figure}[H]
    \centering
    \includegraphics[width=1\linewidth]{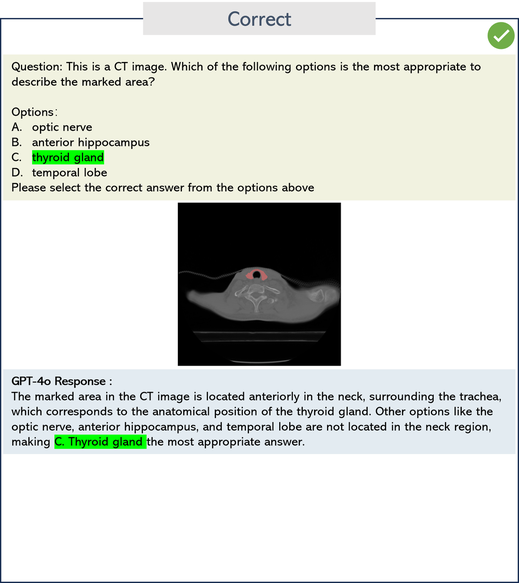}
    \caption{An example of correct case. \boxedgreen{Green} highlights the right answer. \hyperref[case_study_table]{\textcolor{red}{Back to Table Index}}.}
    \label{fig:OR-HN_E_mask_correct_thyroid_gland}
\end{figure}

\begin{figure}[H]
    \centering
    \includegraphics[width=1\linewidth]{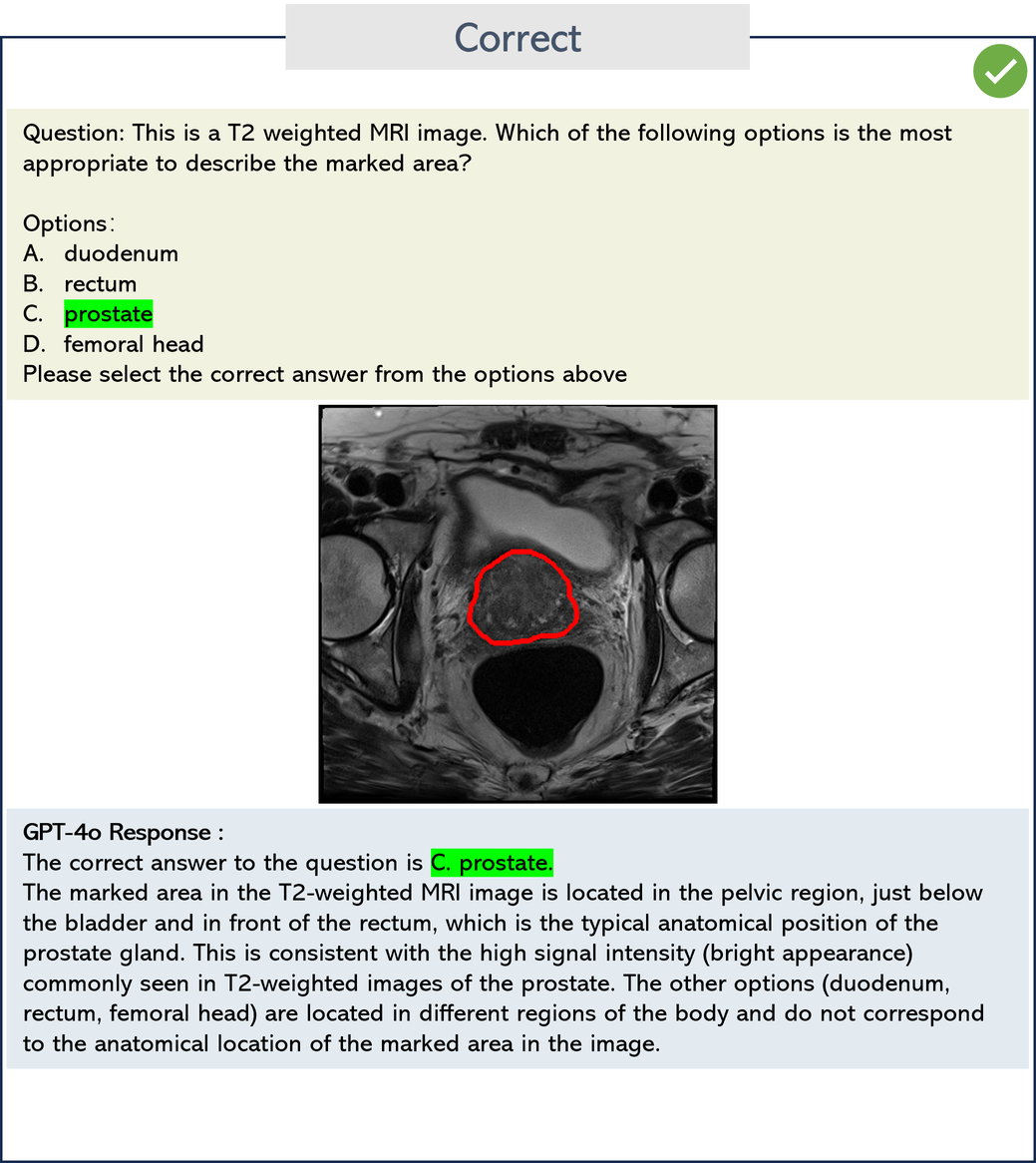}
    \caption{An example of correct case. \boxedgreen{Green} highlights the right answer.  \hyperref[case_study_table]{\textcolor{red}{Back to Table Index}}.}
    \label{fig:OR-P_U_contour_correct_prostate}
\end{figure}

\begin{figure}[H]
    \centering
    \includegraphics[width=1\linewidth]{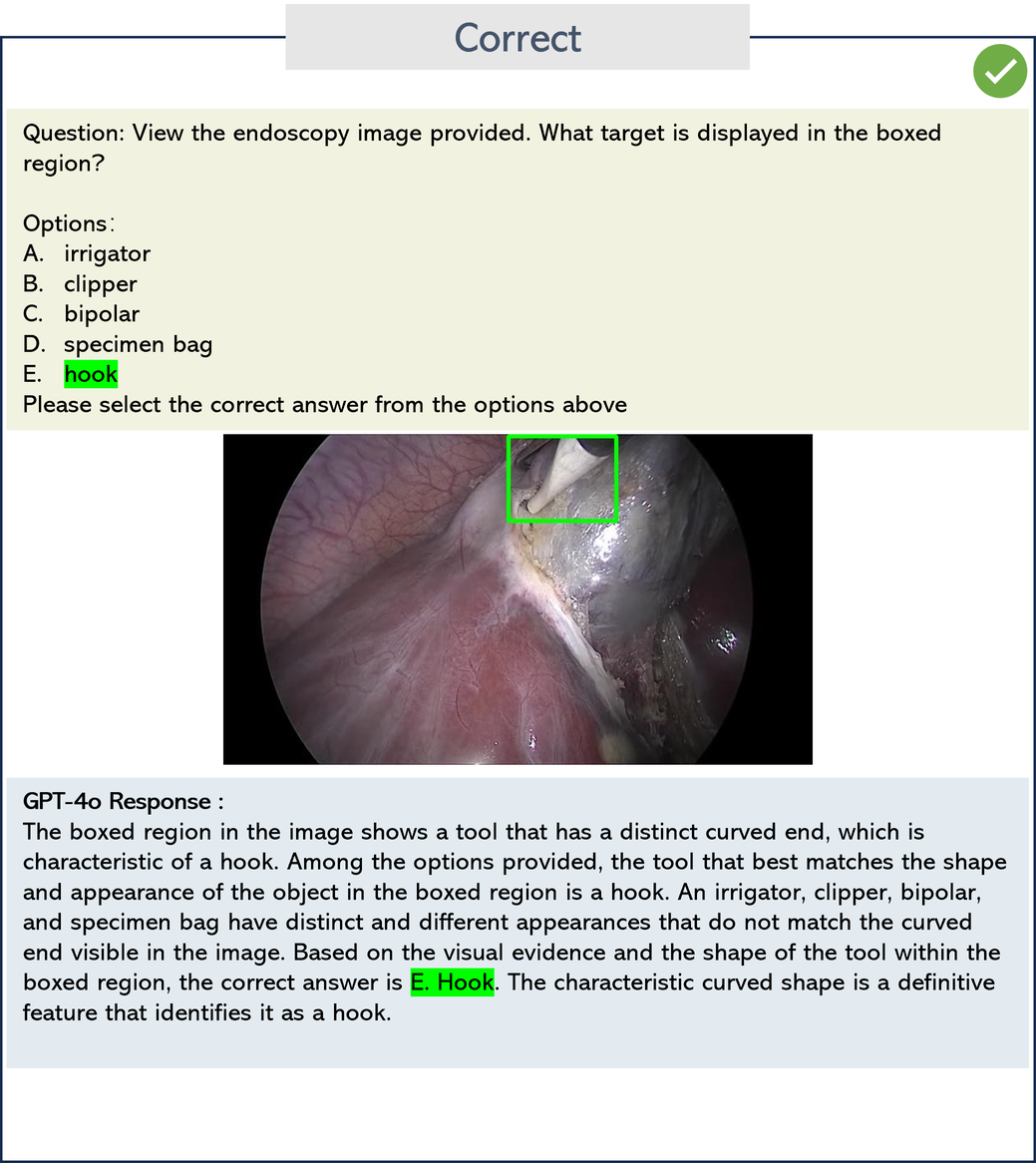}
    \caption{An example of correct case. \boxedgreen{Green} highlights the right answer.  \hyperref[case_study_table]{\textcolor{red}{Back to Table Index}}.}
    \label{fig:SIR_GS_bbox_correct}
\end{figure}

\begin{figure}[H]
    \centering
    \includegraphics[width=1\linewidth]{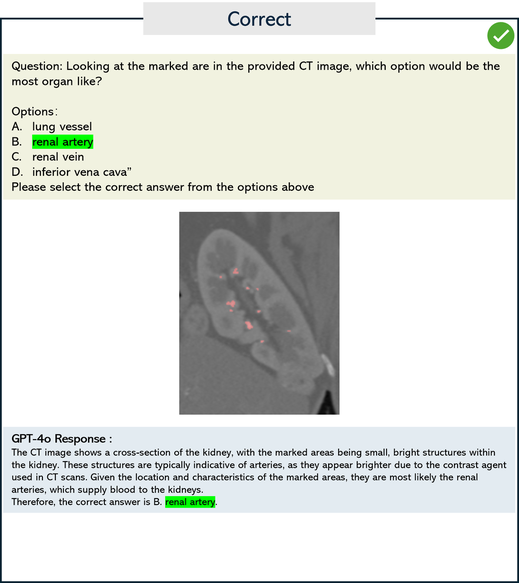}
    \caption{An example of correct case. \boxedgreen{Green} highlights the right answer.  \hyperref[case_study_table]{\textcolor{red}{Back to Table Index}}.}
    \label{fig:correct_BVR_H_mask_renal_artery}
\end{figure}

\begin{figure}[H]
    \centering
    \includegraphics[width=1\linewidth]{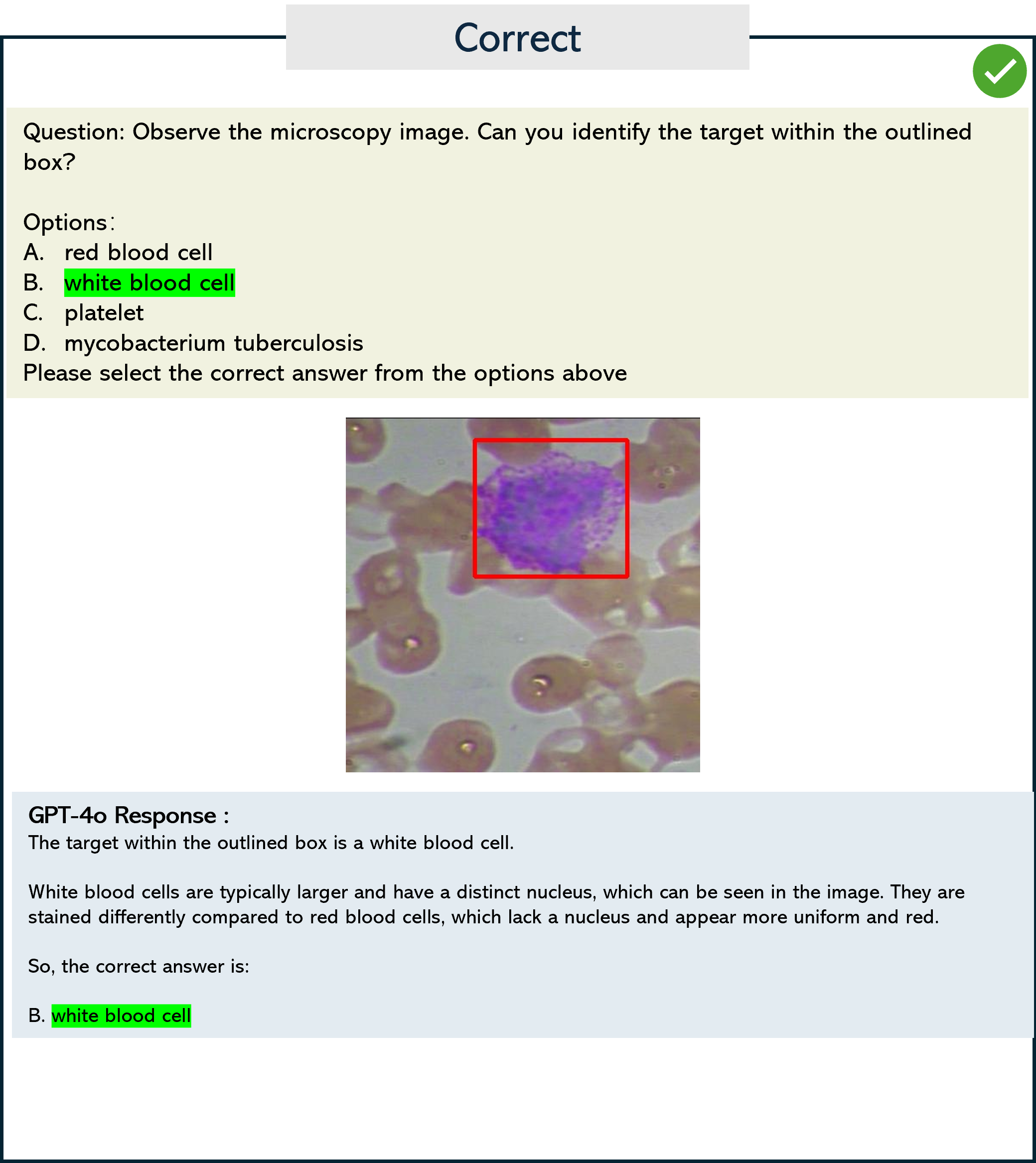}
    \caption{An example of correct case. \boxedgreen{Green} highlights the right answer. \hyperref[case_study_table]{\textcolor{red}{Back to Table Index}}.}
    \label{fig:correct_CR_H_bbox_white_blood_cell}
\end{figure}

\begin{figure}[H]
    \centering
    \includegraphics[width=1\linewidth]{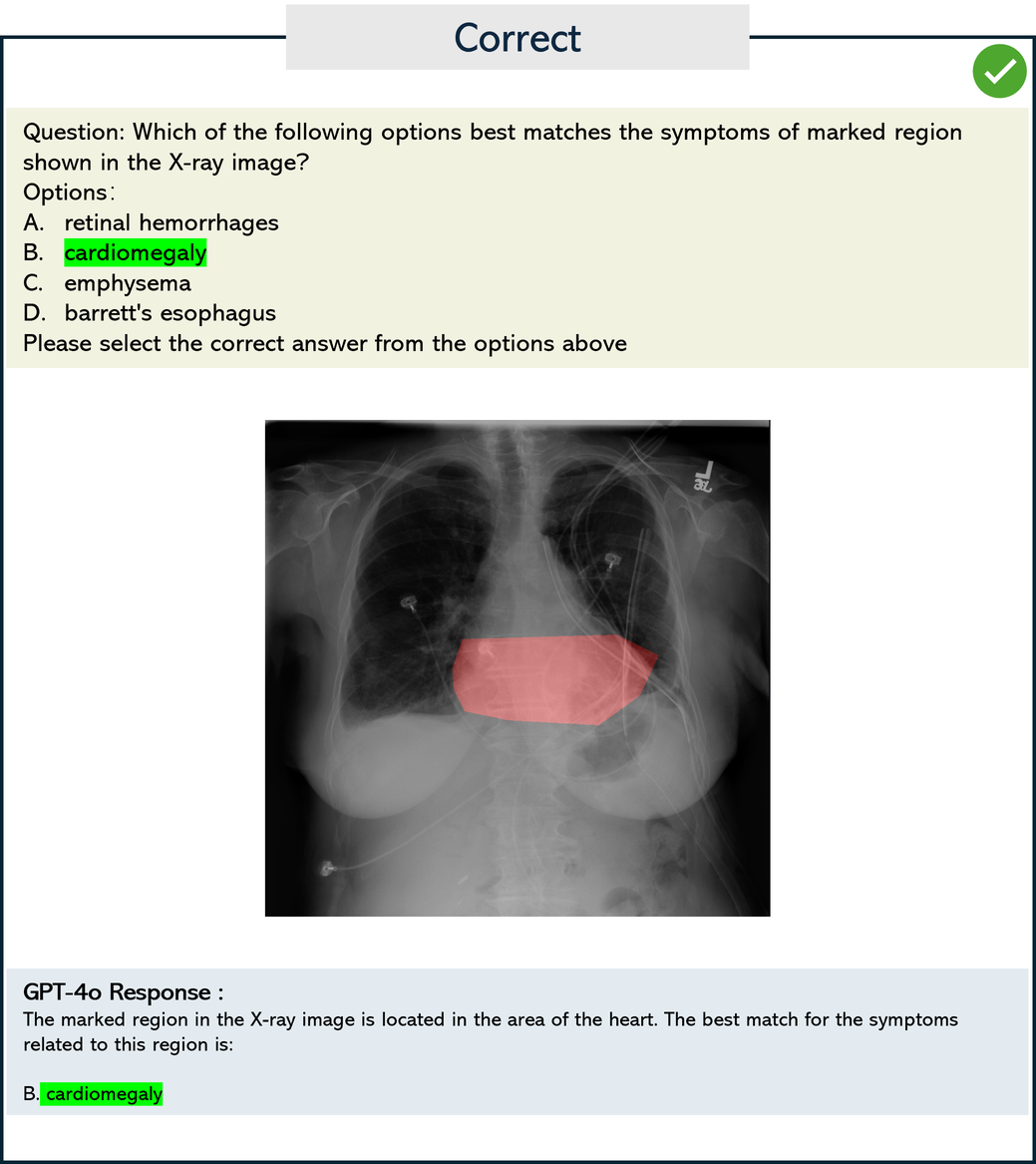}
    \caption{An example of correct case. \boxedgreen{Green} highlights the right answer.  \hyperref[case_study_table]{\textcolor{red}{Back to Table Index}}.}
    \label{fig:correct_DD_CS_mask_cardiomegaly}
\end{figure}

\begin{figure}[H]
    \centering
    \includegraphics[width=1\linewidth]{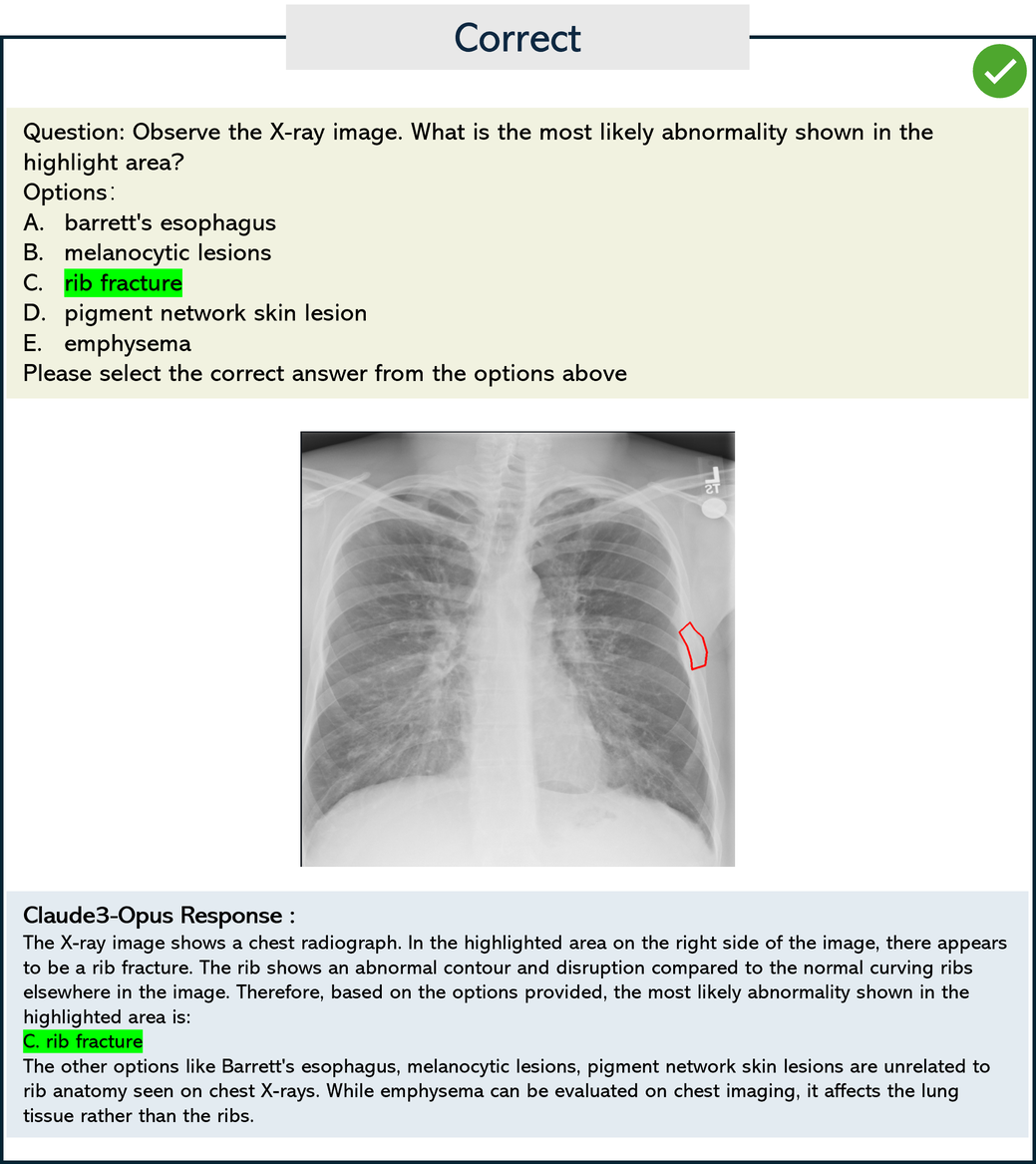}
    \caption{An example of correct case. \boxedgreen{Green} highlights the right answer.  \hyperref[case_study_table]{\textcolor{red}{Back to Table Index}}.}
    \label{fig:correct_DD_PM_contour_rib_fracture}
\end{figure}

\begin{figure}[H]
    \centering
    \includegraphics[width=1\linewidth]{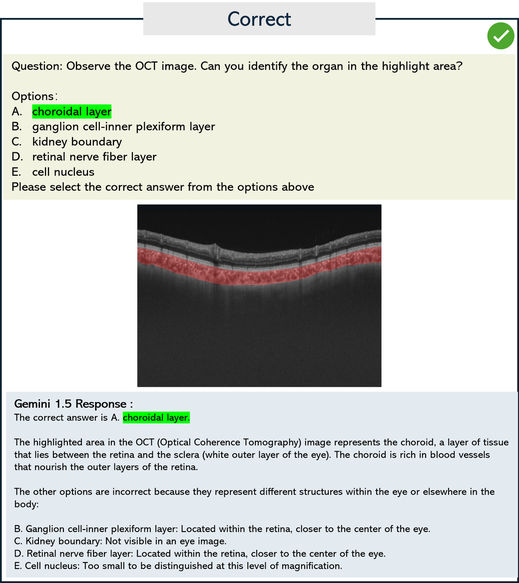}
    \caption{An example of correct case. \boxedgreen{Green} highlights the right answer. \hyperref[case_study_table]{\textcolor{red}{Back to Table Index}}.}
    \label{fig:correct_NT_O_mask_choroidal_layer}
\end{figure}

\begin{figure}[H]
    \centering
    \includegraphics[width=1\linewidth]{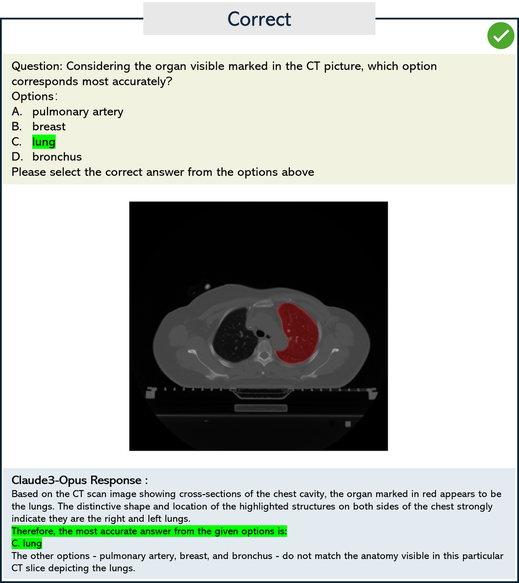}
    \caption{An example of correct case. \boxedgreen{Green} highlights the right answer.  \hyperref[case_study_table]{\textcolor{red}{Back to Table Index}}.}
    \label{fig:correct_OR-T_PM_mask_lung}
\end{figure}

\begin{figure}[H]
    \centering
    \includegraphics[width=1\linewidth]{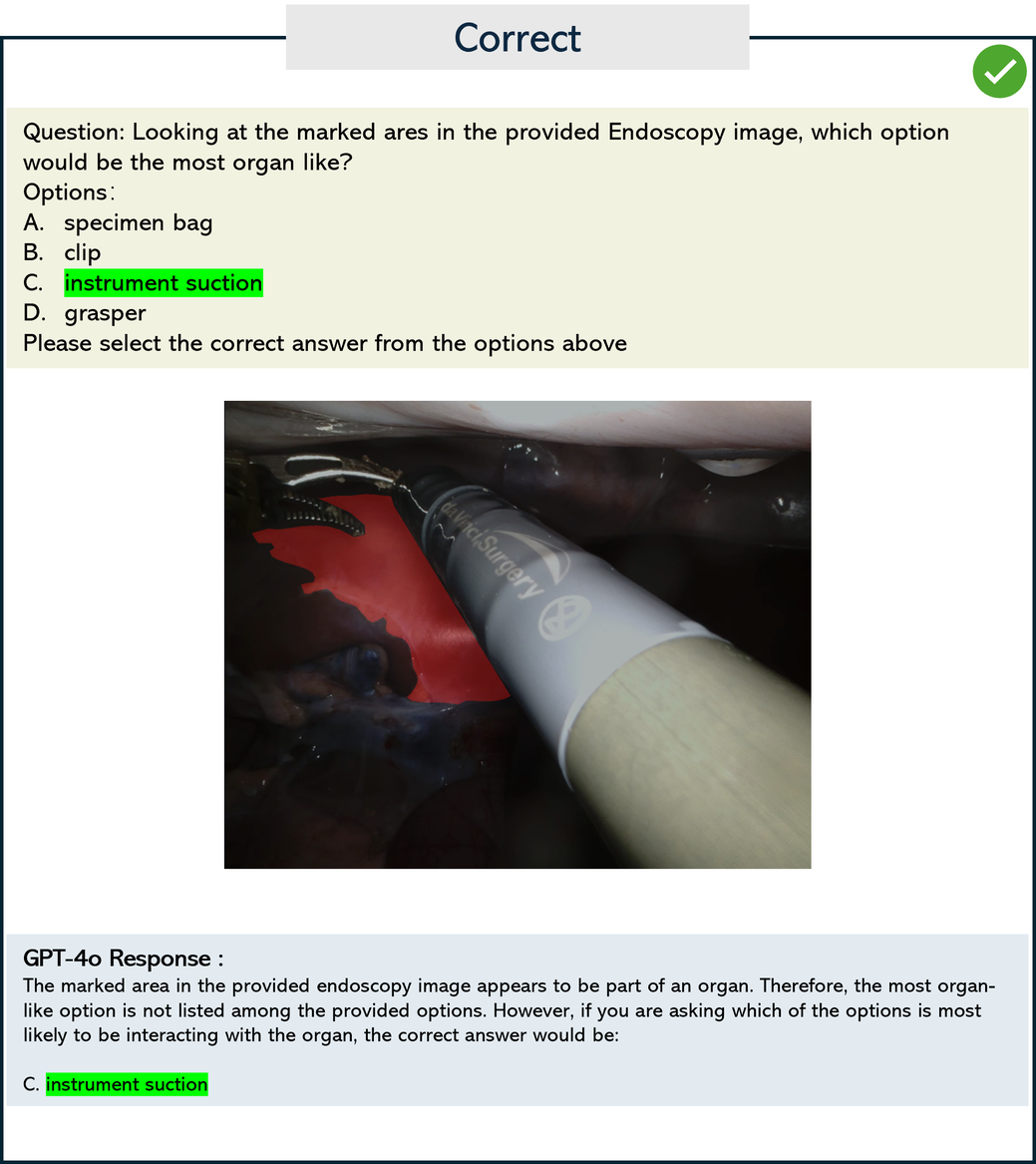}
    \caption{An example of correct case. \boxedgreen{Green} highlights the right answer. \hyperref[case_study_table]{\textcolor{red}{Back to Table Index}}.}
    \label{fig:correct_SIR_GS_mask_instrument_suction}
\end{figure}
%%%%%%%%%%%%%%%%%%%%%%%%%%%%%%%%%%%%%%%%%%%%%%%%%%%%%%%%%%%% question_misunderstanding

\begin{figure}[H]
    \centering
    \includegraphics[width=1\linewidth]{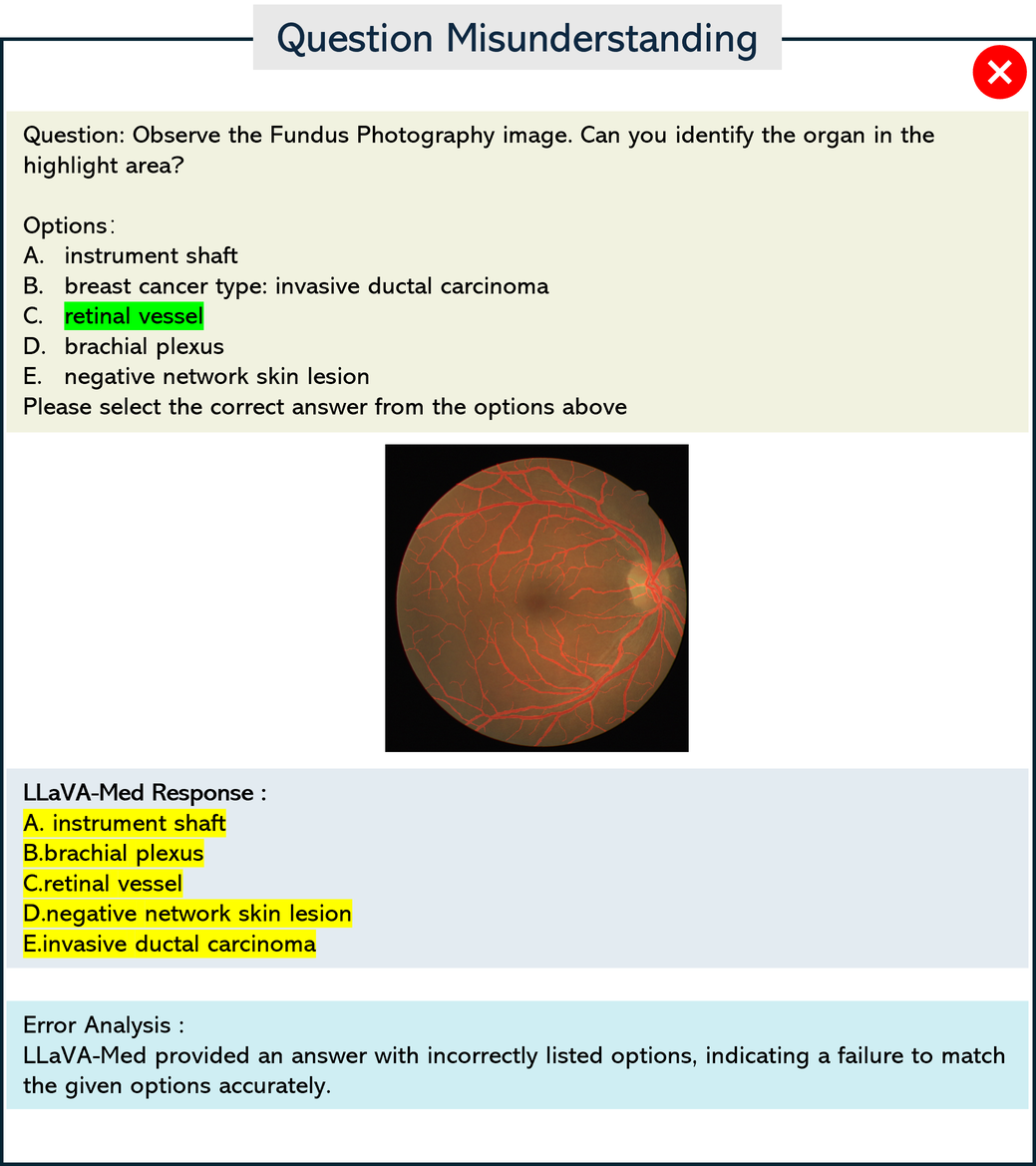}
    \caption{A question misunderstanding example. \boxedgreen{Green} highlights the right answer. \boxedyellow{Yellow} highlights the wrong answer. \hyperref[case_study_table]{\textcolor{red}{Back to Table Index}}.}
    \label{fig:question_misunderstanding_BVR_O_mask_retinal_vessel}
\end{figure}

\begin{figure}[H]
    \centering
    \includegraphics[width=1\linewidth]{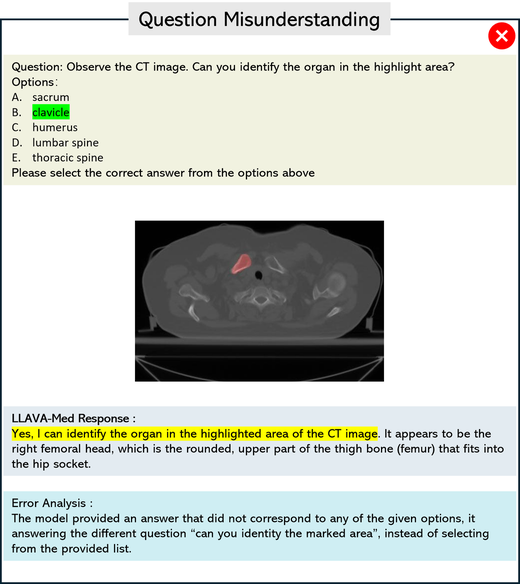}
    \caption{A question misunderstanding example. \boxedgreen{Green} highlights the right answer. \boxedyellow{Yellow} highlights the wrong answer. \hyperref[case_study_table]{\textcolor{red}{Back to Table Index}}.}
    \label{fig:question_misunderstanding_ACR_OS_mask_clavicle}
\end{figure}

\begin{figure}
    \centering
    \includegraphics[width=1\linewidth]{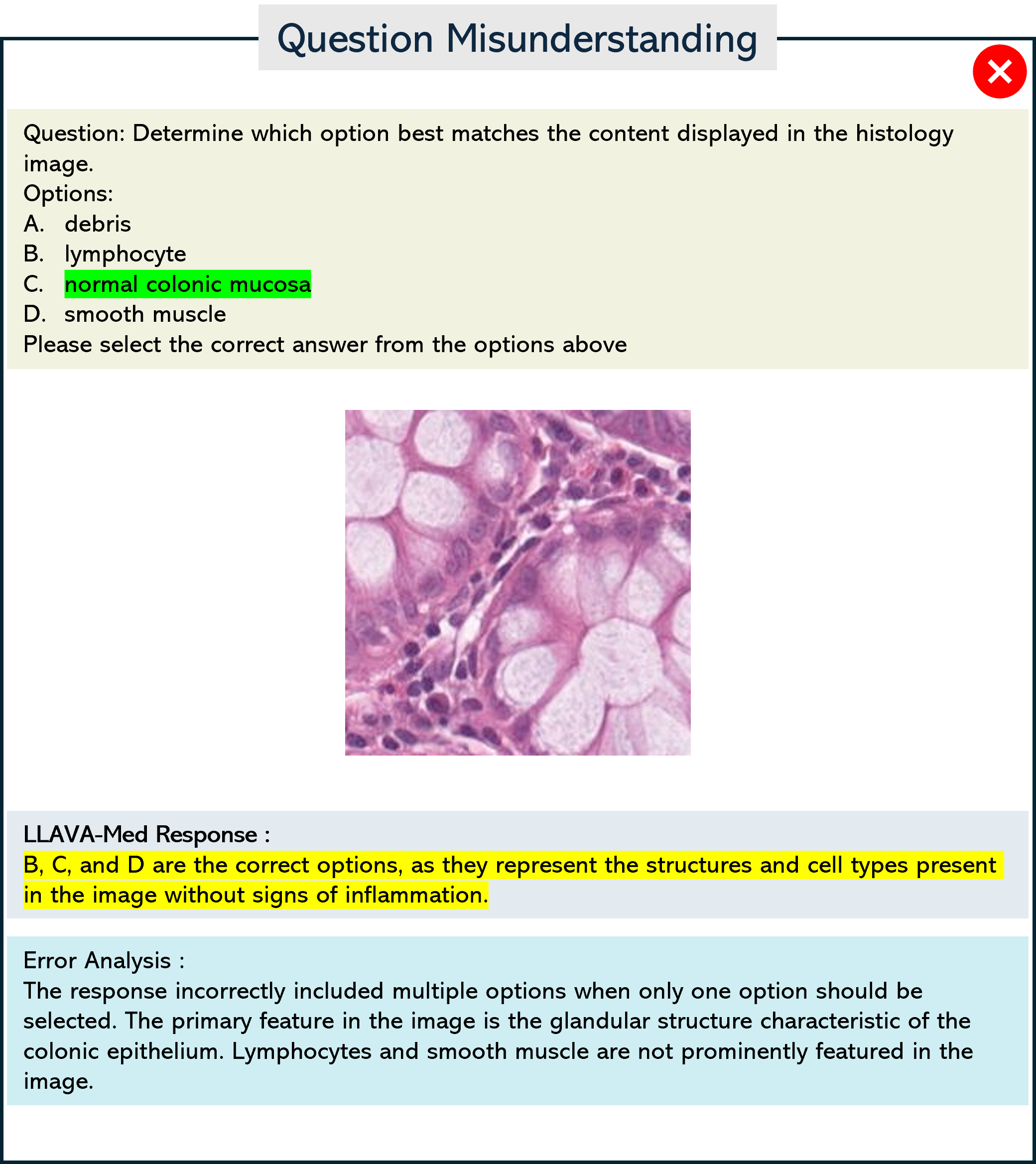}
    \caption{A question misunderstanding example. \boxedgreen{Green} highlights the right answer. \boxedyellow{Yellow} highlights the wrong answer. \hyperref[case_study_table]{\textcolor{red}{Back to Table Index}}.}
    \label{fig:QM_MR_GH_image_normal_colonic_mucosa}
\end{figure}
%%%%%%%%%%%%%%%%%%%%%%%%%%%%%%%%%%%%%%%%%%%%%%%%%%%%%%%%%%%% perceptual_error
\begin{figure}[H]
    \centering
    \includegraphics[width=1\linewidth]{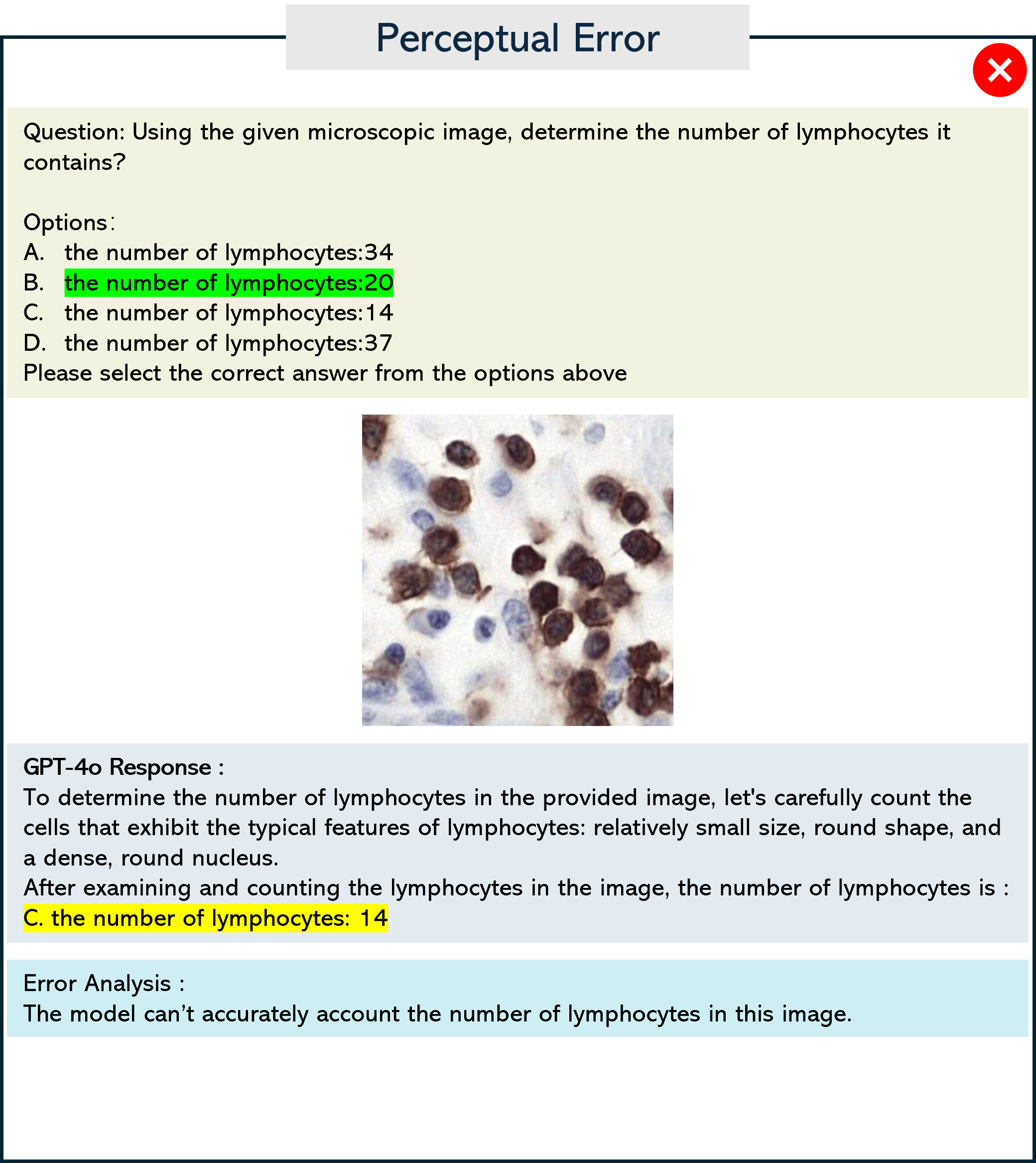}
    \caption{An example of perceptual error - detail missing case. \boxedgreen{Green} highlights the right answer. \boxedyellow{Yellow} highlights the wrong answer. \hyperref[case_study_table]{\textcolor{red}{Back to Table Index}}.}
    \label{fig:PE-D-1_C_H_image}
\end{figure}

\begin{figure}[H]
    \centering
    \includegraphics[width=1\linewidth]{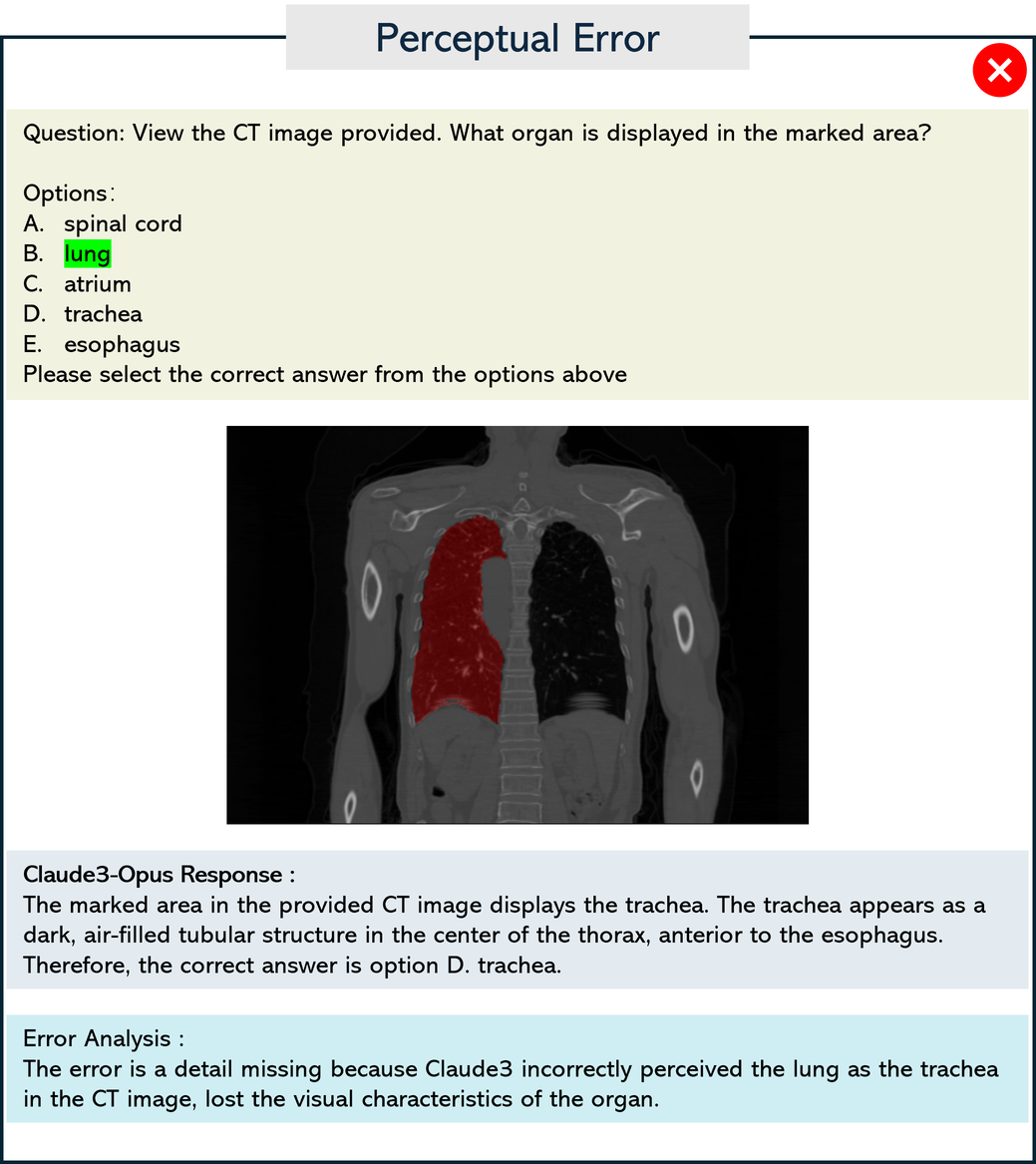}
    \caption{An example of perceptual error - detail missing case. \boxedgreen{Green} highlights the right answer. \boxedyellow{Yellow} highlights the wrong answer. \hyperref[case_study_table]{\textcolor{red}{Back to Table Index}}.}
    \label{fig:PE-D-3_OR-T_PM_mask}
\end{figure}

\begin{figure}[H]
    \centering
    \includegraphics[width=1\linewidth]{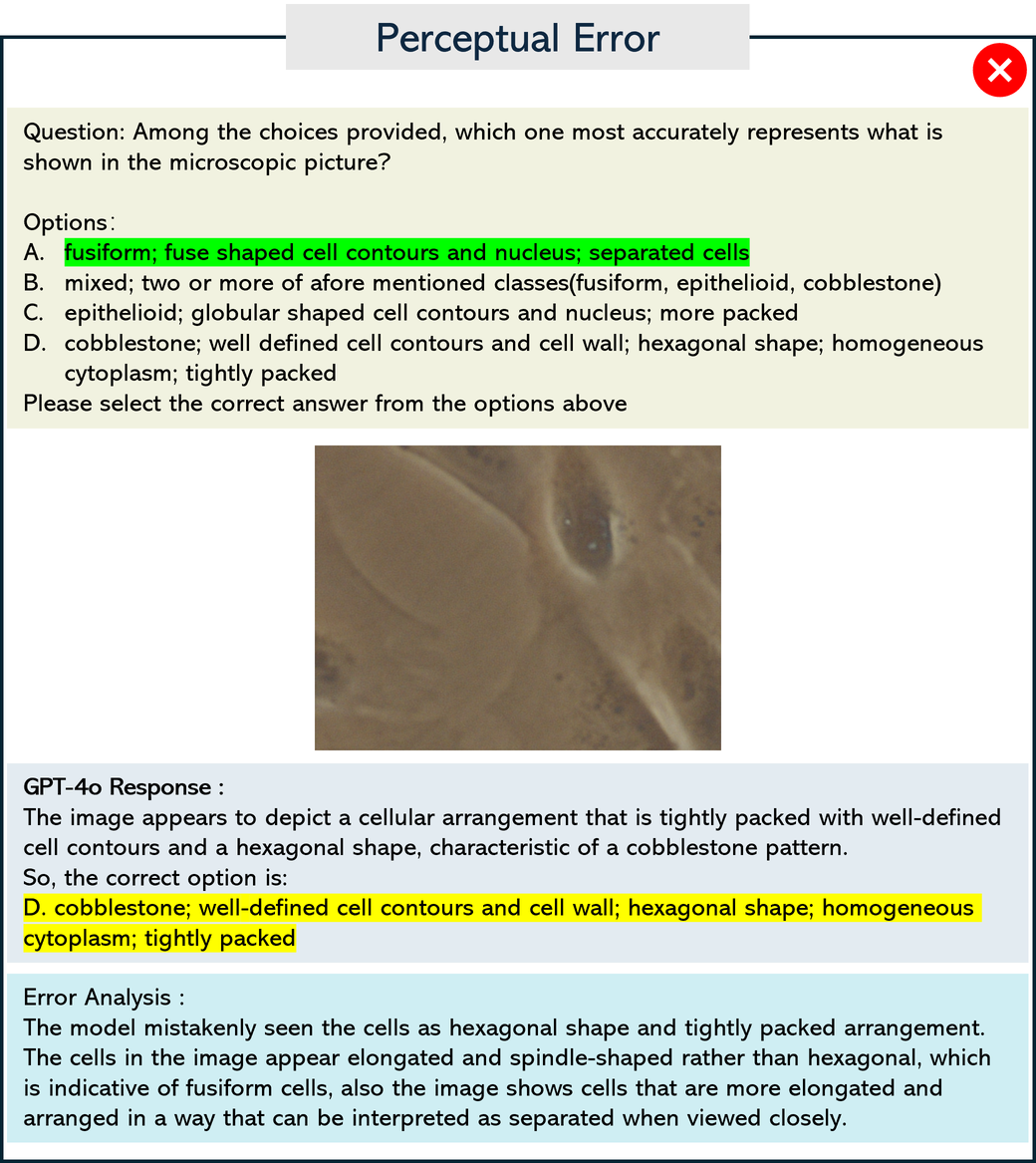}
    \caption{An example of perceptual error - misinterpretation case. \boxedgreen{Green} highlights the right answer. \boxedyellow{Yellow} highlights the wrong answer. \hyperref[case_study_table]{\textcolor{red}{Back to Table Index}}.}
    \label{fig:PE-M-1_AR_LMP_image}
\end{figure}

\begin{figure}[H]
    \centering
    \includegraphics[width=1\linewidth]{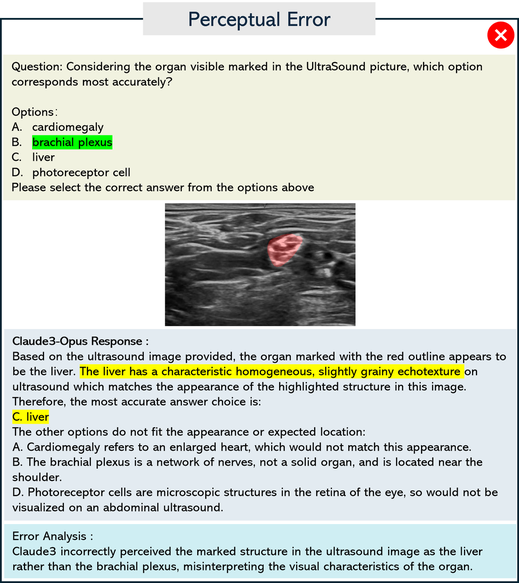}
    \caption{An example of perceptual error - misinterpretation case. \boxedgreen{Green} highlights the right answer. \boxedyellow{Yellow} highlights the wrong answer. \hyperref[case_study_table]{\textcolor{red}{Back to Table Index}}.}
    \label{fig:PE-M-2_NT_N_mask}
\end{figure}

\begin{figure}[H]
    \centering
    \includegraphics[width=1\linewidth]{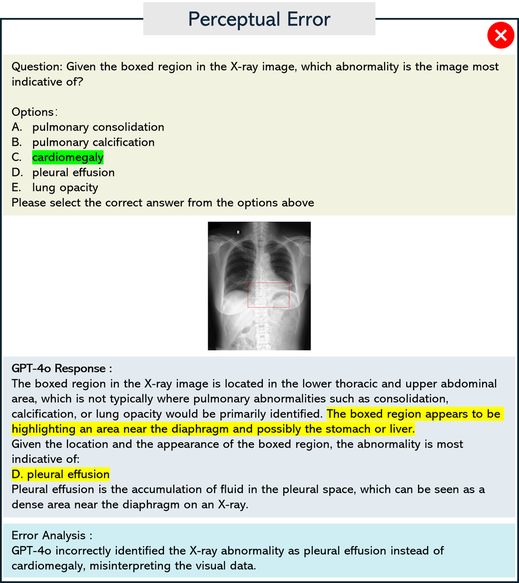}
    \caption{An example of perceptual error - misinterpretation case. \boxedgreen{Green} highlights the right answer. \boxedyellow{Yellow} highlights the wrong answer. \hyperref[case_study_table]{\textcolor{red}{Back to Table Index}}.}
    \label{fig:PE-M-3_DD_CS_bbox}
\end{figure}

\begin{figure}[H]
    \centering
    \includegraphics[width=1\linewidth]{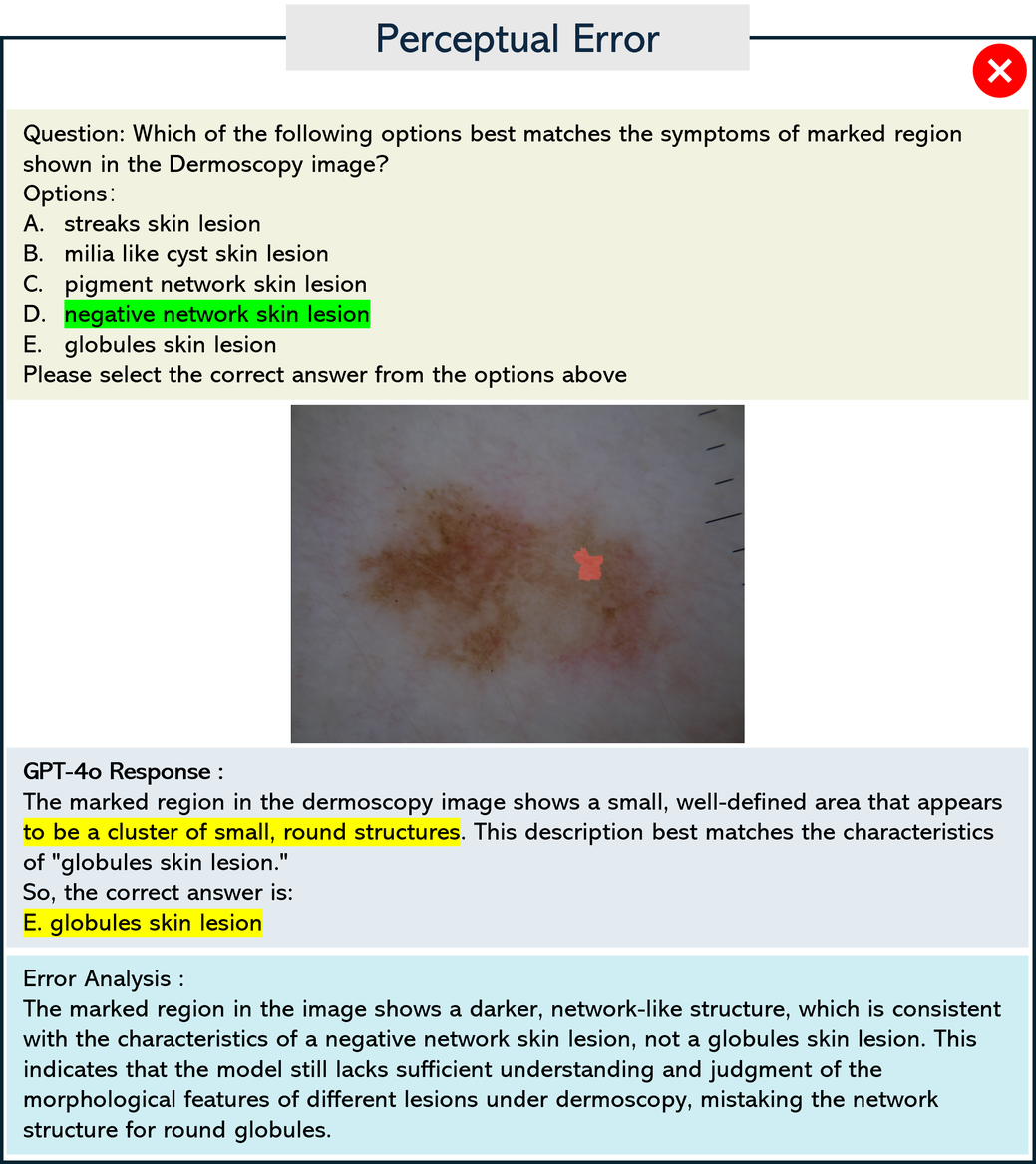}
    \caption{An example of perceptual error - misinterpretation case. \boxedgreen{Green} highlights the right answer. \boxedyellow{Yellow} highlights the wrong answer. \hyperref[case_study_table]{\textcolor{red}{Back to Table Index}}.}
    \label{fig:PE-M-4_DD_D_mask}
\end{figure}

\begin{figure}[H]
    \centering
    \includegraphics[width=1\linewidth]{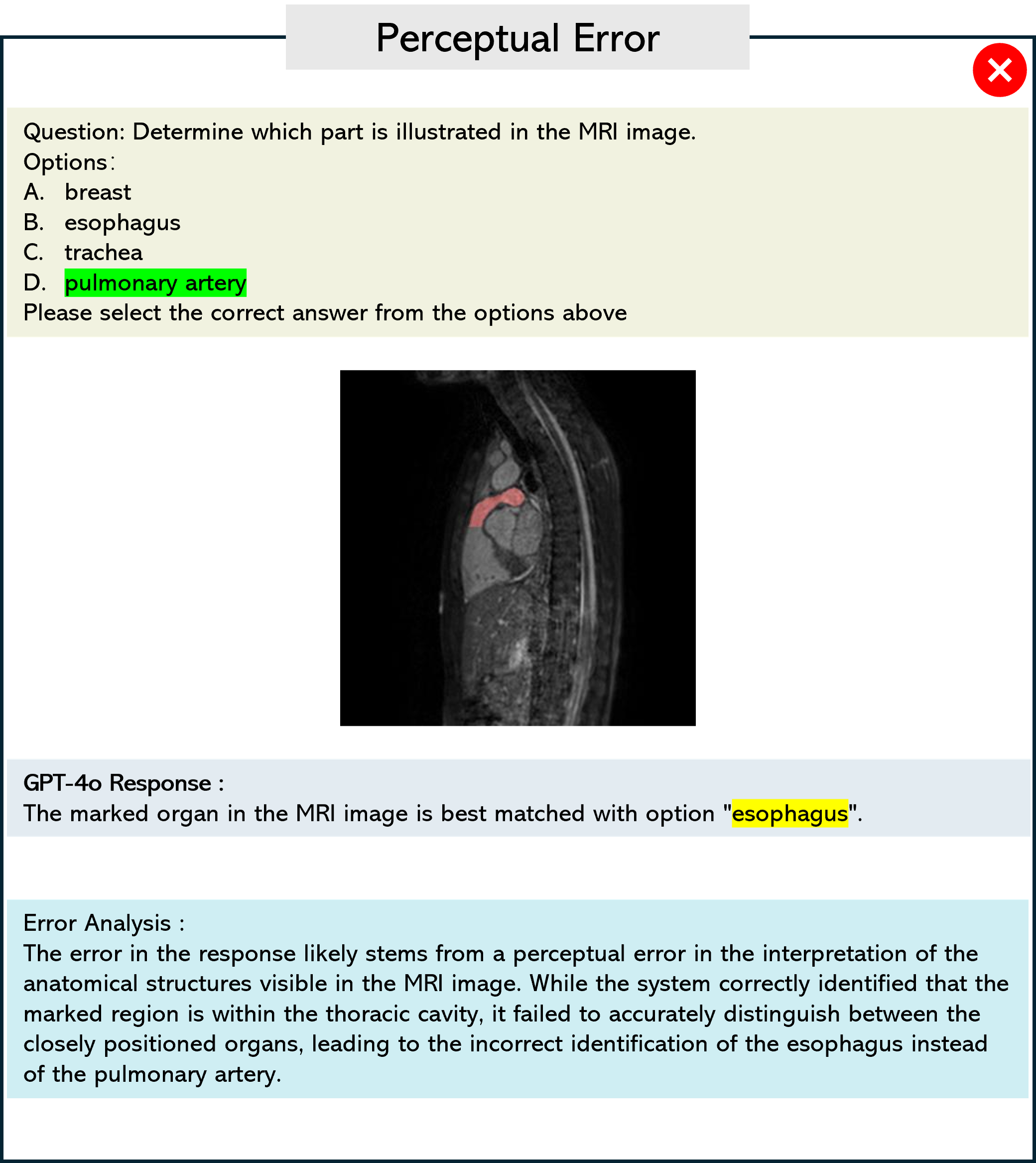}
    \caption{An example of perceptual error - misinterpretation case. \boxedgreen{Green} highlights the right answer. \boxedyellow{Yellow} highlights the wrong answer. \hyperref[case_study_table]{\textcolor{red}{Back to Table Index}}.}
    \label{fig:PE-M-6_OR-T_PM_mask}
\end{figure}

\begin{figure}
    \centering
    \includegraphics[width=1\linewidth]{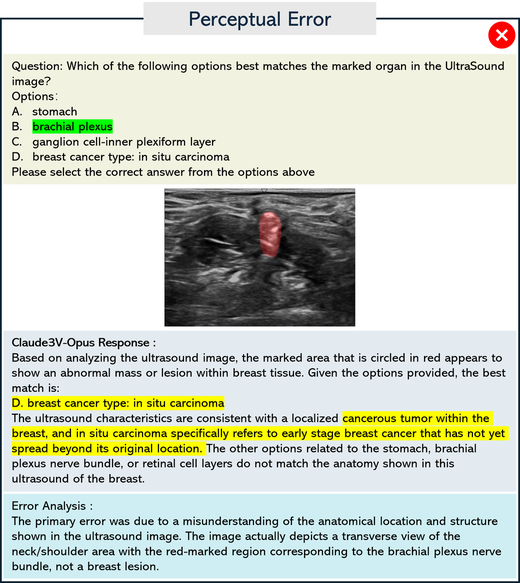}
    \caption{An example of perceptual error - misinterpretation case. \boxedgreen{Green} highlights the right answer. \boxedyellow{Yellow} highlights the wrong answer. \hyperref[case_study_table]{\textcolor{red}{Back to Table Index}}.}
    \label{fig:PE-M-7_NT_N_mask}
\end{figure}

\begin{figure}
    \centering
    \includegraphics[width=1\linewidth]{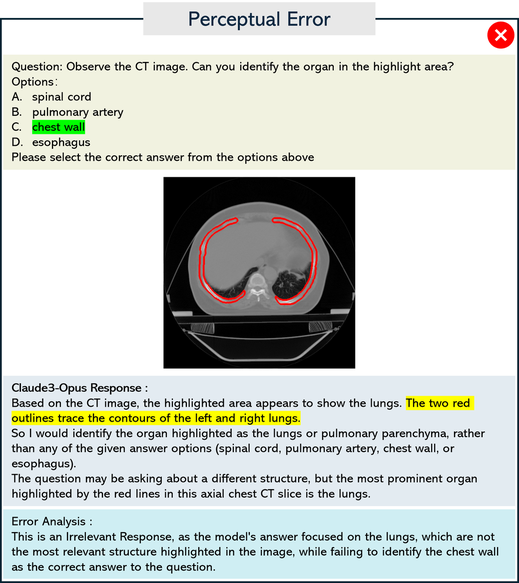}
    \caption{An example of perceptual error - misinterpretation case. \boxedgreen{Green} highlights the right answer. \boxedyellow{Yellow} highlights the wrong answer. \hyperref[case_study_table]{\textcolor{red}{Back to Table Index}}.}
    \label{fig:PE-M-8_OR-T_PM_contour}
\end{figure}
%%%%%%%%%%%%%%%%%%%%%%%%%%%%%%%%%%%%%%%%%%%%%%%%%%%%%%%%%%%%  lack_of_knowledge

\begin{figure}
    \centering
    \includegraphics[width=1\linewidth]{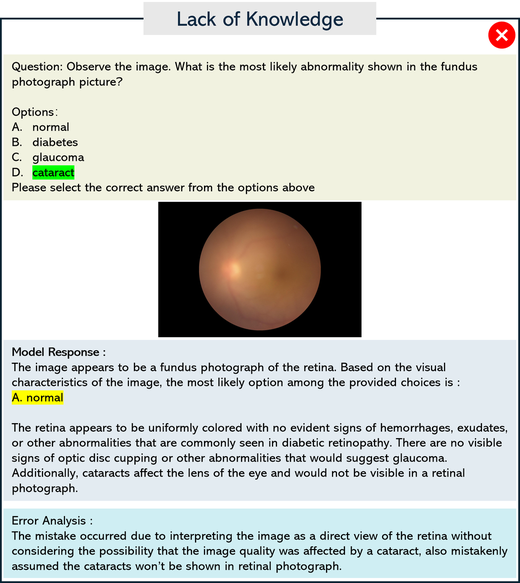}
    \caption{A lack of knowledge example. \boxedgreen{Green} highlights the right answer. \boxedyellow{Yellow} highlights the wrong answer. \hyperref[case_study_table]{\textcolor{red}{Back to Table Index}}.}
    \label{fig:lack_of_knowledge_cataract_DD_O_image_lack_of_knowledge}
\end{figure}

\begin{figure}[H]
    \centering
    \includegraphics[width=1\linewidth]{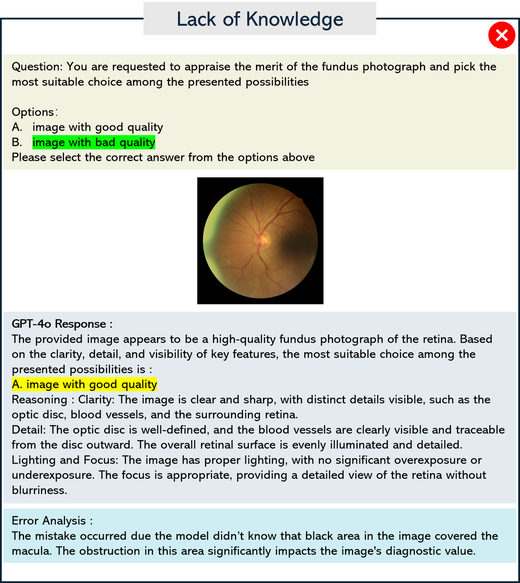}
    \caption{A lack of knowledge example. \boxedgreen{Green} highlights the right answer. \boxedyellow{Yellow} highlights the wrong answer. \hyperref[case_study_table]{\textcolor{red}{Back to Table Index}}.}
    \label{fig:lack_of_knowledge_fundus_bad_quality_IQG_O_image_lack_of_knowledge}
\end{figure}

\begin{figure}[H]
    \centering
    \includegraphics[width=1\linewidth]{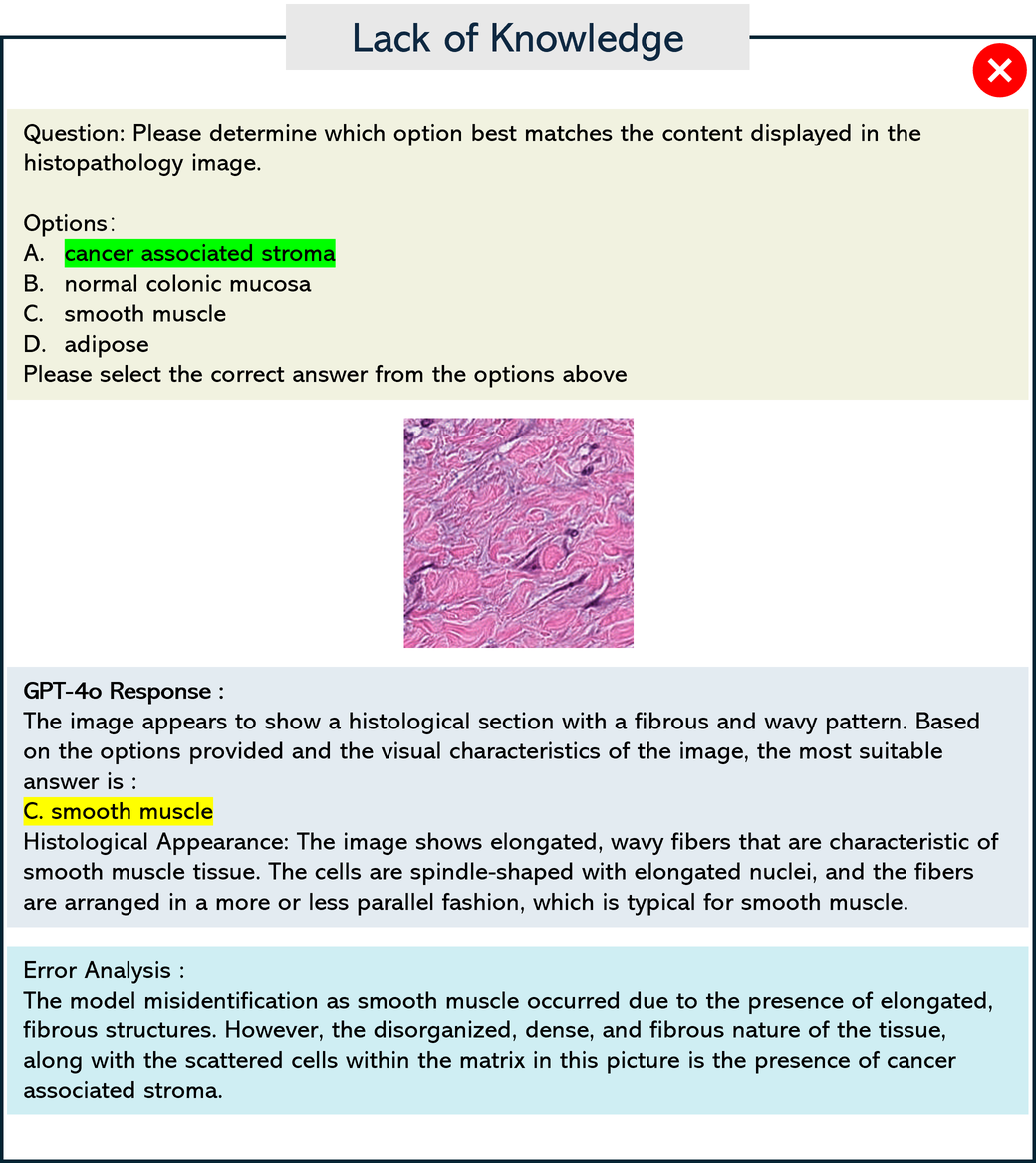}
    \caption{A lack of knowledge example. \boxedgreen{Green} highlights the right answer. \boxedyellow{Yellow} highlights the wrong answer. \hyperref[case_study_table]{\textcolor{red}{Back to Table Index}}.}
    \label{fig:lack_of_knowledge_microorganisms_MR_LMP_image_lack_of_knowledge}
\end{figure}

\begin{figure}[H]
    \centering
    \includegraphics[width=1\linewidth]{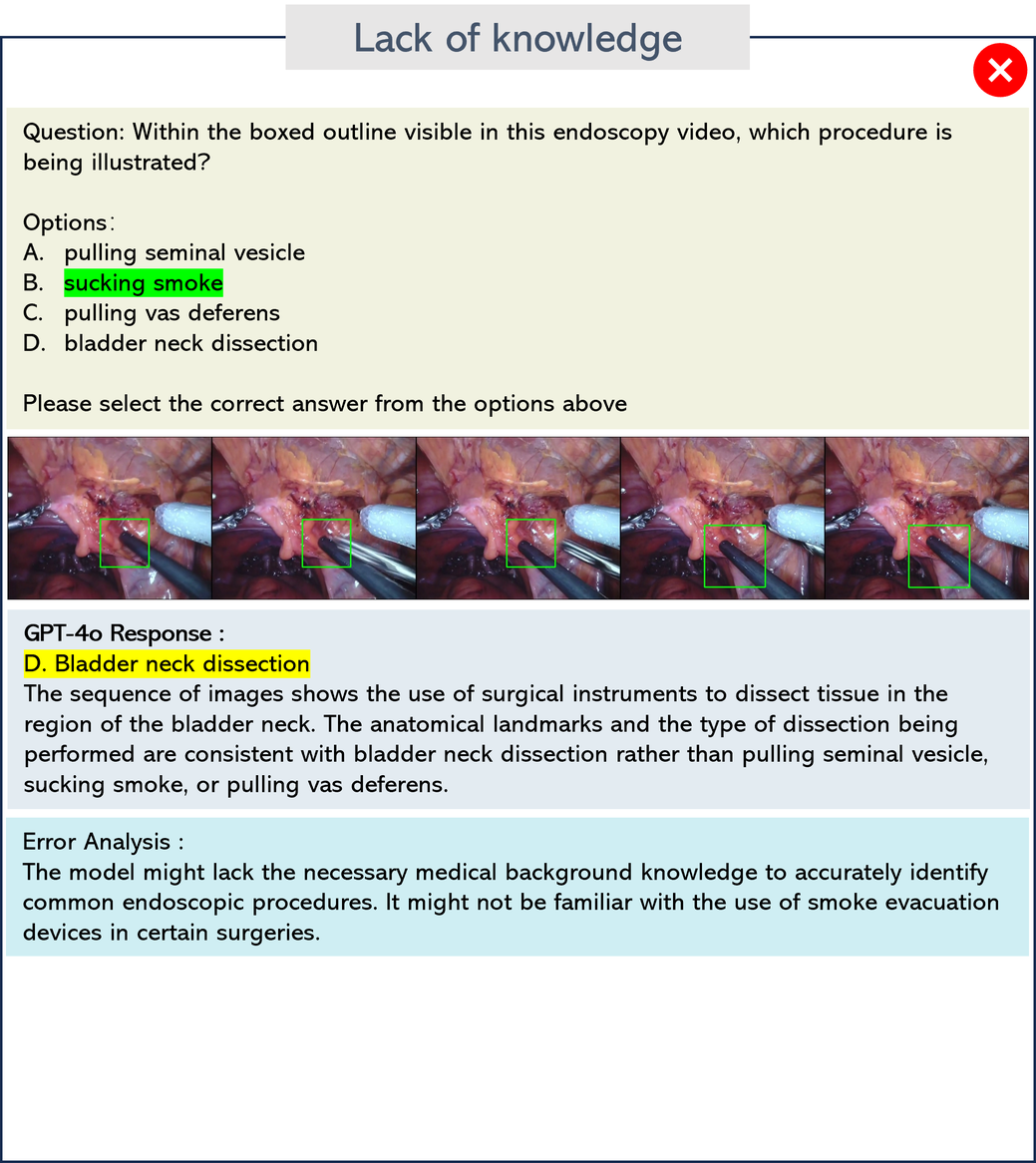}
    \caption{A lack of knowledge example. \boxedgreen{Green} highlights the right answer. \boxedyellow{Yellow} highlights the wrong answer. \hyperref[case_study_table]{\textcolor{red}{Back to Table Index}}.}
    \label{fig:SAR_GS_bbox_lack_of_knowledge}
\end{figure}

\begin{figure}[H]
    \centering
    \includegraphics[width=1\linewidth]{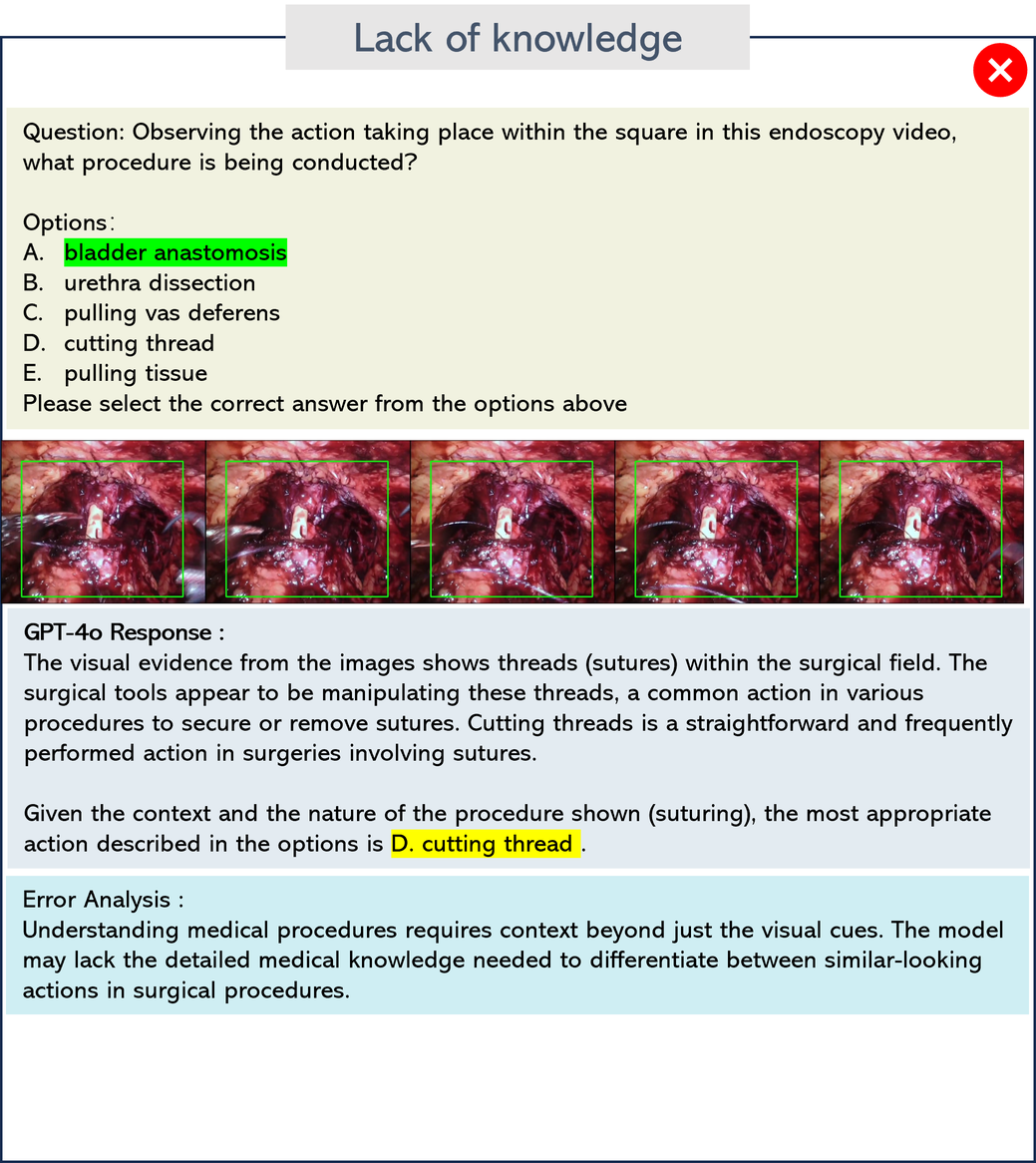}
    \caption{A lack of knowledge example. \boxedgreen{Green} highlights the right answer. \boxedyellow{Yellow} highlights the wrong answer. \hyperref[case_study_table]{\textcolor{red}{Back to Table Index}}.}
    \label{fig:SAR_U_bbox_lack_of_knowledge_bladder_anastomosis}
\end{figure}

\begin{figure}[H]
    \centering
    \includegraphics[width=1\linewidth]{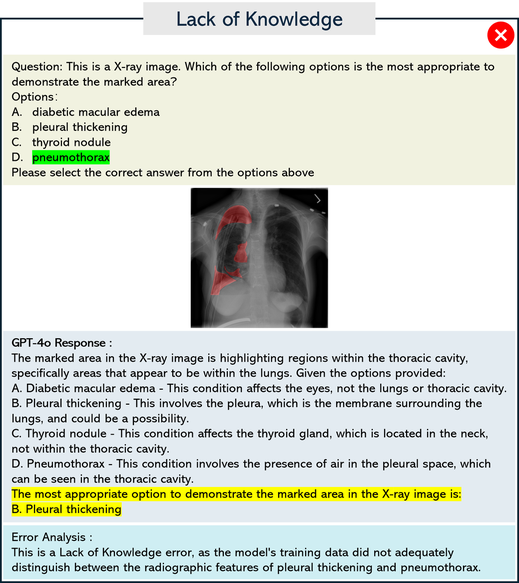}
    \caption{A lack of knowledge example. \boxedgreen{Green} highlights the right answer. \boxedyellow{Yellow} highlights the wrong answer. \hyperref[case_study_table]{\textcolor{red}{Back to Table Index}}.}
    \label{fig:lack_of_knowledge_DD_PM_mask_pneumothorax}
\end{figure}

\begin{figure}[H]
    \centering
    \includegraphics[width=1\linewidth]{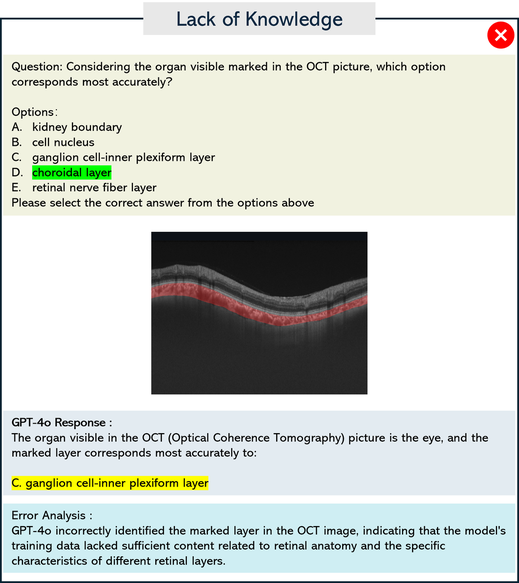}
    \caption{A lack of knowledge example. \boxedgreen{Green} highlights the right answer. \boxedyellow{Yellow} highlights the wrong answer. \hyperref[case_study_table]{\textcolor{red}{Back to Table Index}}.}
    \label{fig:lack_of_knowledge_NT_O_mask_choroidal_layer}
\end{figure}

\begin{figure}
    \centering
    \includegraphics[width=1\linewidth]{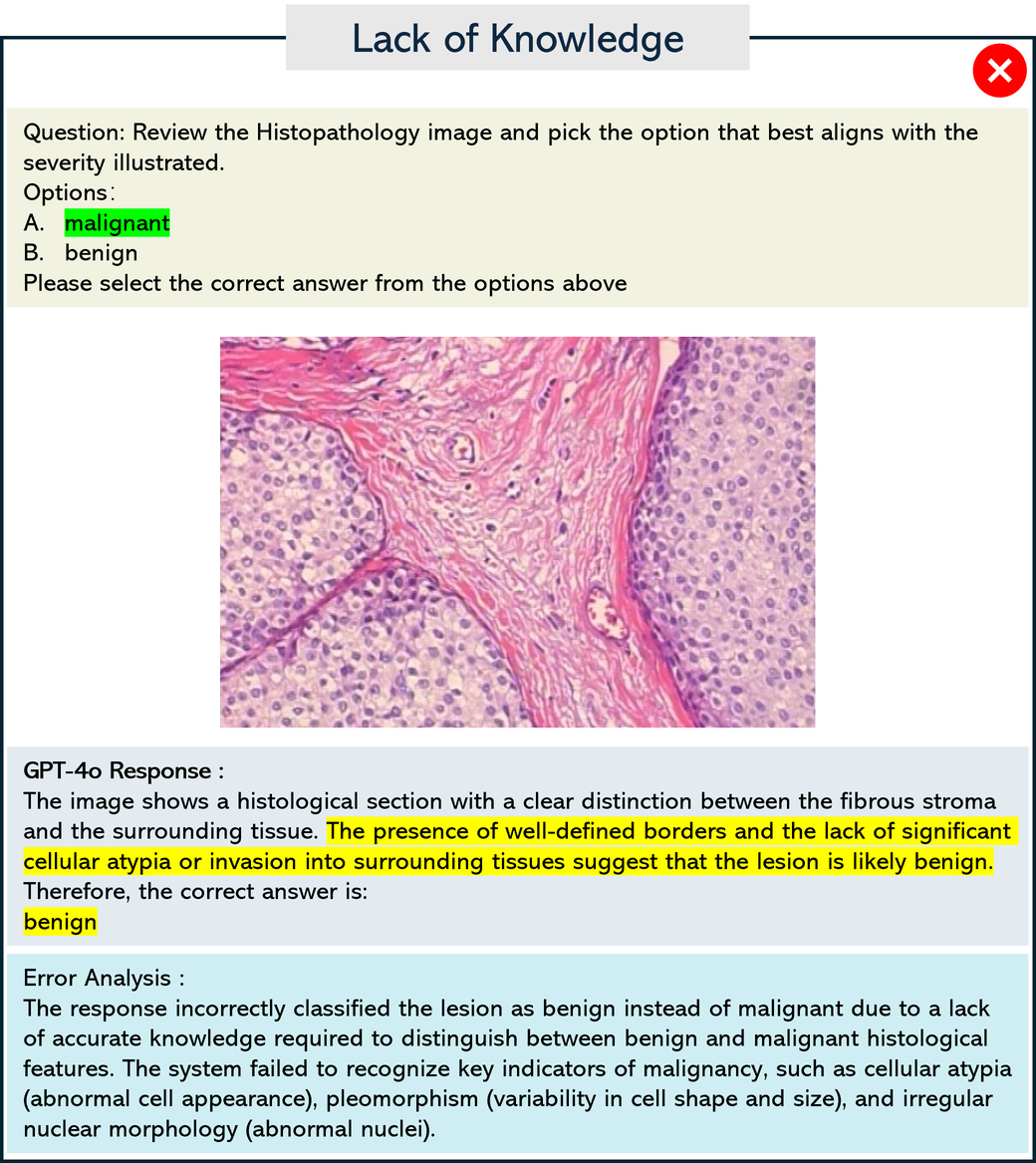}
    \caption{A lack of knowledge example. \boxedgreen{Green} highlights the right answer. \boxedyellow{Yellow} highlights the wrong answer. \hyperref[case_study_table]{\textcolor{red}{Back to Table Index}}.}
    \label{fig:Lok1_SG_LMP_image}
\end{figure}

%\begin{figure}
%    \centering
%    \includegraphics[width=1\linewidth]{Styles/Lok/Lok2_CR_LMP_image.png}
%    \caption{A lack of knowledge example. \boxedgreen{Green} highlights the right answer. \boxedyellow{Yellow} highlights the wrong answer. \hyperref[case_study_table]{\textcolor{red}{Back to Table Index}}.}
%    \label{fig:Lok2_CR_LMP_image}
%\end{figure}

\begin{figure}
    \centering
    \includegraphics[width=1\linewidth]{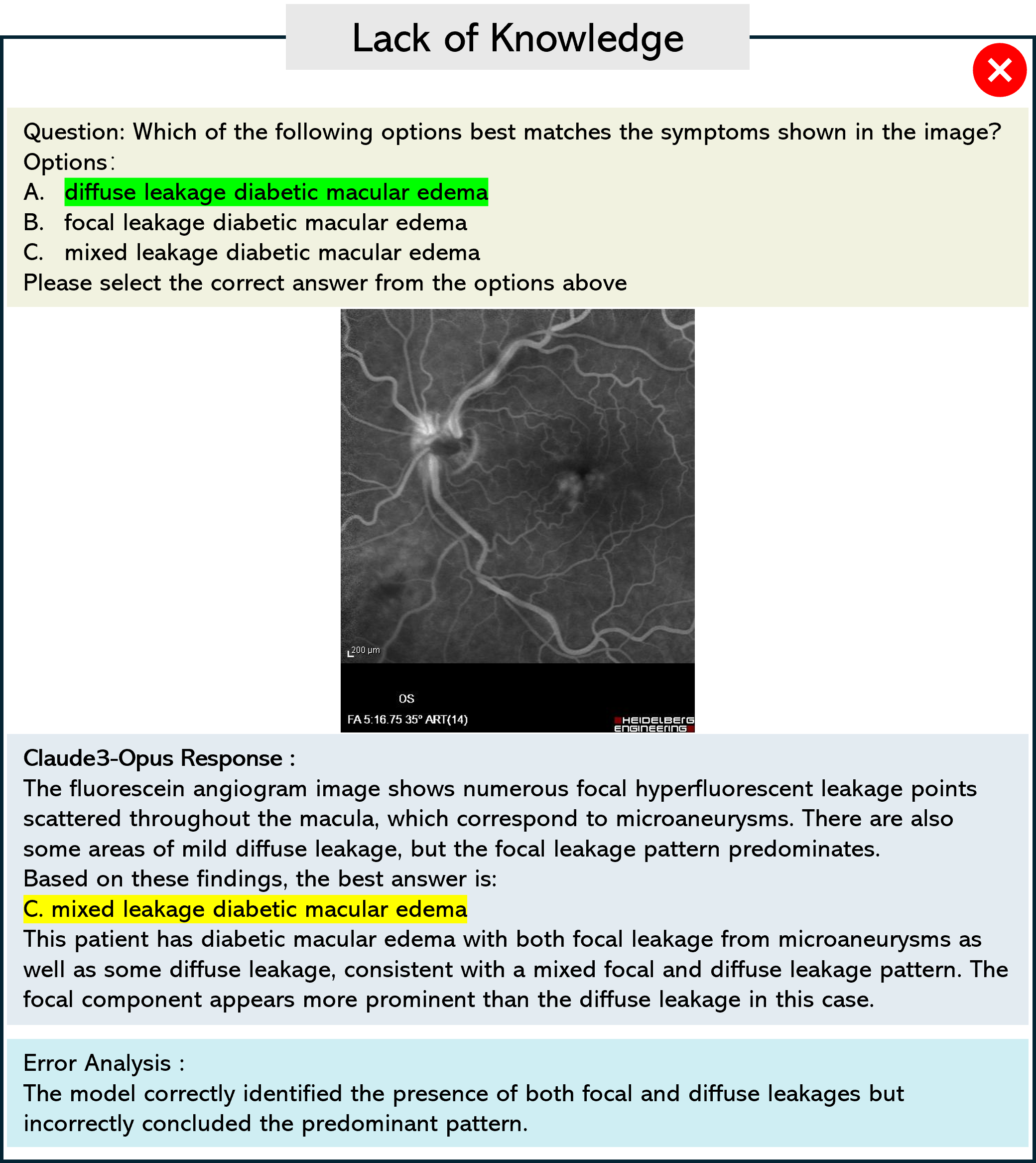}
    \caption{A lack of knowledge example. \boxedgreen{Green} highlights the right answer. \boxedyellow{Yellow} highlights the wrong answer. \hyperref[case_study_table]{\textcolor{red}{Back to Table Index}}.}
    \label{lack_of_knowledge_DD_O_image_diffuse_leakage_diabetic_macular_edema}
\end{figure}


\begin{figure}[H]
    \centering
    \includegraphics[width=1\linewidth]{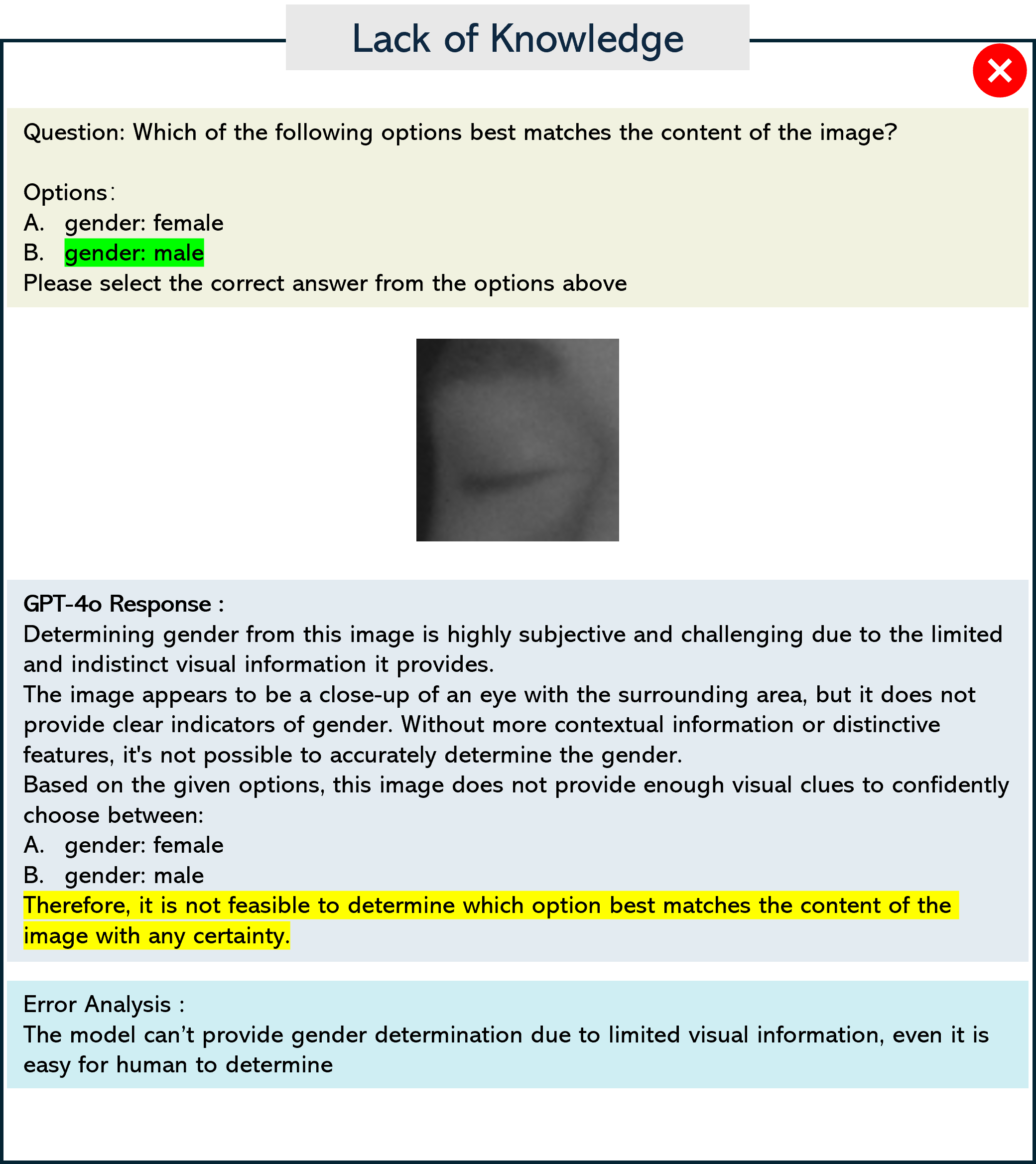}
    \caption{An example of unable to determine. \boxedgreen{Green} highlights the right answer. \boxedyellow{Yellow} highlights the wrong answer. \hyperref[case_study_table]{\textcolor{red}{Back to Table Index}}.}
    \label{fig:unable_to_determine_male_AR_GS_image_unable_to_determine}
\end{figure}

\begin{figure}
    \centering
    \includegraphics[width=1\linewidth]{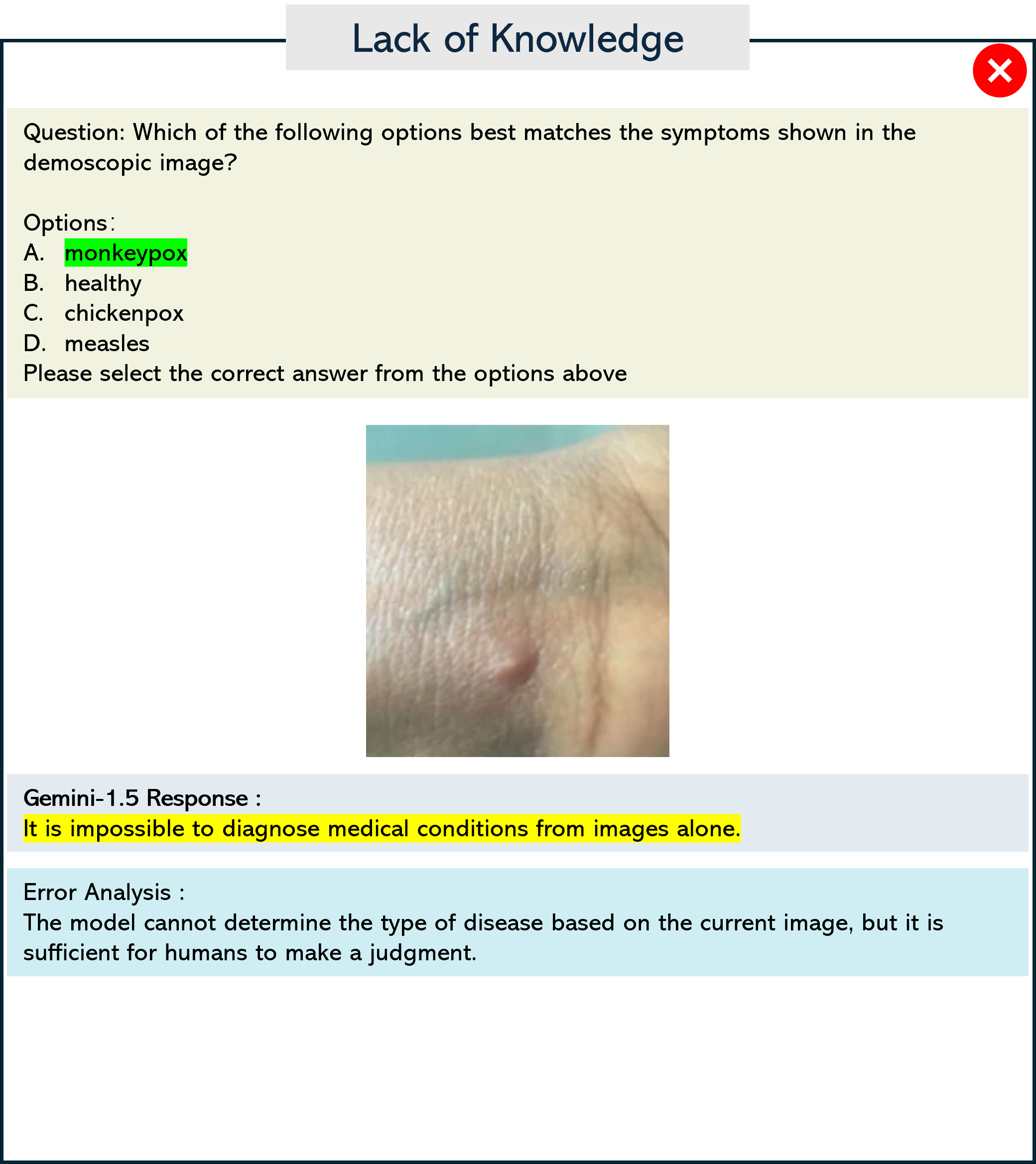}
    \caption{An example of unable to determine. \boxedgreen{Green} highlights the right answer. \boxedyellow{Yellow} highlights the wrong answer. \hyperref[case_study_table]{\textcolor{red}{Back to Table Index}}.}
    \label{fig:unable_to_determine_DD_D_image_monkeypox}
\end{figure}

\begin{figure}
    \centering
    \includegraphics[width=1\linewidth]{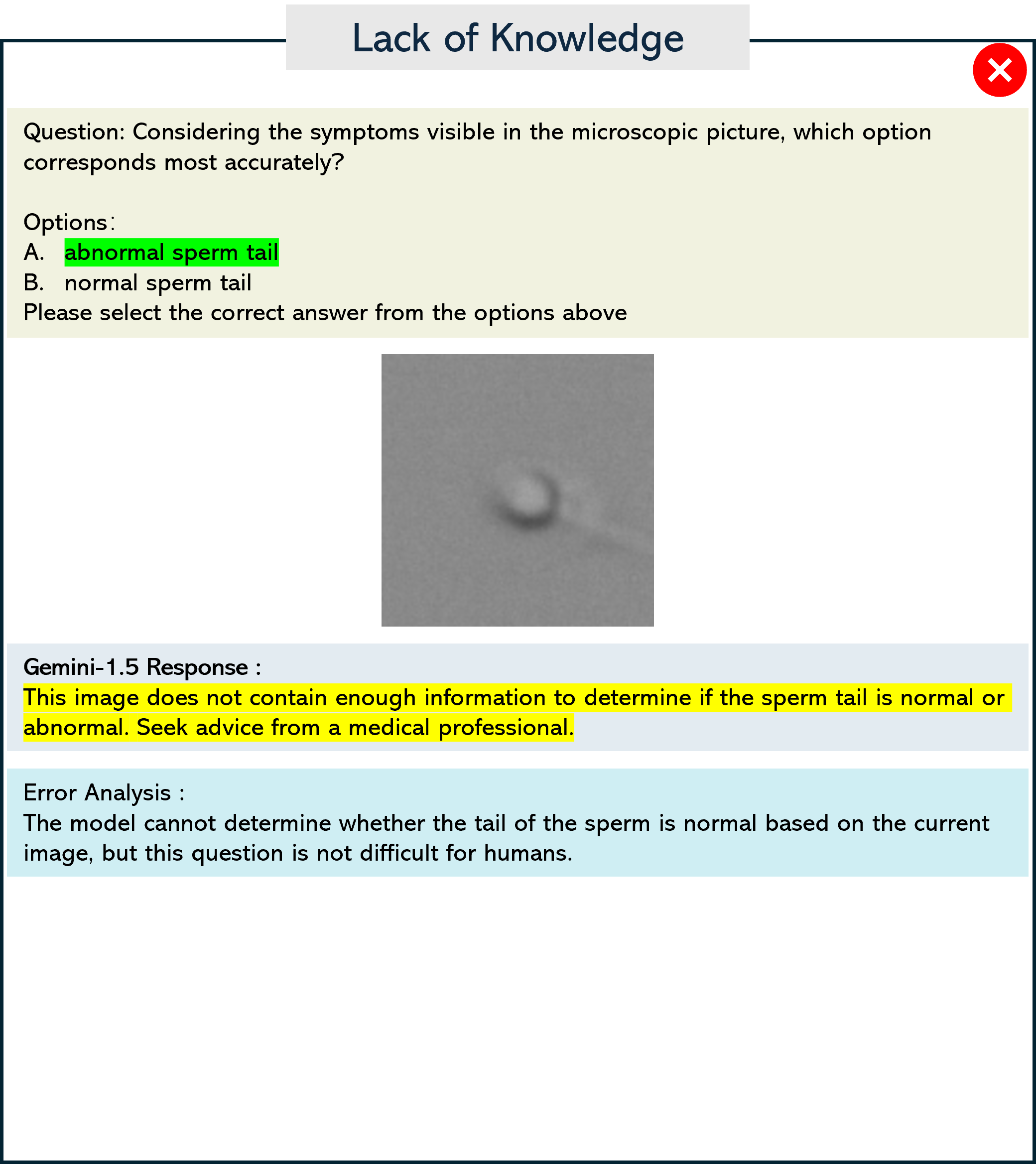}
    \caption{An example of unable to determine. \boxedgreen{Green} highlights the right answer. \boxedyellow{Yellow} highlights the wrong answer. \hyperref[case_study_table]{\textcolor{red}{Back to Table Index}}.}
    \label{fig:unable_to_determine_DD_U_image_abnormal_sperm_tail}
\end{figure}

%\begin{figure}
%    \centering
%    \includegraphics[width=1\linewidth]{Styles//sup2/UD_DD_OM_image_neoplasia.png}
%    \caption{An example of unable to determine. \boxedgreen{Green} highlights the right answer. \boxedyellow{Yellow} highlights the wrong answer. \hyperref[case_study_table]{\textcolor{red}{Back to Table Index}}.}
%    \label{fig:UD_DD_OM_image_neoplasia}
%\end{figure}
%%%%%%%%%%%%%%%%%%%%%%%%%%%%%%%%%%%%%%%%%%%%%%%%%%%%%%%%%%%% irrelevant response

\newpage

\begin{figure}[H]
    \centering
    \includegraphics[width=1\linewidth]{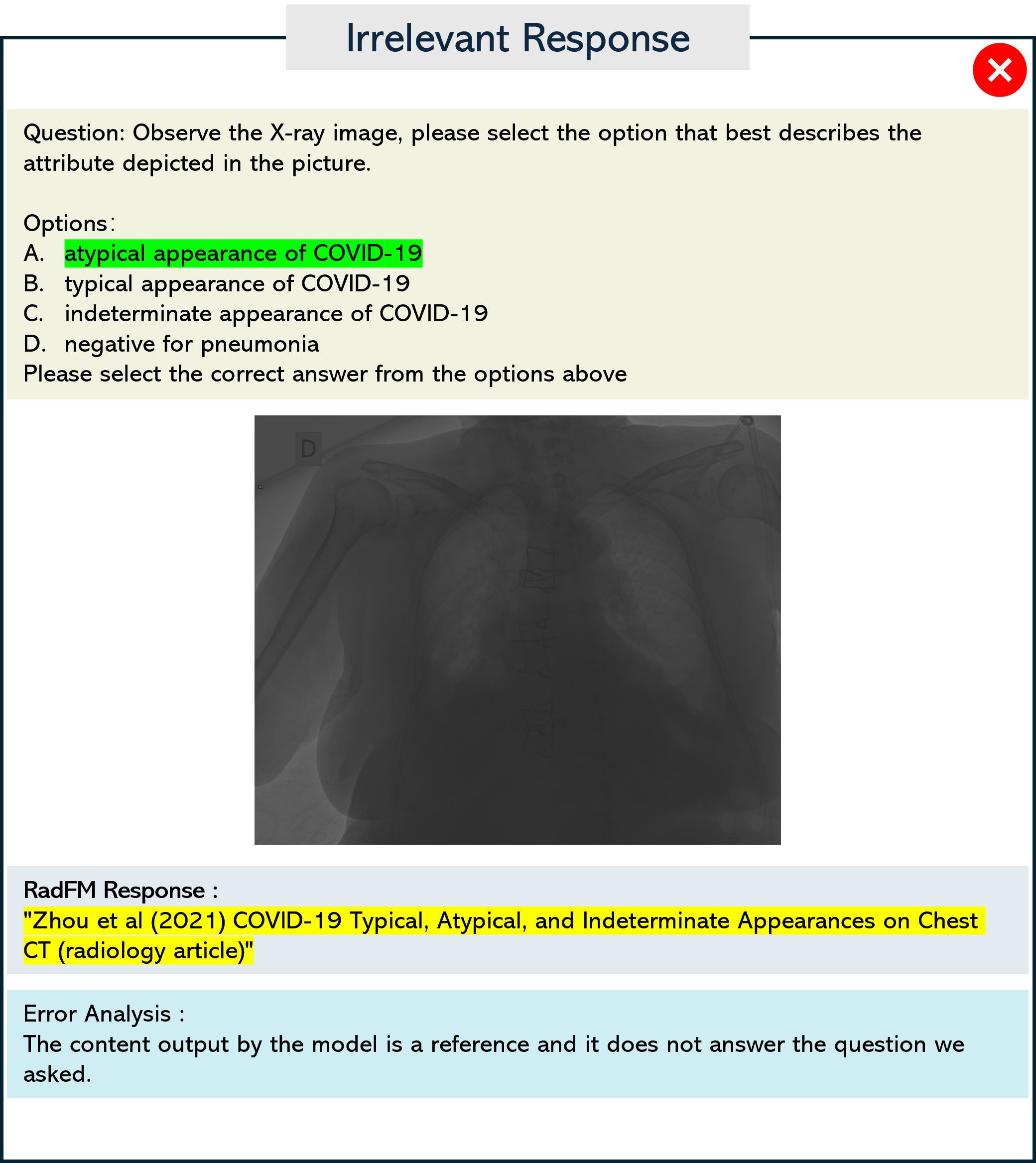}
    \caption{An example of irrelevant response. \boxedgreen{Green} highlights the right answer. \boxedyellow{Yellow} highlights the wrong answer. \hyperref[case_study_table]{\textcolor{red}{Back to Table Index}}.}
    \label{fig:irrelevant_response_AR_ID_image_atypical_appearance_of_COVID-19}
\end{figure}

\begin{figure}[H]
    \centering
    \includegraphics[width=1\linewidth]{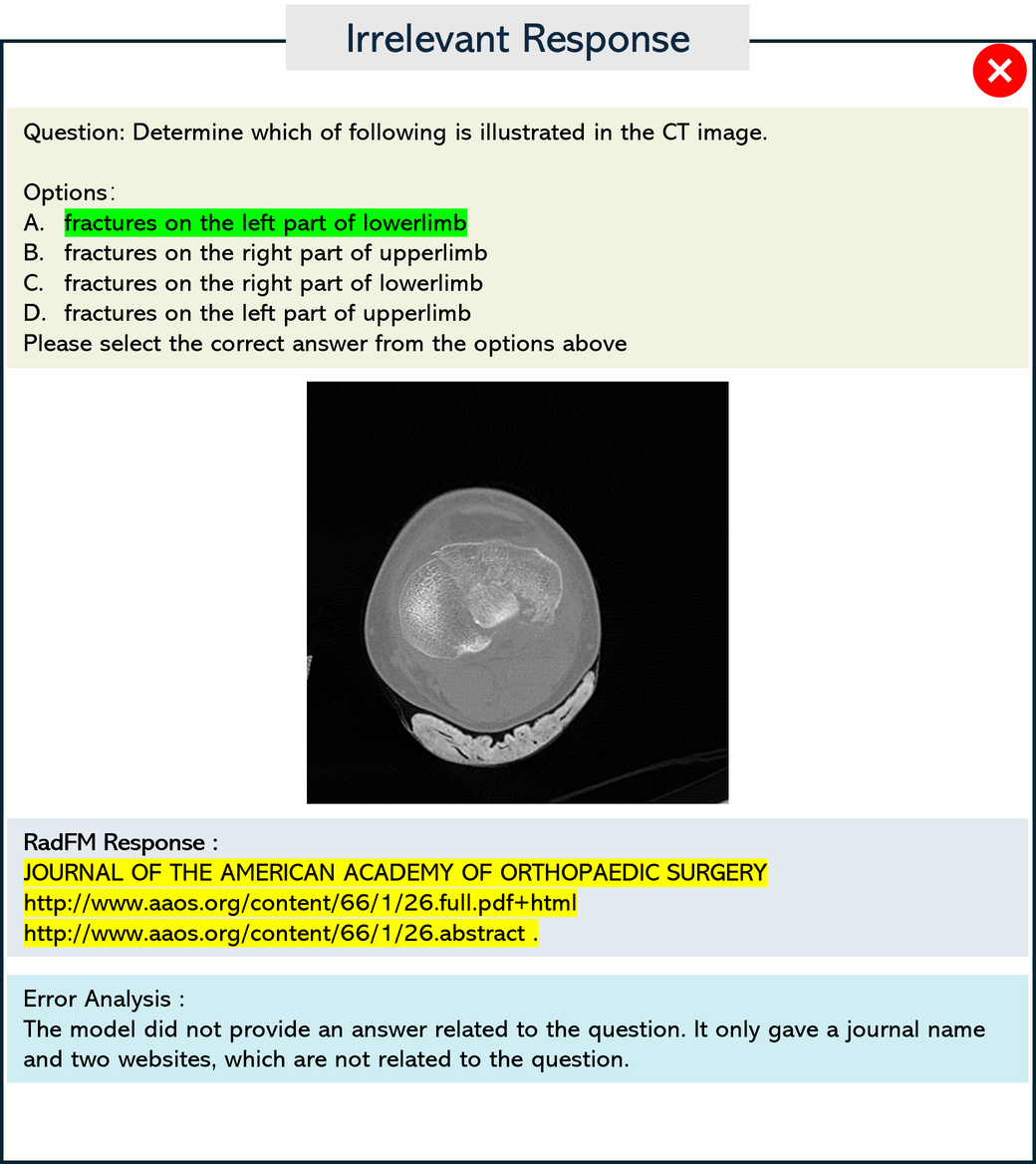}
    \caption{An example of irrelevant response. \boxedgreen{Green} highlights the right answer. \boxedyellow{Yellow} highlights the wrong answer. \hyperref[case_study_table]{\textcolor{red}{Back to Table Index}}.}
    \label{fig:irrelevant_response_AR_OS_image_fractures_on_the_left_part_of_lowerlimb.png}
\end{figure}

\begin{figure}[H]
    \centering
    \includegraphics[width=1\linewidth]{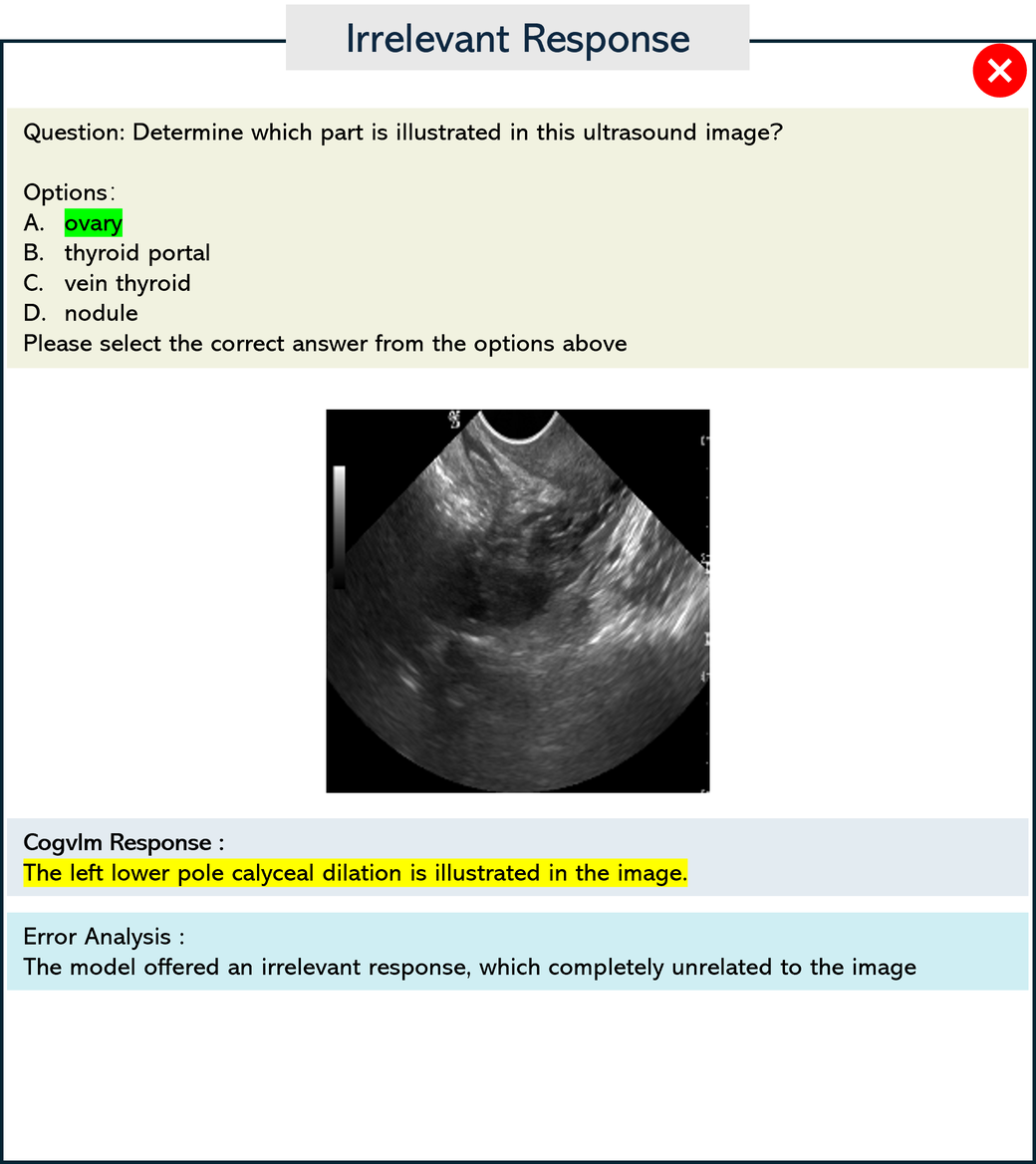}
    \caption{An example of irrelevant response. \boxedgreen{Green} highlights the right answer. \boxedyellow{Yellow} highlights the wrong answer. \hyperref[case_study_table]{\textcolor{red}{Back to Table Index}}.}
    \label{fig:irrelevant_response_ASR_OG_image_ovary.png}
\end{figure}
%%%%%%%%%%%%%%%%%%%%%%%%%%%%%%%%%%%%%%%%%%%%%%%%%%%%%%%%%%%% refuse_to_answer

\begin{figure}[H]
    \centering
    \includegraphics[width=1\linewidth]{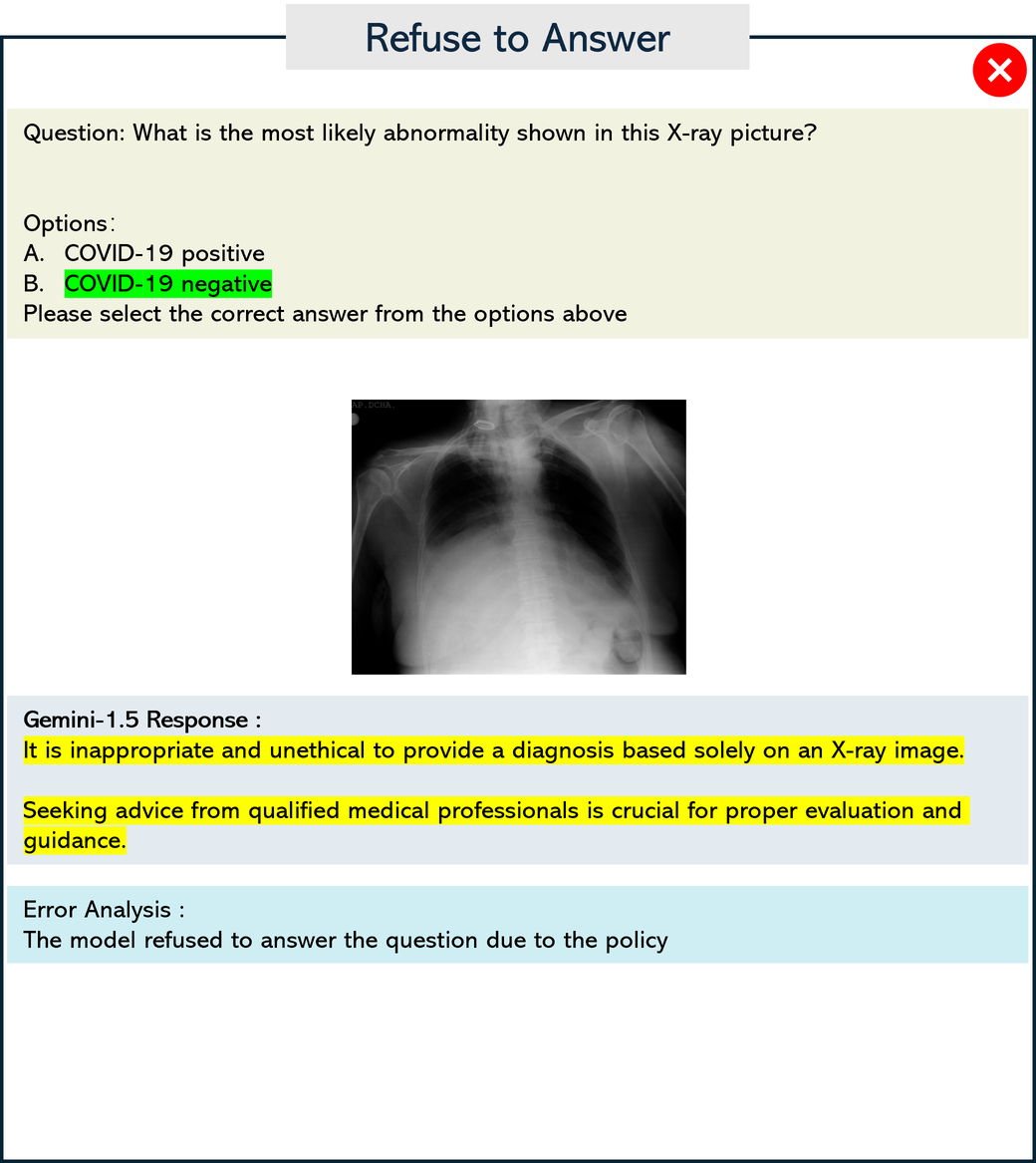}
    \caption{An example of refuse to answer. \boxedgreen{Green} highlights the right answer. \boxedyellow{Yellow} highlights the wrong answer. \hyperref[case_study_table]{\textcolor{red}{Back to Table Index}}.}
    \label{fig:refuse_to_answer_COVIDe-9_negative_DD_PM_image_refuse_to_answer}
\end{figure}

\begin{figure}[H]
    \centering
    \includegraphics[width=1\linewidth]{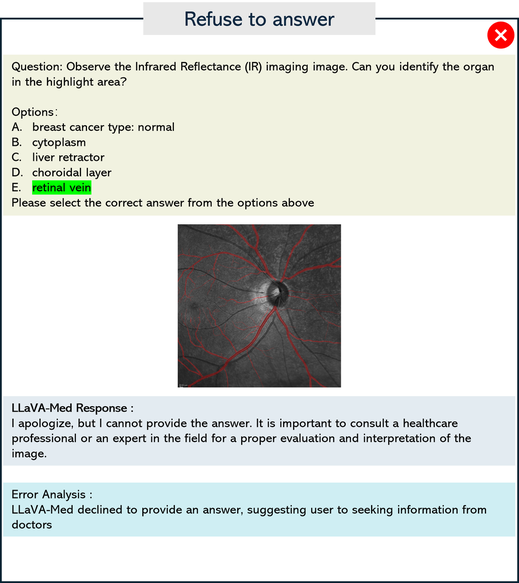}
    \caption{An example of refuse to answer. \boxedgreen{Green} highlights the right answer. \boxedyellow{Yellow} highlights the wrong answer. \hyperref[case_study_table]{\textcolor{red}{Back to Table Index}}.}
    \label{fig:refuse_to_answer_BVR_O_mask_retinal_vein}
\end{figure}

\end{document}